%% file: paper-MiLo.tex
\documentclass[showpacs,aps,superscriptaddress,letterpaper,nofootinbib]{revtex4}

\usepackage{xfrac}
\usepackage{xspace}
\usepackage{graphicx}
\usepackage{epsfig}
\usepackage{float}
\usepackage[nointegrals]{wasysym}
\usepackage{hyperref}
\usepackage{url}
\usepackage{subfig}
\usepackage{mwe}
\usepackage{amsmath}
\usepackage{xcolor}
\usepackage{amssymb,array}
\usepackage{orcidlink}

\newcommand{\be}{\begin{equation}}
\newcommand{\ee}{\end{equation}}
\newcommand{\ba}{\begin{eqnarray}}
\newcommand{\ea}{\end{eqnarray}}

\begin{document}

\vspace{0.6cm}

\title{Maximizing Returns: Optimizing Experimental Observables at the LHC}
\author{Jeffrey Davis\orcidlink{0000-0001-6488-6195}\footnote{e-mail: jdavi231@jhu.edu}}
\affiliation{Department of Physics and Astronomy, Johns Hopkins University, Baltimore, MD 21218, USA}
\author{Andrei V. Gritsan\orcidlink{0000-0002-3545-7970}\footnote{e-mail: gritsan@jhu.edu}}
\affiliation{Department of Physics and  Astronomy, Johns Hopkins University, Baltimore, MD 21218, USA}
\author{Lucas~S.~Mandacar\'{u}~Guerra\orcidlink{0000-0002-4858-0396}\footnote{e-mail: lm4604@princeton.edu}}
\affiliation{Department of Physics and Astronomy, Johns Hopkins University, Baltimore, MD 21218, USA}
\affiliation{Department of Physics, Princeton University, Princeton, NJ 08544, USA} 
\author{Lucas Kang\orcidlink{0000-0002-0941-4512}\footnote{e-mail: lkang12@jhu.edu}}
\affiliation{Department of Physics and Astronomy, Johns Hopkins University, Baltimore, MD 21218, USA}
\author{Michalis Panagiotou\orcidlink{0009-0000-0692-523X}\footnote{e-mail: mpanagi1@jh.edu}}
\affiliation{Department of Physics and Astronomy, Johns Hopkins University, Baltimore, MD 21218, USA}
\author{Jeffrey Roskes\orcidlink{0000-0001-8761-0490}\footnote{e-mail: hroskes@jhu.edu}}
\affiliation{Department of Physics and Astronomy, Johns Hopkins University, Baltimore, MD 21218, USA}
\author{Mohit Srivastav\orcidlink{0000-0003-3603-9102}\footnote{e-mail: msrivas6@jhu.edu}}
\affiliation{Department of Physics and Astronomy, Johns Hopkins University, Baltimore, MD 21218, USA}

\date{January 15, 2026}

\begin{abstract}
\vspace{2mm}
\noindent 
We introduce a framework that integrates both analytical and machine-learning approaches for calculating 
observables optimal for EFT and broader applications at the LHC. 
A new metric for evaluating the performance of these approaches has been introduced.
In addition, we demonstrate how the majority of relevant information can be effectively stored in a limited 
number of bins, allowing for efficient data analysis, data preservation, and global data combination, 
while also providing tools to achieve these benefits.
A key feature of this approach is the reduction in the dimensionality of the observable information, 
which enhances both the effectiveness and practicality of the data analysis while maximizing gains 
within limited resources. These features have been demonstrated through simulated analyses 
of the Higgs boson production and decay processes at the~LHC.
\end{abstract}


\maketitle
\small
\tableofcontents
\normalsize
\clearpage 


\section{Introduction}
\label{sect:input_intro}

\noindent The program of the Large Hadron Collider (LHC), featuring the four major particle experiments
ATLAS, CMS, LHCb, and ALICE, seeks to explore the fundamental processes that occur during 
the collisions of partons, gluons or quarks, within the two colliding protons.
A prominent example of these investigations is the discovery of the $H$ (Higgs) boson~\cite{CMS:2012qbp,ATLAS:2012yve}.
Understanding the laws of interactions of fundamental particles, 
whether described by the Standard Model (SM) of particle physics 
or theories beyond it (BSM), is of significant importance. Optimizing experimental techniques 
for analyzing LHC data will maximize the returns on this significant investment.

In this paper, we provide an overview of our experience in developing an optimal sequence of  data analysis approaches 
using the event generator and likelihood-driven approaches, specifically {\tt JHUGen} and {\tt MELA} 
framework~\cite{Gao:2010qx,Bolognesi:2012mm,Anderson:2013afp,Gritsan:2016hjl,Gritsan:2020pib,Martini:2021uey,Davis:2021tiv}.
This process spans from the initial stages of $H$ boson discovery and property characterization
to the subsequent Effective Field Theory (EFT)~\cite{Georgi:1993mps,Weinberg:1995mt} analysis of its 
interactions~\cite{Nelson:1986ki,Soni:1993jc,Plehn:2001nj,Choi:2002jk,Buszello:2002uu,Hankele:2006ma,Accomando:2006ga,
Godbole:2007cn,Hagiwara:2009wt,Gao:2010qx,DeRujula:2010ys,Christensen:2010pf,Bolognesi:2012mm,Ellis:2012xd,Chen:2012jy,
Artoisenet:2013puc,Anderson:2013afp,Chen:2013waa,Maltoni:2013sma,
Azatov:2014jga,Cacciapaglia:2014rla,Denner:2014cla,Dolan:2014upa,Englert:2014ffa,Gonzalez-Alonso:2014eva,
Ballestrero:2015jca,Greljo:2015sla,Hespel:2015zea,Kauer:2015dma,Kauer:2015hia,Kilian:2015opv,Mimasu:2015nqa,
Degrande:2016dqg,Dwivedi:2016xwm,Gritsan:2016hjl,deFlorian:2016spz,Azatov:2016xik,
Denner:2017vms,Deutschmann:2017qum,Greljo:2017spw,Goncalves:2017gzy,Jager:2017owh,
Brass:2018hfw,Gomez-Ambrosio:2018pnl,Goncalves:2018pkt,Harlander:2018yns,Harlander:2018yio,Lee:2018fxj,Kalinowski:2018oxd,Perez:2018kav,
Jaquier:2019bfs,Denner:2019fcr,Banerjee:2019twi,
Gritsan:2020pib,Martini:2021uey,Davis:2021tiv}.
We also introduce several new approaches for data analysis that may have broader applications,
utilizing simulation, analytical calculations, and machine-learning techniques.
The EFT framework is used to capture the observable effects 
of higher-energy BSM phenomena that cannot be directly detected at the LHC, 
and it serves as an illustrative example for our data analysis objectives.

\begin{figure}[b!]
  \begin{center}
    \captionsetup{justification=centerlast}
    \includegraphics[width=0.45\linewidth]{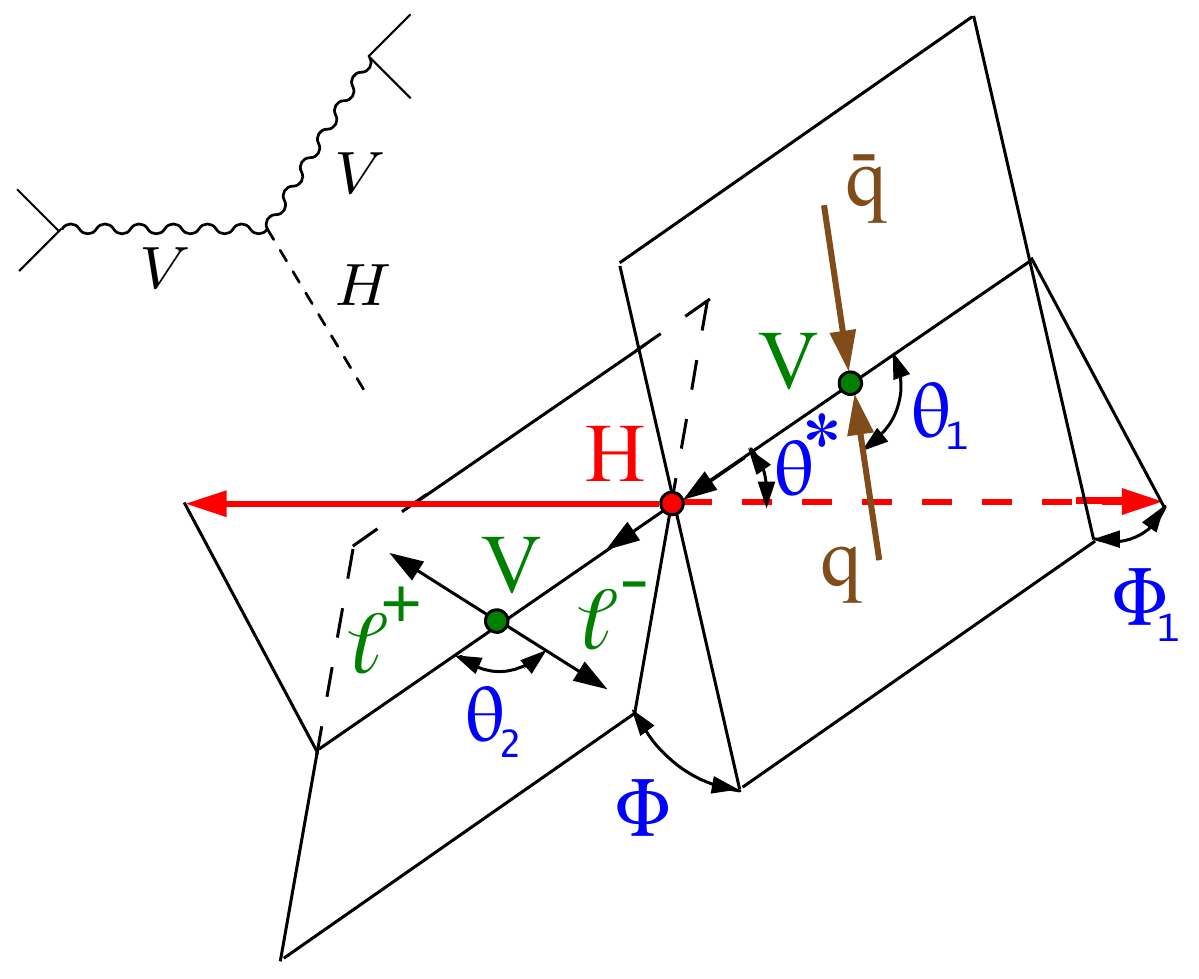}
    ~~~~~
    \includegraphics[width=0.45\linewidth]{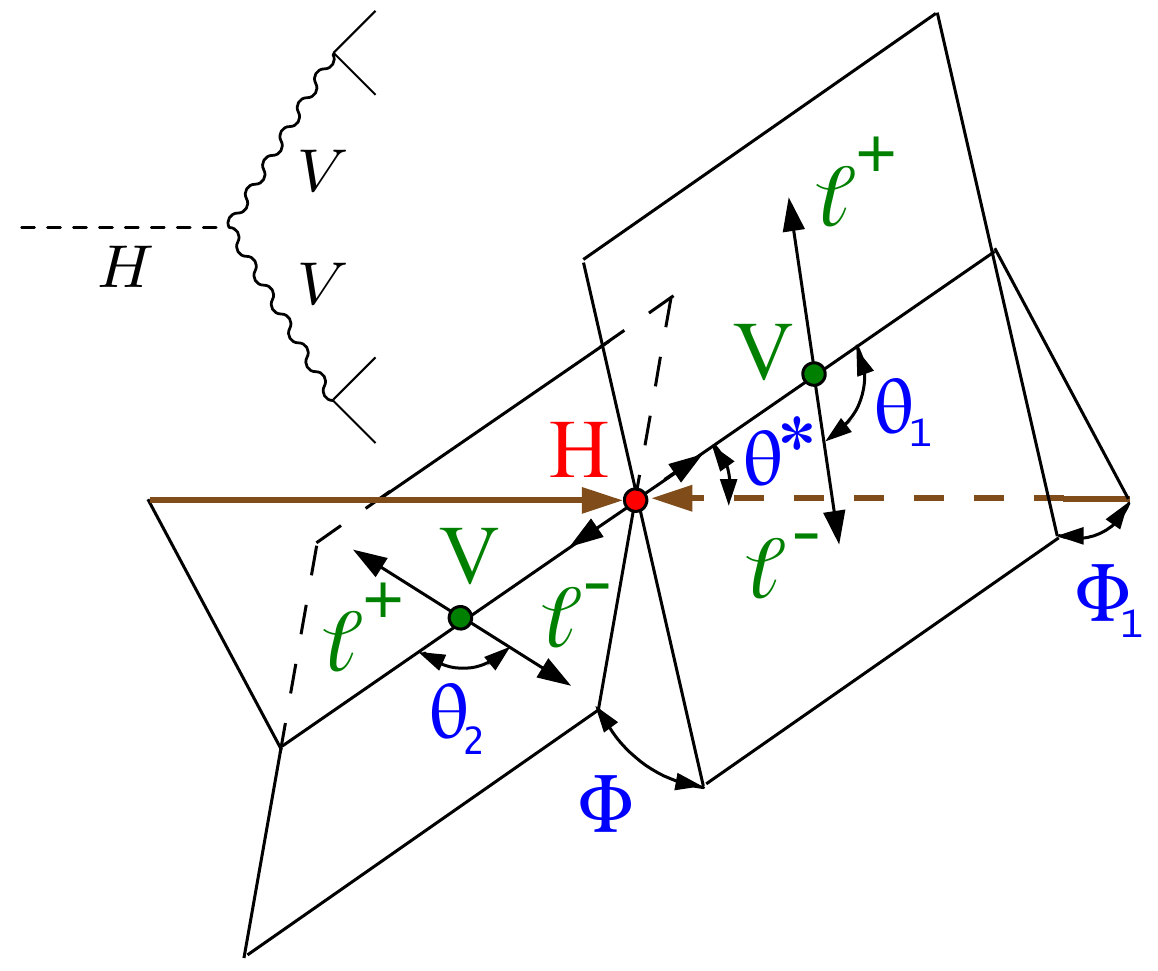}
    \caption{
Diagrams illustrating the $H$ boson production and decay processes at the LHC:
$q\bar{q}\to (Z/\gamma^*) \to H(Z/\gamma^*) \to H(\ell^+\ell^-)$ and $H\to(Z/\gamma^*)(Z/\gamma^*) \to 4\ell$
with angular observables defined in the rest frames of $H$ and 
$V=Z$\,or\,$\gamma^*$~\cite{Gao:2010qx,Bolognesi:2012mm,Anderson:2013afp}. 
     }
    \label{fig:process}
  \end{center}
\end{figure}

Two specific $H$ boson production and decay processes occurring at the LHC are introduced as examples, 
but the discussion applies to a much broader range of scenarios.
In one case, we examine the Higgs-strahlung process
$q\bar{q}\to (Z/\gamma^*) \to H(Z/\gamma^*) \to H(\ell^+\ell^-)$,
as illustrated in the left diagrams of Fig.~\ref{fig:process}~\cite{Anderson:2013afp,Gritsan:2020pib}. 
In the other case, we analyze the decay into four charged leptons $H\to(Z/\gamma^*)(Z/\gamma^*) \to 4\ell$,
as depicted in the right diagrams of Fig.~\ref{fig:process}~\cite{Gao:2010qx,Bolognesi:2012mm}. 
Both processes have been extensively studied in the literature, both experimentally at the LHC
and from a theoretical standpoint.
It is the clear and well-understood kinematical description of these processes that 
led us to use them to illustrate more general methods.
We also note that the first topology corresponds to a Higgs factory design, 
based on electron and positron beams with energies of approximately 120 GeV, 
where the dominant $H$ boson production channel is $e^+e^-\to Z\to HZ$.

We support this effort using an event generator ({\tt JHUGen}) and a library for matrix element calculations ({\tt MELA}). 
The {\tt JHUGen} and {\tt MELA} framework has enabled the modeling of BSM/EFT couplings of the $H$ boson in both 
production and decay~\cite{Gao:2010qx,Bolognesi:2012mm,Anderson:2013afp,Gritsan:2016hjl,Gritsan:2020pib,Martini:2021uey,Davis:2021tiv}, 
and has been extensively utilized in analyzing LHC 
data~\cite{CMS:2012qbp,CMS:2012vby,CMS:2013fjq,CMS:2014nkk,Sirunyan:2017tqd,Sirunyan:2018qlb,Sirunyan:2019pqw,Sirunyan:2019nbs,Sirunyan:2019twz,Sirunyan:2020sum,CMS:2021ugl,CMS:2021nnc,CMS:2022ley,CMS:2022uox,CMS:2024eka,CMS:2025xkn,CMS:2025fpt}.
We use this framework to develop both analytical and machine-learning approaches for calculating observables 
that are optimal for EFT applications. A new metric for evaluating the performance of such approaches has been established.
We also demonstrate how most of the relevant information can be efficiently stored in a limited number of bins, 
which is optimal for data analysis, data preservation, and global combination of data,
while also providing tools in a Python package ({\tt MiLoMege}) to achieve these benefits.
In the subsequent sections, we present the details of these methods,
with code available in Ref.~\cite{mela}.


\section{Measurements and observables}
\label{sect:input_obs}

\begin{figure}[t]
  \begin{center}
    \captionsetup{justification=centerlast}
    \includegraphics[width=0.65\linewidth]{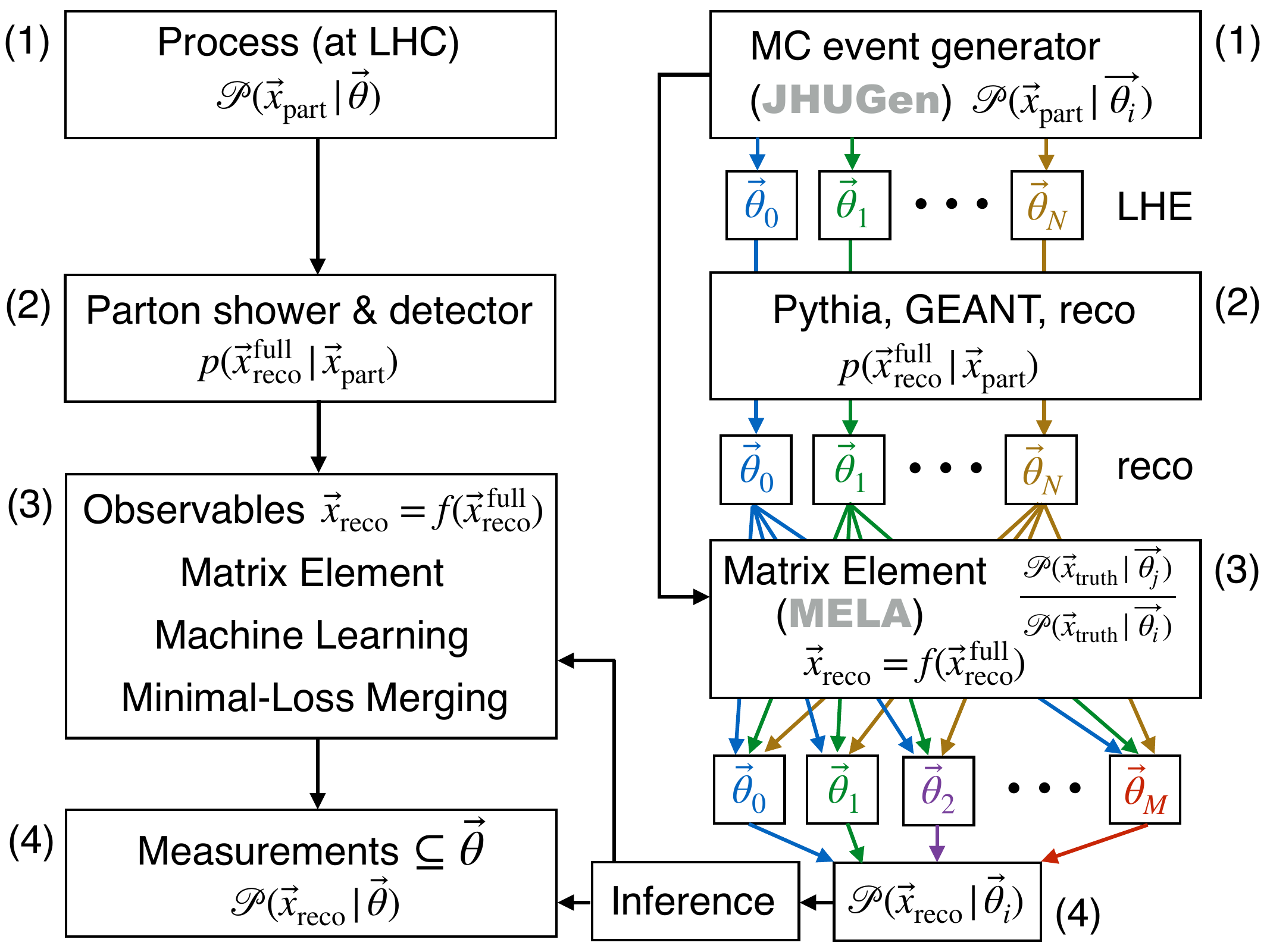}
    \caption{
Left: A schematic view of event analysis in an LHC experiment for a set of parameters $\vec{\theta}$
that define a fundamental process in proton collisions at the LHC, 
characterized by the partonic kinematic quantities $\vec{x}_\mathrm{part}$. 
Right: The sequence of MC modeling of the probability density ${\cal P}(\vec{x}_\mathrm{reco}|\vec{\theta})$
which depends on the reconstructed observables $\vec{x}_\mathrm{reco}$ and the test parameters $\vec{\theta}_i$.
Refer to the text for further details.
     }
    \label{fig:templates}
  \end{center}
\end{figure}

\noindent In a typical LHC data analysis, the objective is to characterize a parton-level {process} using parameters 
$\vec{\theta}$, which describe either the SM or BSM.
This process is characterized by quantities $\vec{x}_\mathrm{part}$, which usually consist of the four-momenta 
of all partons involved, including both incoming and outgoing particles.
In the context of the processes shown in Fig.~\ref{fig:process}, these include the incoming quark-antiquark pair 
and the decay products of the $H$ boson and associated $V$ (left), or two incoming gluons and four outgoing 
leptons (right).
This, along with the following steps in LHC data processing, is depicted in the left portion of the diagram 
in Fig.~\ref{fig:templates}, where ${\cal P}(\vec{x}_\mathrm{part}|\vec{\theta})$ represents the squared matrix 
element of the hard process, incorporating the parton luminosities, propagators, and phase-space factors.

The {parton shower} and hadronization that take place during collision events, along with the subsequent 
{detector response} and reconstruction algorithms, give rise to the reconstructed quantities 
$\vec{x}_\mathrm{reco}$, which differ from $\vec{x}_\mathrm{part}$.
The challenges in experimental analysis arise from the interaction between the two.
The probability density function (pdf) for a reconstructed process can be derived from the parton-level process 
pdf using the transfer function as
\begin{equation}
{\cal P}(\vec{x}_\mathrm{reco}|\vec{\theta}) = 
\int \mathrm{d}\vec{x}_\mathrm{part} ~p(\vec{x}_\mathrm{reco}|\vec{x}_\mathrm{part}) {\cal P}(\vec{x}_\mathrm{part}|\vec{\theta}) \,,
\label{eq:P} 
\end{equation}
where $p(\vec{x}_\mathrm{reco}|\vec{x}_\mathrm{part})$ is the transfer function. 
Depending on the use case, by convention, the effects of parton shower and hadronization may be assigned either to 
${\cal P}(\vec{x}_\mathrm{part}|\vec{\theta})$ or $p(\vec{x}_\mathrm{reco}|\vec{x}_\mathrm{part})$.

It is important to emphasize that, in many cases, it is not possible to invert the relationship in Eq.~(\ref{eq:P}) 
and recover ${\cal P}(\vec{x}_\mathrm{part}|\vec{\theta})$ from the observed reconstructed distributions, 
due to information loss during the transfer. This loss can occur either because of detector inefficiencies, 
such as particles lost, or because $\vec{x}_\mathrm{reco}$ does not correspond to the complete set of 
$\vec{x}_\mathrm{part}$.
In an ideal situation, we would be able to measure the complete set of {observables} 
$\vec{x}_\mathrm{reco}^\mathrm{\,full}=\vec{x}_\mathrm{part}$, which would fully describe the process 
of interest in Eq.~(\ref{eq:P}) without any information loss: 
$p(\vec{x}_\mathrm{reco}^\mathrm{\,full}|\vec{x}_\mathrm{part})=\delta(\vec{x}_\mathrm{reco}^\mathrm{\,full}-\vec{x}_\mathrm{part})$. 
In practice, this is either impossible or impractical, 
and some information is inevitably lost or obstructed due to detector effects.
Such an approach is often regarded as impractical because it would involve a highly multi-dimensional 
space of observables, making it difficult to utilize.
A central goal of this work is to reduce the dimensionality of the observable information while preserving near-optimal performance.

As an example, when applied to the two processes shown in Fig.~\ref{fig:process}, the full set of observables 
describing the $HVV$ vertex in either production (left) or decay (right) includes five observables: three angular variables, 
$\cos\Theta_1$, $\cos\Theta_2$, and $\Phi$, as well as the invariant masses of the two vector bosons, $m_1$ and $m_2$.
For a spin-zero $H$ boson, the angular distributions of $\cos\Theta^*$ and $\Phi_1$ are uniform,
as these observables characterize correlations between production and decay, and
a scalar particle cannot induce spin correlations between the initial and final states.
A simple approach to characterize the five-dimensional distribution with a histogram 
using only six bins along each dimension, already subject to information loss, would result in $6^5 = 7\,776$ bins. 
This number of bins is prohibitively large for practical data analysis and must be reduced by at least two orders of magnitude.
However, reducing the number of bins per dimension unavoidably leads to significant information loss.
Therefore, it is essential to pursue a more sophisticated approach to address this challenge.

In the following, we will investigate several methods for computing a reduced set of observables, 
including matrix-element-based calculations, machine learning techniques, 
and histogramming observables with a limited number of bins.
We denote $\vec{x}_\mathrm{reco}$ as a reduced set of observables that characterize an event. 
This can be illustrated through the following transformation, where 
$\vec{x}_\mathrm{reco}^\mathrm{\,full}$ contains the maximum amount of information about the process 
that can be reconstructed from the event:
\begin{equation}
\vec{x}_\mathrm{reco} = f(\vec{x}_\mathrm{reco}^\mathrm{\,full})  \,.
\label{eq:xreco} 
\end{equation}

In the end, a {measurement} is performed, which is a broad term referring to the deliverable results of 
analysis of LHC data. A measurement may involve constraining a set of parameters of interest, \(\vec{\theta}\),
or determining the production cross section (equivalently, the probability) within specific bins of observables.
A measurement is performed when an observed distribution of events ${\cal P}(\vec{x}_\mathrm{reco})$ 
is compared to predictions of a model through the inference process. 
As an example, the main objective of the measurement described in this work 
is to constrain eight parameters of the model introduced in Sec.~\ref{sect:input_eft}.

Modeling of the transfer encoded in Eq.~(\ref{eq:P}) is often performed with Monte Carlo (MC) techniques,
shown in the right diagram of Fig.~\ref{fig:templates}.
The hard process ${\cal P}(\vec{x}_\mathrm{part}|\vec{\theta})$ would be modeled with a MC event generator,
for which we use the {\tt JHUGen} framework for illustration in this work,
producing events in the Les Houches Event (LHE)~\cite{Alwall:2006yp} data format. 
The parton shower is modeled with e.g. Pythia~\cite{Sjostrand:2014zea}, detector response with e.g. GEANT~\cite{GEANT4:2002zbu},
and the reconstruction software is specific to an experimental approach that would be used for calculation of $\vec{x}_\mathrm{reco}$
and is identical to that used in the left diagram of Fig.~\ref{fig:templates}.
Data-driven techniques can be employed either to calibrate or to substitute for MC, but the general idea 
remains the same: some approximation is used to predict ${\cal P}(\vec{x}_\mathrm{reco}|\vec{\theta}_i)$,
which are later used in the inference process. 

One important aspect in the inference process is that multiple models $i$ of ${\cal P}(\vec{x}_\mathrm{reco}|\vec{\theta}_i)$
must be generated with MC techniques, and often a large number of models $M$ beyond the reference model $\vec{\theta}_0$ 
are required. This can be achieved by generating $N$ models covering the extreme variation of the phase-space 
of $\vec{x}_\mathrm{part}$. Within the {\tt JHUGen} framework, 
the {\tt MELA} matrix-element re-weighting is applied to each of the $N$ samples to calculate $M$
weights to reproduce each model. As a typical example, $N\sim10^1$ and $M\sim10^2$ may appear in the EFT modeling, 
discussed below. This way, all generated events are re-used to model each distribution, optimally using CPU resources. 

In the end, a typical particle physics analysis of LHC data requires construction of a likelihood function ${\cal L}$, 
or other means of inference, which describes a sum of contributing {processes} defined for
a set of {observables} reconstructed in an experiment $\vec{x}_\mathrm{reco}$ as a function of parameters 
of interest~$\vec{\theta}$, which are the target of a {measurement}. A formulation based on quantum mechanics
allows ${\cal L}$ to be parameterized as a linear combination of ${\cal P}(\vec{x}_\mathrm{reco}|\vec{\theta}_i)$
multiplied by parameters of interest. While an unbinned formulation is possible, it is more practical to represent the data
in a finite number of bins, leading to pdf's ${\cal P}(\vec{x}_\mathrm{reco}|\vec{\theta}_i)$ 
described as templates (histograms). The goal of this paper is to illustrate how to make the process 
described in Fig.~\ref{fig:templates} optimal for data analysis, data preservation, and global combination of data. 
We demonstrate this through a series of representative examples.


\section{Application to Effective Field Theory}
\label{sect:input_eft}

\noindent A common application in EFT
involves a parameter set $\vec{\theta}$ representing
$(K+1)$ operators, which include the SM coupling $\theta_0$, such that 
$\vec{\theta} = (\theta_0, \theta_1, \ldots, \theta_K)$. We consider a set of models $\vec{\theta}_i$, 
where each model corresponds to a single non-zero coupling $\theta_i$. 
To illustrate a specific example of a parameter set \(\vec{\theta}\), we focus on the Standard Model Effective Field Theory 
(SMEFT)~\cite{Buchmuller:1985jz,Grzadkowski:2010es,Dedes:2017zog,Brivio:2017vri}
in which new operators are added to the Lagrangian governing the Higgs potential and gauge boson interactions:
\begin{eqnarray}
 {\cal L} = &&
 -\frac{1}{4}W^I_{\mu\nu}W^{I\mu\nu}
 -\frac{1}{4}B_{\mu\nu}B^{\mu\nu}
 +({D_\mu}\varphi)^\dagger({D_\mu}\varphi) 
 +m^2(\varphi^\dagger\varphi)
 -\frac{\lambda}{2}(\varphi^\dagger\varphi)^2
\nonumber \\
 &&  +C^{\varphi}(\varphi^\dagger\varphi)^3
  +C^{\varphi\Box}(\varphi^\dagger\varphi)\Box(\varphi^\dagger\varphi)
  +C^{\varphi{D}}(\varphi^\dagger D_\mu \varphi)^*(\varphi^\dagger D^\mu \varphi)
\nonumber \\
 && +C^{\varphi W}(\varphi^\dagger\varphi)W^I_{\mu\nu}W^{I\mu\nu}
 +C^{\varphi B}(\varphi^\dagger\varphi)B_{\mu\nu}B^{\mu\nu}
 +C^{\varphi WB}(\varphi^\dagger \tau^I \varphi)W^I_{\mu\nu}B^{\mu\nu}
  \nonumber \\
 && +C^{\varphi\widetilde{W}}(\varphi^\dagger\varphi)\widetilde{W}^I_{\mu\nu}W^{I\mu\nu}
 +C^{\varphi\widetilde{B}}(\varphi^\dagger\varphi)\widetilde{B}_{\mu\nu}B^{\mu\nu}
 +C^{\varphi\widetilde{W}{B}}(\varphi^\dagger \tau^I \varphi)\widetilde{W}^I_{\mu\nu}{B}^{\mu\nu}
   \nonumber \\
 && + \cdots
    \label{eq:Lagrangiansmeft}
\end{eqnarray}
where the \(\tau^I\) are the Pauli matrices acting on the Higgs doublet \(\varphi\) with index \(I=1,2,3\), 
\(W^{I}_{\mu\nu}\) and \(\widetilde{W}^{I}_{\mu\nu}\) denote the CP-even and CP-odd field strength tensors 
of the \(SU(2)_L\) gauge fields, respectively, and similarly \(B_{\mu\nu}\) corresponds to the field strength 
tensor of the \(U(1)_Y\) gauge field.

Although there are a total of 2,499 new operators in SMEFT, in Eq.~(\ref{eq:Lagrangiansmeft}) we include 
only those operators relevant to Higgs--gauge boson interactions, with the exception of \( C^{\varphi} \), 
that contribute to the processes illustrated in Fig.~\ref{fig:process}. Other SMEFT operators can also contribute. 
For example, the same processes, involving identical initial and final state particles as shown 
in Fig.~\ref{fig:process}, can also originate from four-particle point interaction vertices present in the SMEFT, 
which lead to dipole-type operators. Examples include \( C_{f_1f_2}^{eW} \) and \( C_{f_1f_2}^{eB} \), 
which can mediate direct decays such as \( H \to \gamma^* \ell^+ \ell^- \) and \( H \to Z \ell^+ \ell^- \) 
via four-particle interactions. However, these same operators also generate three-particle interaction 
without the $H$ boson, like \( \gamma^* \to \ell^+ \ell^- \) and \( Z \to \ell^+ \ell^- \)~\cite{Dedes:2017zog}, 
which are far more commonly studied and can provide stronger constraints than Higgs-related processes. 
Therefore, we omit these operators from our analysis of the processes shown in Fig.~\ref{fig:process}.
We adopt this approach for Higgs-related processes, as it enables a significant simplification by reducing 
the potentially large set of operators to just eight that are relevant for the analysis of these specific processes.

Another observation we make is that working in the mass eigenstate basis of the gauge bosons 
is more practical than using the weak eigenstate basis, particularly in the context of experimental analyses:
\begin{eqnarray}
\begin{pmatrix}
Z_\mu \\
A_\mu
\end{pmatrix}
=
\begin{pmatrix}
\cos\theta_W & -\sin\theta_W \\
\sin\theta_W & \cos\theta_W
\end{pmatrix}
\begin{pmatrix}
W^3_\mu \\
B_\mu
\end{pmatrix}
    \label{eq:totateWB}
\end{eqnarray}
This is especially relevant for the \( HZZ \), \( HZ\gamma \), and \( H\gamma\gamma \) interactions, 
which contribute to the processes shown in Fig.~\ref{fig:process}.
For example, the \( H\gamma\gamma \) interaction can be described by a single operator coefficient 
\( c_{\gamma\gamma} \) in the mass eigenstate basis, replacing the need for the three separate coefficients
\( C^{\varphi W} \), \( C^{\varphi B} \), and \( C^{\varphi WB} \) defined in the weak eigenstate basis.

Ultimately, the following operator coefficients in the mass-eigenstate basis, which include the coefficient \( \delta c_w \) 
that does not affect the processes shown in Fig.~\ref{fig:process}, can be used as substitutes for the operators in 
the weak-eigenstate basis~\cite{Falkowski:2015wza}:
\begin{eqnarray}
\delta c_z &=& \frac{v^2}{\Lambda^2}\left(C^{\varphi \Box} - \frac{1}{4}C^{\varphi D} -3\delta_{v}\right) \nonumber \\
c_{z\Box} &=& \frac{v^2}{\Lambda^2} \left( \frac{s_{w}^2}{2e^2} C^{\varphi D} + \frac{2s_{w}^2}{e^2} \delta_{v}\right)  \nonumber \\
c_{zz} &=& \frac{v^2}{\Lambda^2} \left( \frac{4c_{w}^4s_{w}^2}{e^2} C^{\varphi W} + \frac{4c_{w}^3s_{w}^3}{e^2} C^{\varphi WB} + \frac{4c_{w}^2s_{w}^4}{e^2} C^{\varphi B}\right)  \nonumber \\
c_{z\gamma} &=& \frac{v^2}{\Lambda^2} \left( \frac{4c_{w}^2s_{w}^2}{e^2} C^{\varphi W} - \frac{2c_{w}s_{w}(c_{w}^2-s_{w}^2)}{e^2} C^{\varphi WB} - \frac{4c_{w}^2s_{w}^2}{e^2} C^{\varphi B} \right)  \nonumber \\
c_{\gamma\gamma} &=& \frac{v^2}{\Lambda^2} \left( \frac{4s_{w}^2}{e^2}  C^{\varphi W} - \frac{4c_{w}s_{w}}{e^2} C^{\varphi WB} + \frac{4c_{w}^2}{e^2} C^{\varphi B} \right) \nonumber \\
\tilde{c}_{zz} &=& \frac{v^2}{\Lambda^2} \left(  \frac{4c_{w}^4s_{w}^2}{e^2} C^{\varphi \widetilde{W}} + \frac{4c_{w}^3s_{w}^3}{e^2} C^{\varphi \widetilde{W}B} + \frac{4c_{w}^2s_{w}^4}{e^2} C^{\varphi \widetilde{B}} \right)  \nonumber \\
\tilde{c}_{z\gamma} &=& \frac{v^2}{\Lambda^2} \left( \frac{4c_{w}^2s_{w}^2}{e^2} C^{\varphi \widetilde{W}} - \frac{2c_{w}s_{w}(c_{w}^2-s_{w}^2)}{e^2} C^{\varphi \widetilde{W}B} - \frac{4c_{w}^2s_{w}^2}{e^2} C^{\varphi \widetilde{B}} \right)  \nonumber \\
\tilde{c}_{\gamma\gamma} &=& \frac{v^2}{\Lambda^2} \left( \frac{4s_{w}^2}{e^2}  C^{\varphi \widetilde{W}} - \frac{4c_{w}s_{w}}{e^2} C^{\varphi \widetilde{W}B} + \frac{4c_{w}^2}{e^2} C^{\varphi \widetilde{B}} \right)  \nonumber \\
\delta c_w &=& \frac{v^2}{\Lambda^2} \left(C^{\varphi \Box} + \frac{s_{w}c_{w}}{c_{w}^2-s_{w}^2} C^{\varphi \widetilde{W} B} + \frac{s_{w}^2 - 2}{4(c_{w} ^2 - s_{w}^2)} C^{\varphi D} -\frac{s_{w}^2}{c_{w}^2-s_{w}^2} \delta_{v}\right) 
    \label{eq:massbasis}
\end{eqnarray}
where $\delta_{v} = \frac{1}{2}\left[ C^{\varphi l }_{11} + C^{\varphi l }_{22}\right] - \frac{1}{4}C^{ll}_{1221}$. 
These operators forming $\delta_{v}$ contribute to muon decay at tree level and are therefore expected to be small.
Alternatively, imposing the custodial symmetry condition $\delta c_w = \delta c_z$ leads to the constraint
$\delta_{v} = -({c_{w}^2C^{\varphi D} + 4 e^2 C^{\varphi WB}})/({4 s_{w}^2})$. 
The precise treatment of the $\delta_v$ parameter in relating the mass-eigenstate and weak-eigenstate
parameterizations depends on the adopted assumptions.

The set of coefficients in Eq.~(\ref{eq:massbasis}) will be treated as the parameter set \(\vec{\theta}\) in the following discussion. 
It is important to note that \(\delta c_w\) does not affect the \(HZZ\), \(HZ\gamma\), or \(H\gamma\gamma\) vertices, 
leaving only eight relevant parameters in these cases.
Therefore, the primary objective of an EFT analysis of the processes shown in Fig.~\ref{fig:process} 
is to constrain the complete parameter set \(\vec{\theta}\), given in Eq.~(\ref{eq:massbasis}).

There is an important consideration when performing a measurement in a single $H$ boson channel. 
Let us take an example of the $H \to (Z/\gamma^*)(Z/\gamma^*) \to 4\ell$ decay with five observables
$\vec{x}= (m_1, m_2, \cos\Theta_1, \cos\Theta_2, \Phi)$ fully characterizing its kinematic distributions. 
The differential distributions for these five observables can be expressed as
\begin{equation}
\frac{d\sigma(\vec{x})}{d\vec{x}}=
\frac{ 
\left(\sum \gamma^\mathrm{prod}_{jk}b_j b_k\right) 
\left(\sum \gamma^\mathrm{dec}_{lm}(\vec{x})c_l c_m\right) }
{\Gamma_\mathrm{tot}(\vec{c}\,)} \,,
\label{eq:crosssection1} 
\end{equation}
where the quantities $\gamma^\mathrm{prod}_{jk}$ correspond to the production cross sections, whereas 
the functions $\gamma^\mathrm{dec}_{lm}(\vec{x})$ describe the decay dependence on the decay observables. 
All the corresponding terms are proportional to products of the EFT couplings $b_k$ and $c_m$, where
by convention, \( b_0 \) and \( c_0 \) denote the SM couplings in production and decay, respectively, with \( c_0 = 1 \).
Additionally, $\Gamma_{\mathrm{tot}}$ denotes the total decay width of the $H$ boson, 
which depends on these couplings as well as on others, collectively represented by $\vec{c}$.

Equation~(\ref{eq:crosssection1}) illustrates that, within a single decay channel of the $H$ boson, 
it is not possible to directly constrain the EFT couplings without making additional assumptions. 
This limitation arises because the total width $\Gamma_\mathrm{tot}$ and production cross sections
depend on a larger number of couplings than there are available experimental constraints. 
As a consequence, the constraints on individual couplings become degenerate, 
resulting in many degrees of freedom that are either weakly constrained or entirely unconstrained.
However, by redefining the parameters as \( c_{m}^\prime \), all degenerate couplings can be factored out
and collected into a single group shown in square brackets below, 
while the remaining terms describe the differential distributions as a polynomial in the newly defined parameters:
\begin{equation}
\frac{d\sigma(\vec{x})}{d\vec{x}}=
\left[
\frac{\sum \gamma^\mathrm{prod}_{jk}b_j b_k}{\Gamma_\mathrm{tot}(\vec{c}\,)}
 (1+\delta c_z)^2
\right]
 \left(
 \gamma^\mathrm{dec}_{00}(\vec{x})
 + \sum_{1\le m} \gamma^\mathrm{dec}_{0m}(\vec{x}) c_m^\prime
 + \sum_{1\le m\le l} \gamma^\mathrm{dec}_{lm}(\vec{x}) c_l^\prime c_m^\prime
 \right) \,,
\label{eq:crosssection2} 
\end{equation}
where $\gamma^\mathrm{dec}_{00}(\vec{x})$ describes the SM decay process.
The newly defined parameters $c_m^\prime$ are expressed as ratios of the couplings to the combination \( (1 + \delta c_z) \), 
which controls the SM-like tensor structure of the interactions:
\begin{equation}
c_{m}^\prime = c_{m} / (1+\delta c_z) \,,
\label{eq:coupleratio} 
\end{equation}
where this relationship depends on the fact that \( c_0 = 1 \).
Further discussion can be found in Ref.~\cite{Barducci:2025ati}. 
For instance, constraining the coupling ratios in Eq.~(\ref{eq:coupleratio}) is equivalent to setting bounds 
on the parameters \( f_{ai} \) (see, e.g., Eq.~(43) of Ref.~\cite{Davis:2021tiv}), or, alternatively, 
on the ``mixing angle'' parameters used in experimental analyses, 
as these quantities are defined in terms of the same coupling ratios.
It is also important to note that, within the EFT validity regime where  \(|\delta c_z| \ll 1\) and  \(|c_m| \ll 1\),
$c_{m}^\prime \simeq c_{m}$ and the last terms in Eq.~(\ref{eq:crosssection2}), 
involving the product \( c_l^\prime c_m^\prime \), can generally be neglected.

What we have accomplished with Eq.~(\ref{eq:crosssection2}) is a reparameterization of the eight decay couplings 
defined in Eq.~(\ref{eq:massbasis}), along with numerous couplings in production and \(\Gamma_{\mathrm{tot}}\), 
into just seven effective decay couplings defined by Eq.~(\ref{eq:coupleratio}), plus a single overall factor enclosed 
in square brackets.
In a global EFT analysis combining all experimental data from the LHC and other sources, each coupling appearing 
within the square brackets can be constrained independently. However, when considering a single channel, 
the constraint on the entire expression inside the square brackets effectively translates to a constraint on the 
overall cross section of the process, which must be reported alongside the seven coupling ratios.

Equation~(\ref{eq:crosssection2}) also motivates the experimental parameterization of the probability density function (pdf), 
which, once the overall event rate is factored out, takes the form
\begin{eqnarray}
    {\cal P}(\vec{x}_\mathrm{reco}|\vec{\theta})  \propto
    {\cal P}_0(\vec{x}_\mathrm{reco}) 
    ~~~~~~~~~~~~~~~~~~~~~~~~~~~~~~~~~~~~~~~~~~~~~~~~~~~~~~~
    ~~~~~~~~~~~~~~~~~~~~~~~~~~~~~~~~~~~~~~~~~~~~~~~~~~~~~~~
    \nonumber  \\
    +  \sum_{1\le k \le K} \left(\frac{2\theta_k}{\theta_0}\right) {\cal P}_{0k}(\vec{x}_\mathrm{reco}) 
    + \sum_{1\le k \le K} \left(\frac{\theta_k}{\theta_0}\right)^2 {\cal P}_k(\vec{x}_\mathrm{reco}) 
    + \sum_{1\le i<j\le K} \left(\frac{2\theta_i\theta_j}{\theta_{0}^2}\right) {\cal P}_{ij}(\vec{x}_\mathrm{reco}) \,, 
    \label{eq:probreco} 
\end{eqnarray}
where a single subscript on the pdf denotes the model type $\vec{\theta}_k$, 
with a single non-zero coupling ${\theta}_k$,
while a double subscript represents the interference between two models, $\vec{\theta}_i$ and $\vec{\theta}_j$. 
The couplings $\theta_k$ are assumed to be real to maintain Hermiticity, 
but allowing them to be complex does not significantly alter the results.
Equation~(\ref{eq:probreco}) provides a way to express the pdf in terms of the parameters of interest $\theta_k$,
and we will revisit this equation later.


\section{Matrix elements and machine learning}
\label{sect:input_ml}

\noindent 
To analyze LHC data, the information from each collision event must be processed 
and categorized to determine its relevance to a specific physical process.
Ideally, the full set of reconstructed information describing the event,
$\vec{x}_\mathrm{reco}^\mathrm{\,full}$ as introduced in Sec.~\ref{sect:input_obs}, 
would be used in the analysis of all relevant events. 
However, this is typically neither practical nor feasible, since the high dimensionality of the observable space 
makes effective analysis extremely challenging, as will be demonstrated later. 
Therefore, the objective is to reduce the set of observables to the minimal necessary subset 
and to focus on a small number of observables that are sensitive to the effects of interest.
Therefore, the central theme of this paper is dimensionality reduction of observables. 

A key consideration in the construction of observables sensitive to effects from EFT is the characteristic behavior 
of higher-dimensional operators, which typically introduces enhancements at large values of \( q^2 \), 
the squared four-momentum of particles exchanged in propagators. Therefore, observables that either depend 
explicitly on \( q^2 \), or exhibit strong correlations with it, can serve as effective probes for identifying 
deviations from SM predictions. An example of such a correlated quantity is the transverse momentum 
of reconstructed final-state objects.
Nonetheless, these generic \( q^2 \)-sensitive observables may not provide discriminatory power to distinguish 
among different operators that yield similar enhancements in \( q^2 \). An example could be distinguishing 
between \( CP \)-even and \( CP \)-odd operators, which typically requires observables specifically constructed 
to be sensitive to \( CP \)-violating effects. In general, the design of EFT-sensitive observables depends on the 
structure and symmetry properties of the higher-dimensional operators. 

Well-established methodologies exist for the optimal design of such observables, 
either through direct calculation of the likelihood functions, such as matrix-element calculations, 
or via its approximations, such as implementation using machine learning techniques.
Although this topic has received increasing attention in the
literature~\cite{Kondo:1988yd,Atwood:1991ka,Diehl:1993br,Anderson:2013afp,Brehmer:2018kdj,Brehmer:2019bvj,Gritsan:2020pib}, 
a comprehensive understanding of how to construct EFT-sensitive observables and 
assess their performance remains limited within the community. 
To address this gap, this section offers a review of the existing knowledge, followed by the 
introduction of novel methodologies in the subsequent sections designed to advance the construction and
performance evaluation of such observables. 
Moreover, the presence of a large number of observables sensitive to various EFT operators introduces additional 
complexities in the analysis of experimental data. Addressing this challenge is also a central objective of this work.

\subsection{Optimized observables using matrix-element calculations}
\label{sect:obs-me}

\noindent The concept of optimized observables was first introduced in particle physics in Refs.~\cite{Atwood:1991ka,Diehl:1993br}. 
In the case of a binary hypothesis test, the Neyman-Pearson lemma~\cite{Neyman:1933wgr} establishes that the likelihood ratio 
${\cal P}_1(\vec{x}_\mathrm{reco}^\mathrm{\,full})/{\cal P}_0(\vec{x}_\mathrm{reco}^\mathrm{\,full})$ 
provides the most powerful test statistic for discriminating between hypotheses 0 and 1.
For a continuous family of hypotheses, corresponding to arbitrary quantum-mechanical mixtures of two states, 
one could in principle apply the Neyman-Pearson lemma to each pair of points in the parameter space. 
However, this approach would require evaluating an infinite set of likelihood ratios. Remarkably, the same 
discriminative information can be captured using just two specific likelihood ratios, which may be interpreted 
as a pair of optimized observables~\cite{Anderson:2013afp}:
\begin{eqnarray}
{\cal D}_\mathrm{opt,2} =
\frac{{\cal P}_1(\vec{x}_\mathrm{reco}^\mathrm{\,full})}
{{\cal P}_0(\vec{x}_\mathrm{reco}^\mathrm{\,full})+\alpha_2\cdot{\cal P}_1(\vec{x}_\mathrm{reco}^\mathrm{\,full})} \,,
\label{eq:optimized2} 
\\
{\cal D}^{(\alpha)}_\mathrm{opt,1} =
\frac{2{\cal P}_{01}(\vec{x}_\mathrm{reco}^\mathrm{\,full})}
{{\cal P}_0(\vec{x}_\mathrm{reco}^\mathrm{\,full})+\alpha\cdot{\cal P}_1(\vec{x}_\mathrm{reco}^\mathrm{\,full})} \,,
\label{eq:optimized1} 
\end{eqnarray}
where ${\cal P}$ serves as an effective probability distribution expressed in terms of the squared matrix elements. 
Both ${\cal P}_1$ and ${\cal P}_2$ are normalized to the same area, 
and the indexing convention is consistent with its use in Eq.~(\ref{eq:probreco}).
A constant parameter $\alpha$ ($\alpha_2$) will be utilized below.
The information conveyed by the $\mathcal{D}_{\text{opt,2}}$ observable is independent of the value 
of $\alpha_2$ in Eq.~(\ref{eq:optimized2}), and for improved visualization, we consistently adopt $\alpha_2 = 1$.
While setting \( \alpha = 1 \) yields a symmetric formulation of the observables, an alternative choice 
of \( \alpha = 0 \) also leads to a well-defined and physically motivated construction.

We will analyze the discriminants $\mathcal{D}_{\text{opt,1}}^{(1)}$ and $\mathcal{D}_{\text{opt,1}}^{(0)}$, where the 
difference between them is solely determined by the values of $\alpha = 1$ or $0$ in Eq.~(\ref{eq:optimized1}), respectively.
For any given BSM coupling $\theta_1$, either pair of observables, $(\mathcal{D}_{\text{opt,2}}, \mathcal{D}_{\text{opt,1}}^{(0)})$ 
or $(\mathcal{D}_{\text{opt,2}}, \mathcal{D}_{\text{opt,1}}^{(1)})$, is fully optimal for distinguishing any magnitude 
of the BSM contribution from the SM, regardless of the relative sizes of the linear and quadratic terms in Eq.~(\ref{eq:probreco}).
The complete optimality of either pair is clear from the observation that an optimal observable 
corresponding to a specific relative size of the linear and quadratic terms can always be re-expressed in terms 
of either of these two observables, when calculated analytically following Eqs.~(\ref{eq:optimized2}) and~(\ref{eq:optimized1}).
The equivalence of the two pairs is evident from the fact that one pair can be simply re-expressed in terms of the other.
For a single observable, only $\mathcal{D}_{\text{opt,1}}^{(0)}$ alone is completely optimal 
in the EFT limit when quadratic terms can be neglected.
However, as we demonstrate below for the machine-learning approach, $\mathcal{D}_{\text{opt,1}}^{(0)}$ 
necessitates training and testing with negative probabilities, which introduces technical challenges. 
For this reason, the $\mathcal{D}_{\text{opt,1}}^{(1)}$ observable serves as a stable alternative, ensuring 
that the analysis remains stable and fully optimal when used in conjunction with $\mathcal{D}_{\text{opt,2}}$.

\begin{figure}[t!]
  \begin{center}
    \captionsetup{justification=centerlast}
    \includegraphics[width=0.3\textwidth]{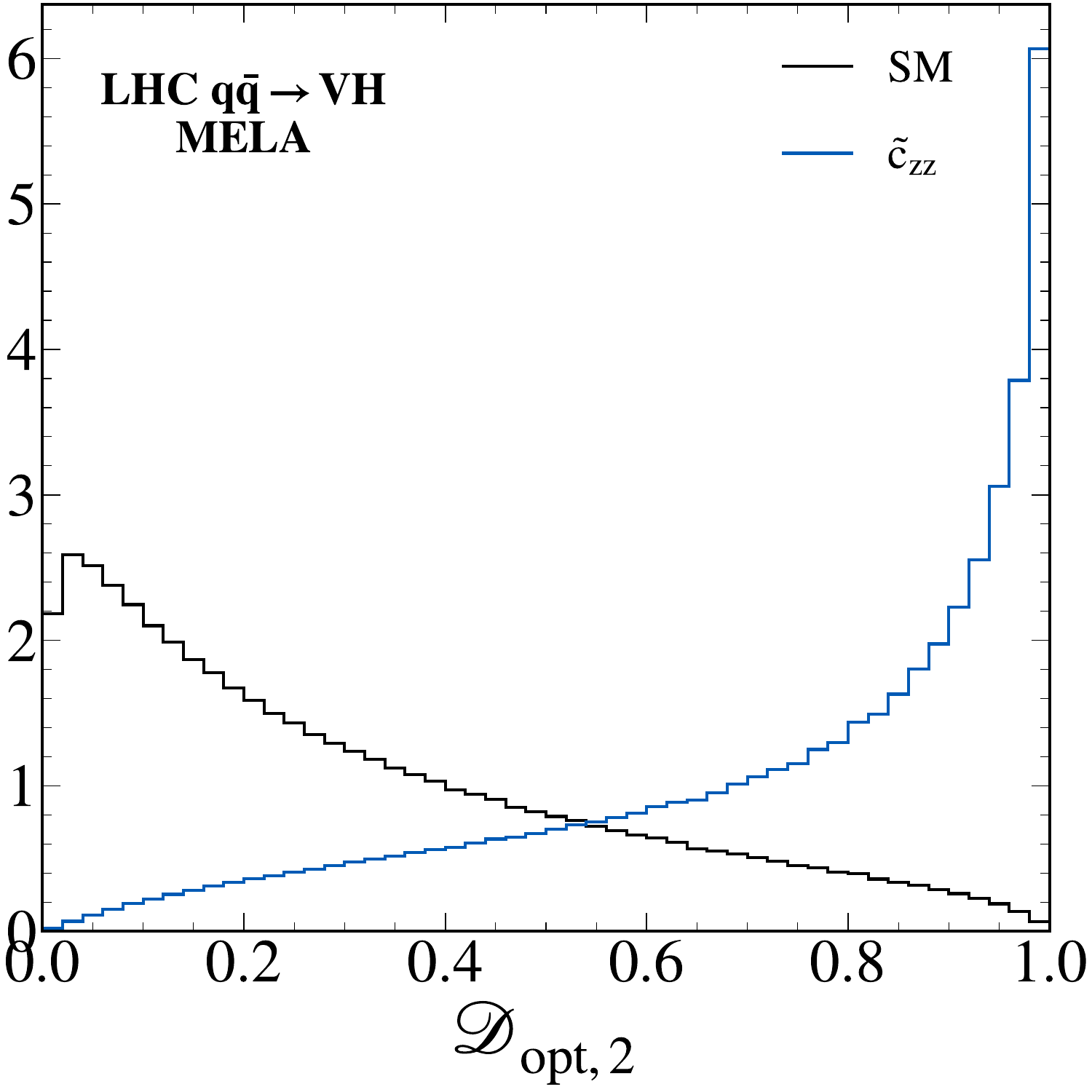}
    \includegraphics[width=0.3\textwidth]{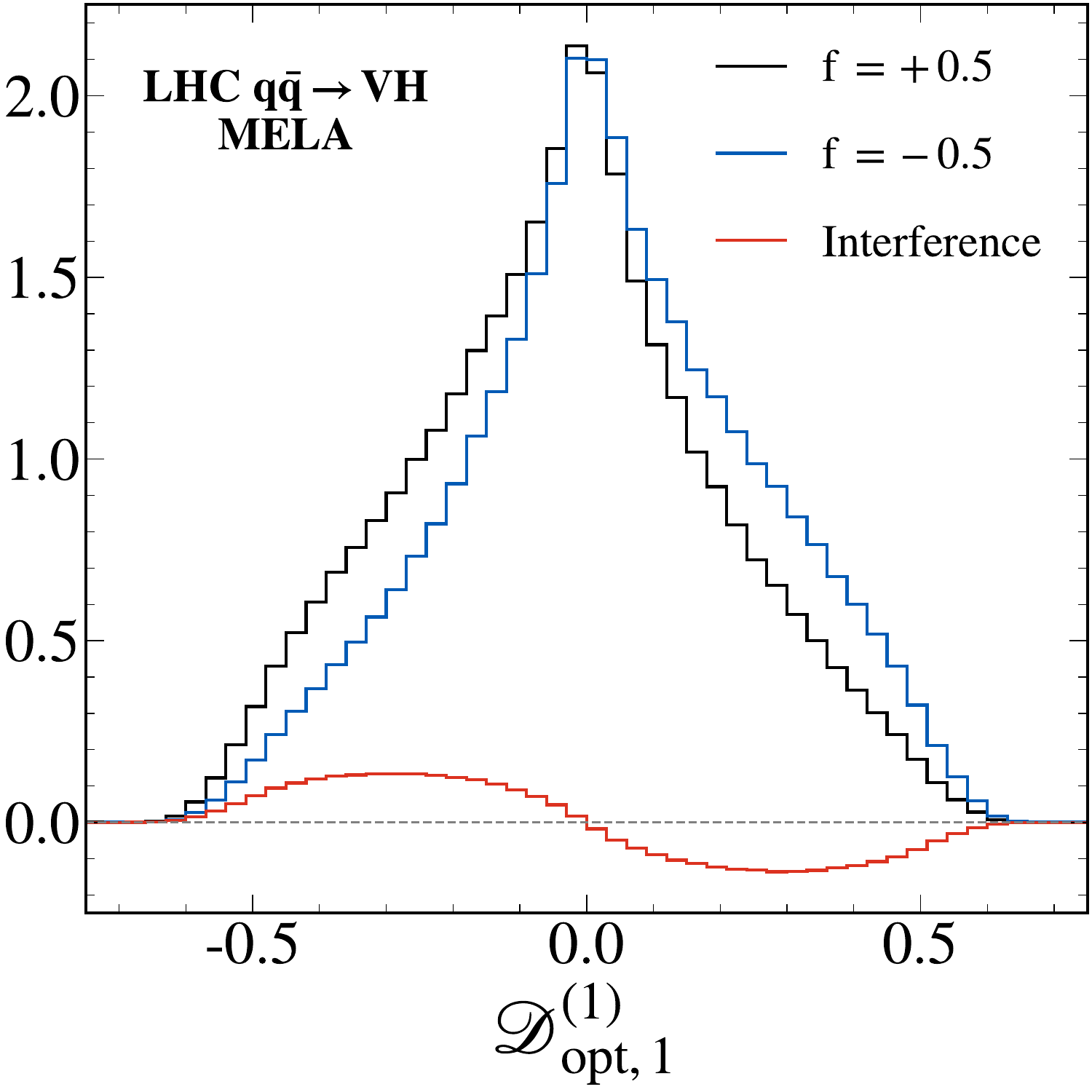}
    \includegraphics[width=0.3\textwidth]{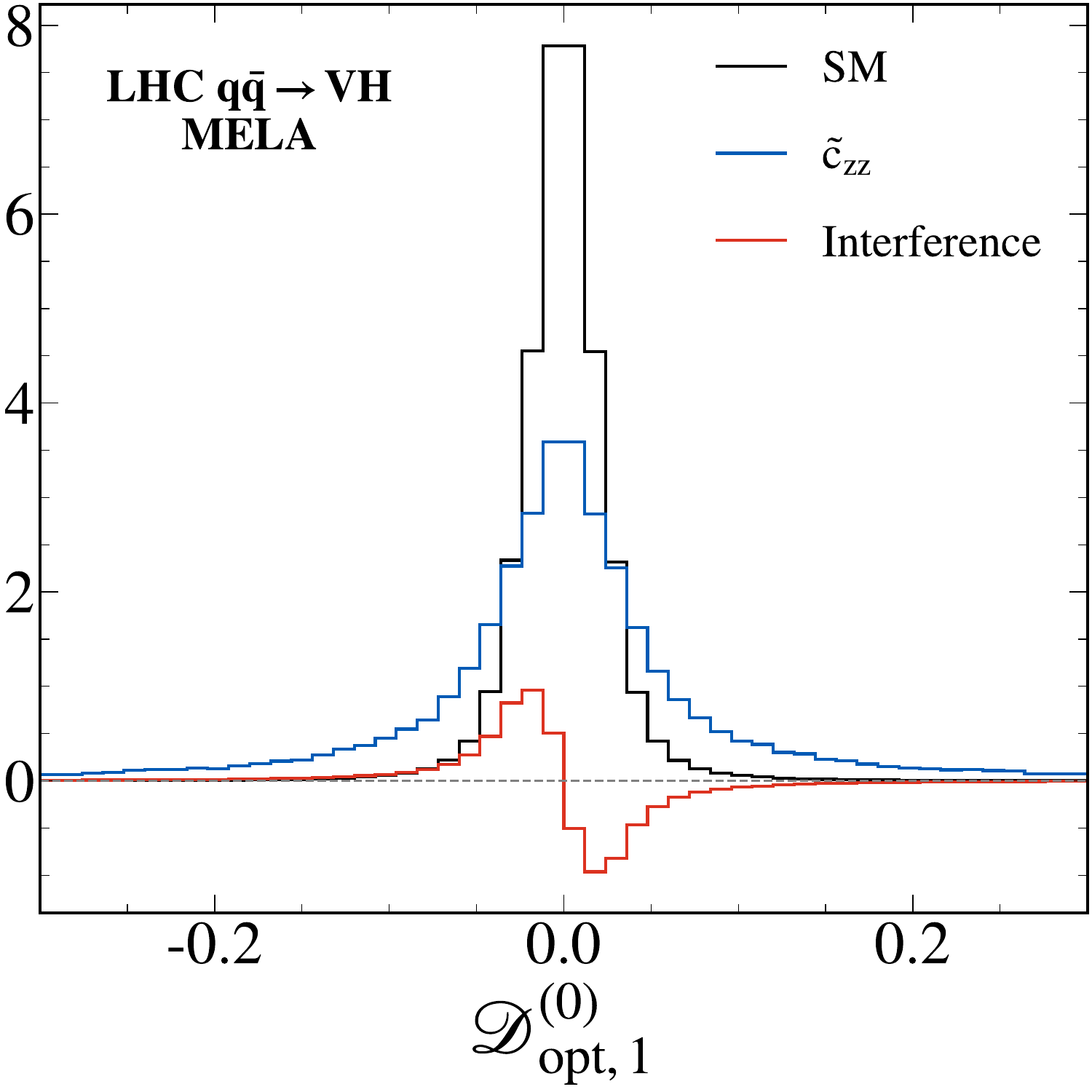}
    \caption{
    The distribution of simulated proton-proton collision events at the LHC for the process
$q\bar{q}\to (Z/\gamma^*) \to H(Z/\gamma^*) \to H(\ell^+\ell^-)$
is shown for the three optimal observables, 
${\cal D}_\mathrm{opt,2}$, ${\cal D}_\mathrm{opt,1}^{(1)}$, and $\mathcal{D}_{\text{opt,1}}^{(0)}$
computed using the matrix-element MELA approach~\cite{Anderson:2013afp}
according to Eqs.~(\ref{eq:optimized2}) and~(\ref{eq:optimized1}).
The coupling $\tilde{c}_{zz}$ is taken as the alternative hypothesis, 
with a $0.5$ contribution to the process cross section in the middle panel, 
where the sign indicates the relative coupling sign.
    }
    \label{fig:MELA_d0m}
  \end{center}
\end{figure}

An illustration of the discriminants ${\cal D}_\mathrm{opt,2}$, ${\cal D}_\mathrm{opt,1}^{(1)}$, and $\mathcal{D}_{\text{opt,1}}^{(0)}$
is shown in Fig.~\ref{fig:MELA_d0m} for the process $q\bar{q}\to (Z/\gamma^*) \to H(Z/\gamma^*) \to H(\ell^+\ell^-)$,
using MC simulations of proton-proton collisions at the LHC with a center-of-mass energy of 13\,TeV.
The $CP$-odd coupling $\tilde{c}_{zz}$, introduced in Eq.~(\ref{eq:massbasis}), is chosen as a representative alternative hypothesis.
The technical details of this simulation and visualization are provided in Sec.~\ref{sect:input_vh}, where the numerical 
evaluation of performance is also presented.

It is important to note that the ratio in Eq.~(\ref{eq:optimized1}) for the case $\alpha \neq 0$ can be re-expressed 
as follows, which is a fact we will utilize in several instances:
\begin{eqnarray}
 {\cal D}^{(\alpha)}_\mathrm{opt,1}(\vec{x})& = &  \left(\frac{1}{2\sqrt{\alpha}}\right) \frac{{\cal R}_\alpha(\vec{x})-1}{{\cal R}_\alpha(\vec{x})+1}
 \label{eq:ratio1}  \,,\\
 \mathrm{where}  && \nonumber \\
 {\cal R}_\alpha(\vec{x}) & = & 
 \frac{{\cal P}(\vec{x};\theta_0=1,\theta_1=+\sqrt{\alpha\sigma_0/\sigma_1}\,)}
        {{\cal P}(\vec{x};\theta_0=1,\theta_1=-\sqrt{\alpha\sigma_0/\sigma_1}\,)}  \nonumber \\
 \mathrm{and}  && \nonumber \\
  {\sigma_0}/{\sigma_1} & = & 
  \frac{\int d\vec{x}~{\cal P}(\vec{x};\theta_0=1,\theta_1=0)}{\int d\vec{x}~{\cal P}(\vec{x};\theta_0=0,\theta_1=1)}\,.  \nonumber
\end{eqnarray}
The normalization factor $\sqrt{\sigma_0/\sigma_1}$ ensures that the integrated probabilities 
(or cross sections in the physical processes) are equal for the couplings $\theta_0=1$ and $\theta_1=\sqrt{\sigma_0/\sigma_1}$.
From Eq.~(\ref{eq:ratio1}), we conclude that when $\alpha \neq 0$, the observable ${\cal D}_\mathrm{opt,1}^{(\alpha)}$ 
is optimal for distinguishing between two hypotheses which differ only in the sign of interference: 
$(\theta_0=1,\theta_1=+\sqrt{\alpha\sigma_0/\sigma_1}\,)$ and $(\theta_0=1,\theta_1=-\sqrt{\alpha\sigma_0/\sigma_1}\,)$, 
since it can be represented through a ratio of their probability distributions. 
For this reason, in Fig.~\ref{fig:MELA_d0m} (middle), we present these two models along with the pure interference 
distribution, which effectively represents the difference between the two.
On the other hand, when $\alpha = 0$, ${\cal D}_\mathrm{opt,1}^{(0)}$ is optimal for differentiating 
the pure interference distribution from the SM, also expressed as a ratio of their probability distributions in Eq.~(\ref{eq:optimized1}). 
Therefore, we show the two in Fig.~\ref{fig:MELA_d0m} (right).
These observations will be significant when we introduce machine learning training in Sec.~\ref{sect:obs-ml}.

Equation~(\ref{eq:probreco}) allows several important observations. In the absence of interference 
terms \({\cal P}_{ij}\), as is typically the case for non-interfering background processes, 
the optimized observable is ${\cal D}_\mathrm{opt,2}$, defined in Eq.~(\ref{eq:optimized2}). 
In this case, a single optimized observable is sufficient to effectively distinguish SM signal from background.
Assuming that the background contribution is independent of the EFT parameters under investigation, 
this observable also effectively distinguishes other signal hypotheses from the background when discriminants 
for signal hypothesis separation are also utilized.

When interference effects are present and only two coupling structures are considered (\(K=1\)) 
with $\theta_0$ and $\theta_1$, 
the number of optimized observables is limited to two, as specified in Eqs.~(\ref{eq:optimized2}) 
and~(\ref{eq:optimized1}). This underscores the power of multivariate analysis techniques, 
whereby the complete information contained within the high-dimensional space of 
\(\vec{x}_\mathrm{reco}^\mathrm{\,full}\) can be effectively captured by only two observables. 

However, when multiple couplings are present with \( K > 1 \), the number of optimized observables increases 
substantially, scaling as \( {(K+2)!}/{(2\,K!)} - 1\). 
For the purpose of an EFT measurement, it is important to note that the last two terms in Eq.~(\ref{eq:probreco}) 
are quadratic in the BSM couplings and can typically be neglected. This approximation reduces the coupling 
expansion to \( K \) linear terms, corresponding to the interference between the BSM and the SM amplitudes. 
As a result, one may select \( K \) observables, each associated with a linear term of the form defined 
in Eq.~(\ref{eq:optimized1}) with \( \alpha=0 \), to achieve optimal discrimination in an EFT analysis.

Although multiple operators typically contribute to a given process, it is often possible to identify a reduced 
subset of $K$ operators that are most effectively constrained within that specific process. The remaining 
operators may be more tightly constrained through other processes. Selecting an appropriate operator basis,
or performing a suitable rotation, is essential to eliminate blind or weakly sensitive directions in parameter 
space. When such a basis can be determined, matrix-element calculations offer a systematic prescription 
for constructing observables that are optimal for EFT measurements. Therefore, prior to performing any 
measurement, it is necessary to establish a mapping between processes and operators, with weak 
directions appropriately resolved.
In Sec.~\ref{sect:input_4l}, we provide an effective example of basis selection with $K=7$ and the application 
of a set of optimal observables for the analysis of the $H \to 4\ell$ process at the LHC.

The computation of optimized observables, expressed as probability ratios in Eqs.~(\ref{eq:optimized2}) 
and~(\ref{eq:optimized1}), can be performed through the matrix-element calculations presented in Eq.~(\ref{eq:P}). 
These calculations rely on the transfer functions defined in Eq.~(\ref{eq:P}), denoted as 
$p(\vec{x}_\mathrm{reco}|\vec{x}_\mathrm{truth})$, which are essential for accurate probability estimation. 
Unlike in the conventional matrix element method, where the transfer functions must be fully modeled 
to ensure unbiased probability density function to describe a process, in the context of probability ratios, 
many detector effects cancel out. For example, effects like detection efficiency, which depend 
similarly on observables across all models, cancel in the ratio, reducing the need for detailed modeling.
Moreover, imperfections in the transfer functions do not introduce bias but merely lead to a marginal 
loss in optimality of a discriminant. As a result, in scenarios where the process is fully reconstructed with 
sufficient resolution, the transfer functions may often be neglected with minimal impact on performance.

\subsection{Optimized observables using machine learning}
\label{sect:obs-ml}

\begin{figure}[t!]
  \begin{center}
    \captionsetup{justification=centerlast}
    \includegraphics[width=0.3\textwidth]{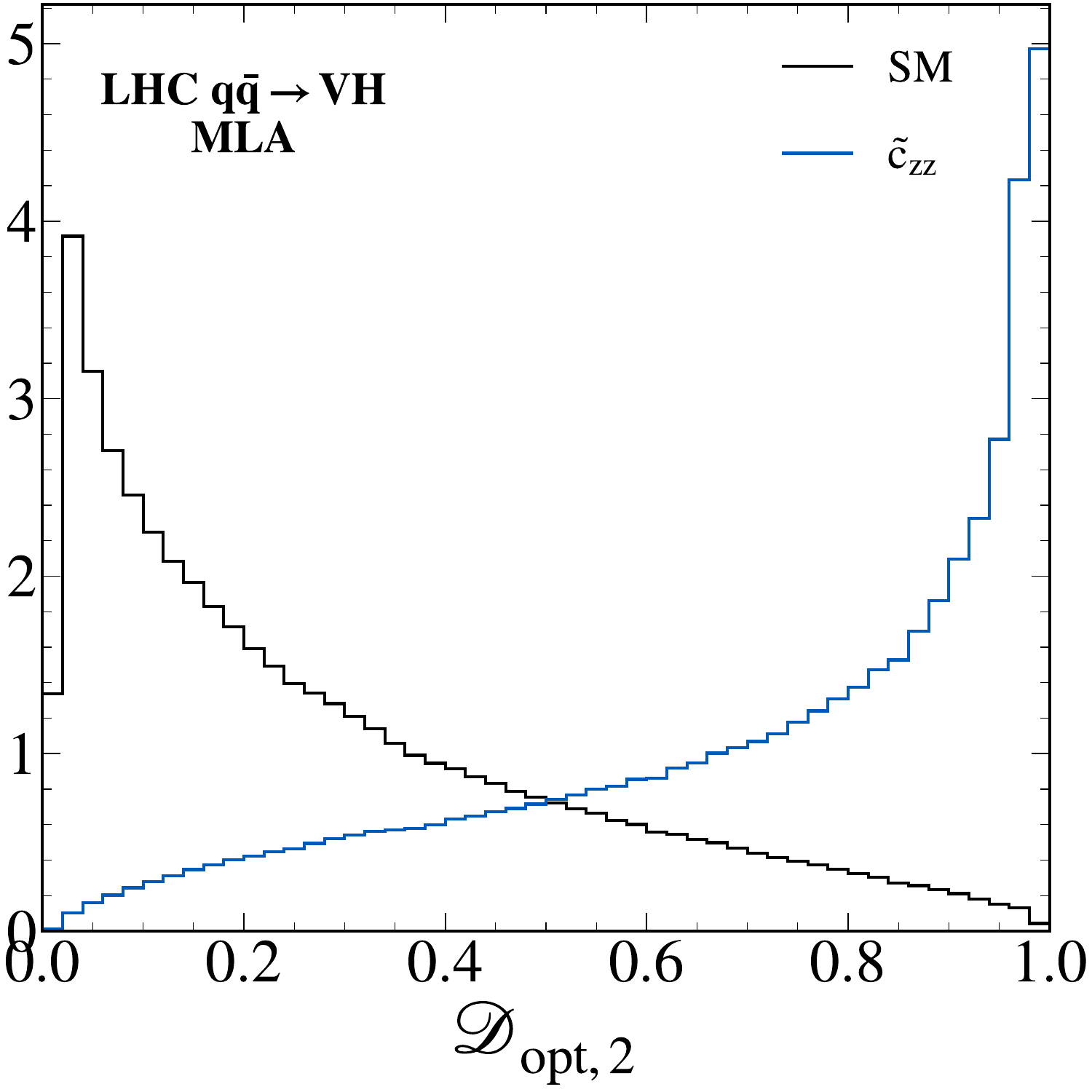}
    \includegraphics[width=0.3\textwidth]{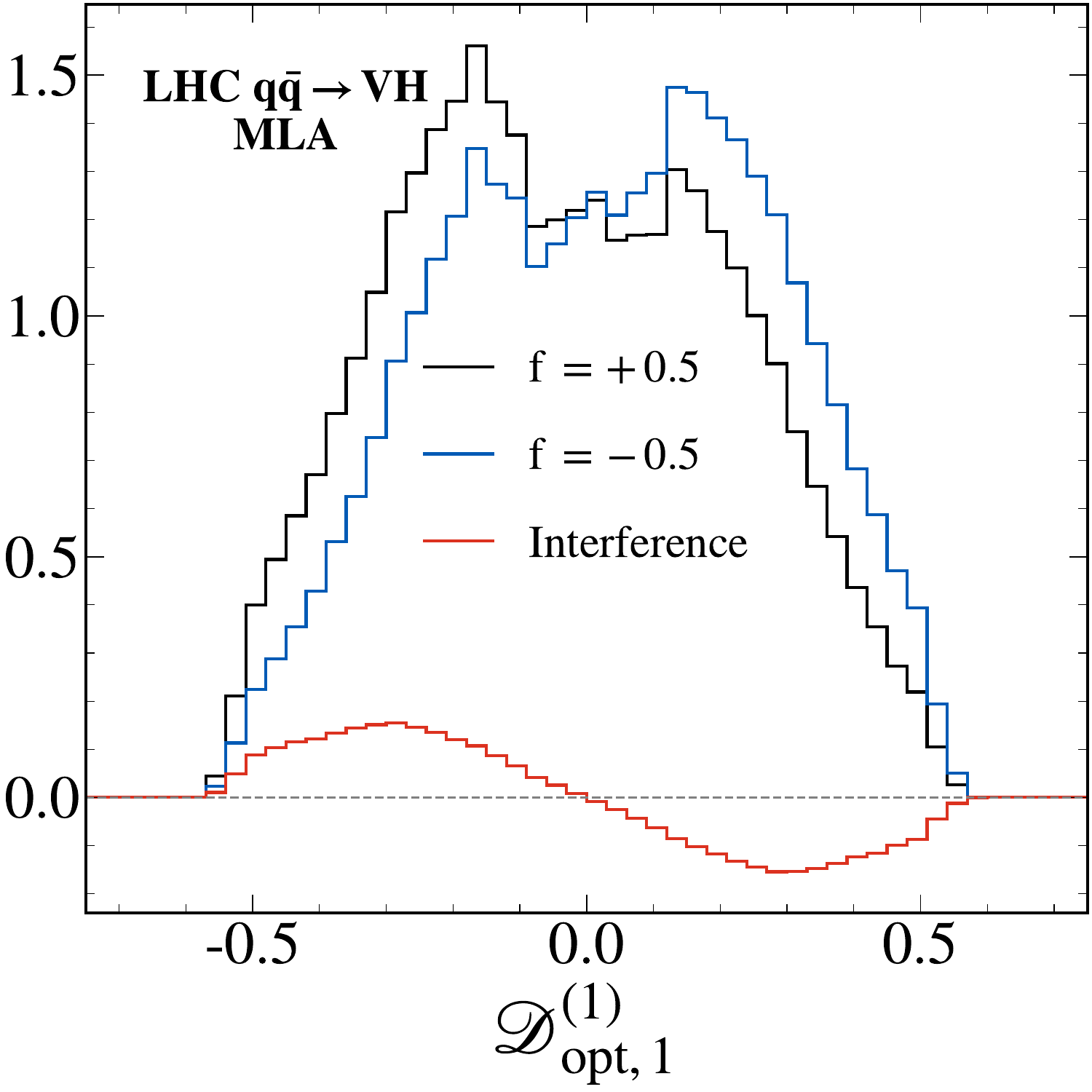}
    \includegraphics[width=0.3\textwidth]{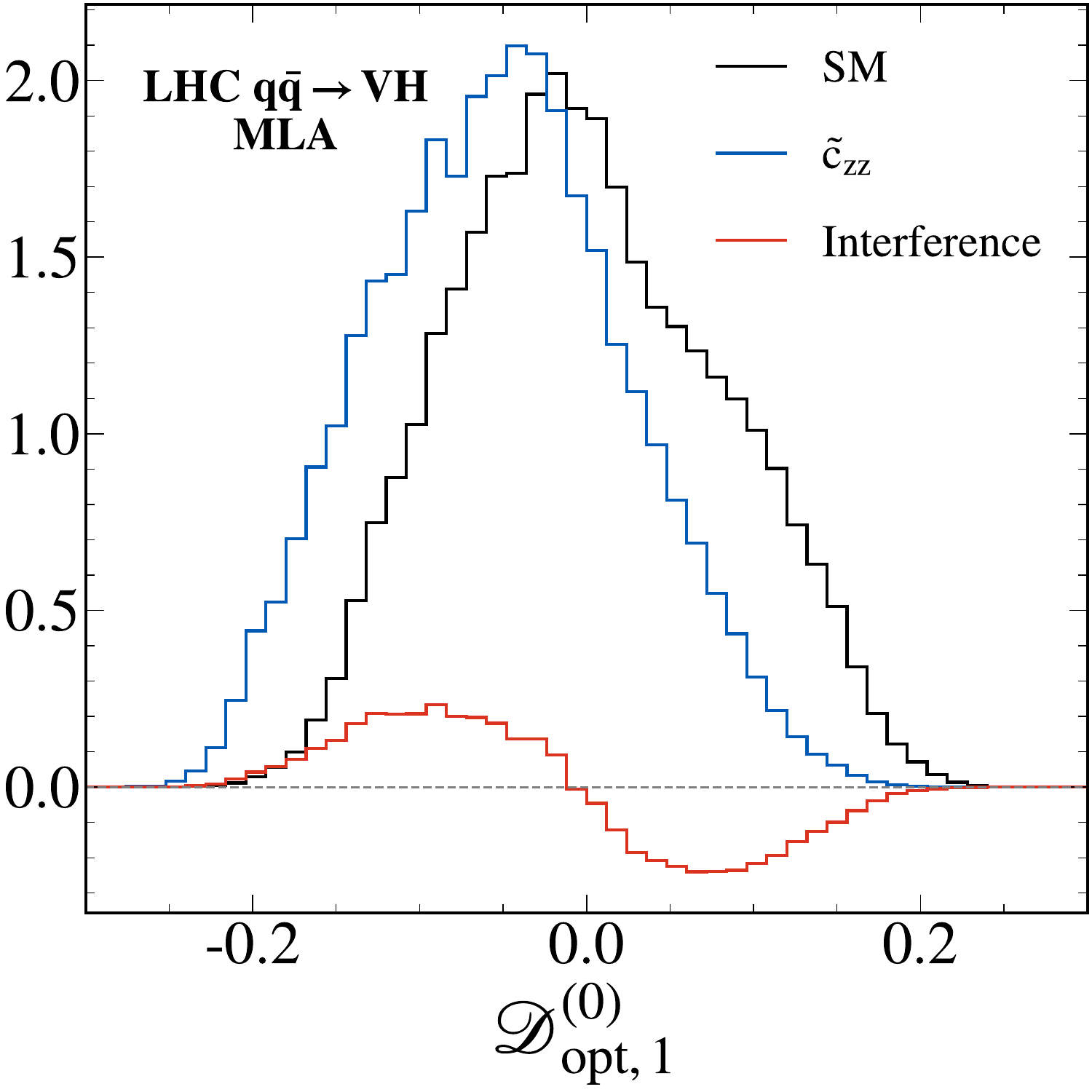}
    \caption{
The distribution of simulated proton-proton collision events at the LHC for the process
$q\bar{q}\to (Z/\gamma^*) \to H(Z/\gamma^*) \to H(\ell^+\ell^-)$
is presented for the three optimal observables, 
${\cal D}_\mathrm{opt,2}$, ${\cal D}_\mathrm{opt,1}^{(1)}$, and $\mathcal{D}_{\text{opt,1}}^{(0)}$,
closely following the presentation of Fig.~\ref{fig:MELA_d0m},
but using the machine-learning approach instead of MELA.
        }
    \label{fig:MLA_d0m}
  \end{center}
\end{figure}

\noindent Matrix-element calculations, though powerful, are often difficult to implement in practice due to several 
substantial challenges. The most critical among these, when applicable, are the transfer functions, 
whose accurate modeling is both technically complex and computationally intensive.
The availability of matrix elements can also be limited in practical applications.
In such cases, machine learning offers a promising alternative. While machine-learning algorithms are trained on 
MC samples that rely on the same underlying matrix elements as those used for the optimal discriminants 
described in Section~\ref{sect:obs-me}, these samples inherently account for parton shower and detector effects,
as illustrated in the diagram in Fig.~\ref{fig:templates}.
As a result, they enable the construction of approximately optimal observables that effectively incorporate 
such realistic experimental effects.

Two key considerations arise in the training of machine-learning algorithms for optimized observable construction: 
(1) the choice of input observables and (2) the selection of training samples. The matrix-element approach offers guidance 
on both fronts and provides a theoretical foundation for integrating machine learning into the construction of optimized 
observables. 
First, the input features should capture the full information content, denoted by $\vec{x}_\mathrm{reco}^\mathrm{\,full}$. 
This input may consist of the reconstructed four-momenta of all relevant particles, as in traditional matrix-element methods, 
or suitably derived physics quantities that are equivalent in terms of information content.
Second, as highlighted in Eqs.~(\ref{eq:optimized2}) and~(\ref{eq:optimized1}), two types of optimal observables 
can be constructed, both of which the machine-learning algorithms must learn to approximate.

The first type corresponds to the standard classification problem between two hypotheses, $\theta_0$ and $\theta_1$. 
In this case, the machine-learning approach (MLA) is trained on two corresponding samples, 
producing an observable that approximates Eq.~(\ref{eq:optimized2}). 
An example of the corresponding observable ${\cal D}_\mathrm{opt,2}$ 
is shown in the left panel of Fig.~\ref{fig:MLA_d0m}, closely reproducing the behavior 
of the MELA-based result displayed in the left panel of Fig.~\ref{fig:MELA_d0m}.
The technical details of the MLA training and it validation are discussed in Sec.~\ref{sect:input_vh}.
At this stage, we simply note that, owing to the full reconstruction of the final state, 
no significant difference in performance of ${\cal D}_\mathrm{opt,2}$ between the two approaches 
is expected or observed, and the comparison is presented purely for illustrative purposes.

The construction of the second type of observable, aimed at isolating 
the interference term, is less straightforward and requires more careful consideration~\cite{Gritsan:2020pib}.
Building on the observations related to Eq.~(\ref{eq:ratio1}), the observable ${\cal D}_\mathrm{opt,1}^{(\alpha)}$ 
with $\alpha \neq 0$ is trained to differentiate between the two models that differ only in the sign of the interference: 
$(\theta_0=1,\theta_1=+\sqrt{\alpha\sigma_0/\sigma_1}\,)$ and $(\theta_0=1,\theta_1=-\sqrt{\alpha\sigma_0/\sigma_1}\,)$. 
This is demonstrated in Fig.~\ref{fig:MLA_d0m} (middle) for ${\cal D}_\mathrm{opt,1}^{(1)}$ with $\alpha = 1$, 
with additional details provided in Sec.~\ref{sect:input_vh}. This sets forth the framework for training such an observable, 
where traditional methods are still applicable for training and validation, as all probability contributions are positive.
A comparison with the direct MELA-based result shown in Fig.~\ref{fig:MELA_d0m} (middle) reveals a reasonable 
agreement, which will be quantified in Sec.~\ref{sect:input_vh}.

In contrast, developing an optimal discriminant ${\cal D}_\mathrm{opt,1}^{(0)}$ for the case \( \alpha = 0 \), 
which is sensitive solely to the interference term, is significantly more complex. This complexity arises because interference 
effects frequently produce both positive and negative probability contributions, making the training of observables and 
the evaluation of their performance more challenging. One can consider training the ${\cal D}_\mathrm{opt,1}^{(\alpha)}$ 
observable with very small values of $\alpha$, effectively approaching the limiting case as $\alpha \to 0$. 
However, in this scenario, the two training samples become nearly indistinguishable, leading to a failure of the training procedure.
The correct procedure involves training to distinguish between the two models: 
the SM and the pure interference contribution, the latter of which can be obtained using the appropriate weights 
derived from the squared matrix elements. This is demonstrated in the observable ${\cal D}_\mathrm{opt,1}^{(0)}$
in Fig.~\ref{fig:MLA_d0m} (right). 
However, training with negative events may present challenges, and it is essential to introduce a methodology 
for evaluating performance with such negative contributions. Consequently, further discussion is deferred to 
Sec.~\ref{sect:input_vh}, after we introduce the necessary tools in Secs.~\ref{sect:input_binary} and ~\ref{sect:input_general}.


\section{Binary classification}
\label{sect:input_binary}

\noindent Let us consider a simple case: binary classification between two processes that both have 
positive probability across the entire phase space, one corresponding to SM and the other driven by the anomalous 
coupling $\theta_1=\tilde{c}_{zz}$ in the process $H\to ZZ\to 2e2\mu$, as defined in Sec.~\ref{sect:input_eft}.
For the purpose of this illustration, we disregard the validity constraints of the EFT and, as a consequence, 
neglect the interference between the amplitudes.
Our goal here is to introduce a method for evaluating the performance of observables using the 
Receiver Operating Characteristic (ROC) curves, and to demonstrate how optimization based on the 
ROC score can achieve near-optimal performance comparable to the likelihood-based construction 
of an observable defined in Eq.~(\ref{eq:optimized2}). This discussion goes beyond a purely academic 
exercise, as we will further develop and apply this approach in the following sections.

\subsection{The classical ROC curve}

\noindent  ROC curves were originally developed in World War II~\cite{RM-753-PR,Peterson1954SignalDetectability} 
to distinguish between radar ``noise'' and actual aircraft detection. In the context of binary classification, 
the ROC curve plots the True Positive Rate (TPR) against the False Positive Rate (FPR), defined as 
\begin{align}
    \text{TPR}(t) = \frac{1}{|B|} \sum_{x \in B} \mathbb{I}[s(x) > t], 
    \label{fig:roc_TPR}
    \\
    \text{FPR}(t) = \frac{1}{|A|} \sum_{x \in A} \mathbb{I}[s(x) > t],
    \label{fig:roc_FPR}
\end{align}
where \( \mathbb{I}=1 \) if the condition is true and 0 otherwise, the function \( s(x) \) represents the classifier score assigned 
to an event \( x \), and \( B \) and \( A \) denote the ``positive" and ``negative" samples of events, with sizes \( |B| \) and \( |A| \), respectively.
Varying the threshold \( t \) across the score range produces the ROC curve.

\begin{figure}[b!]
    \captionsetup{justification=centerlast}
    \centering
	\includegraphics[width=0.3\textwidth]{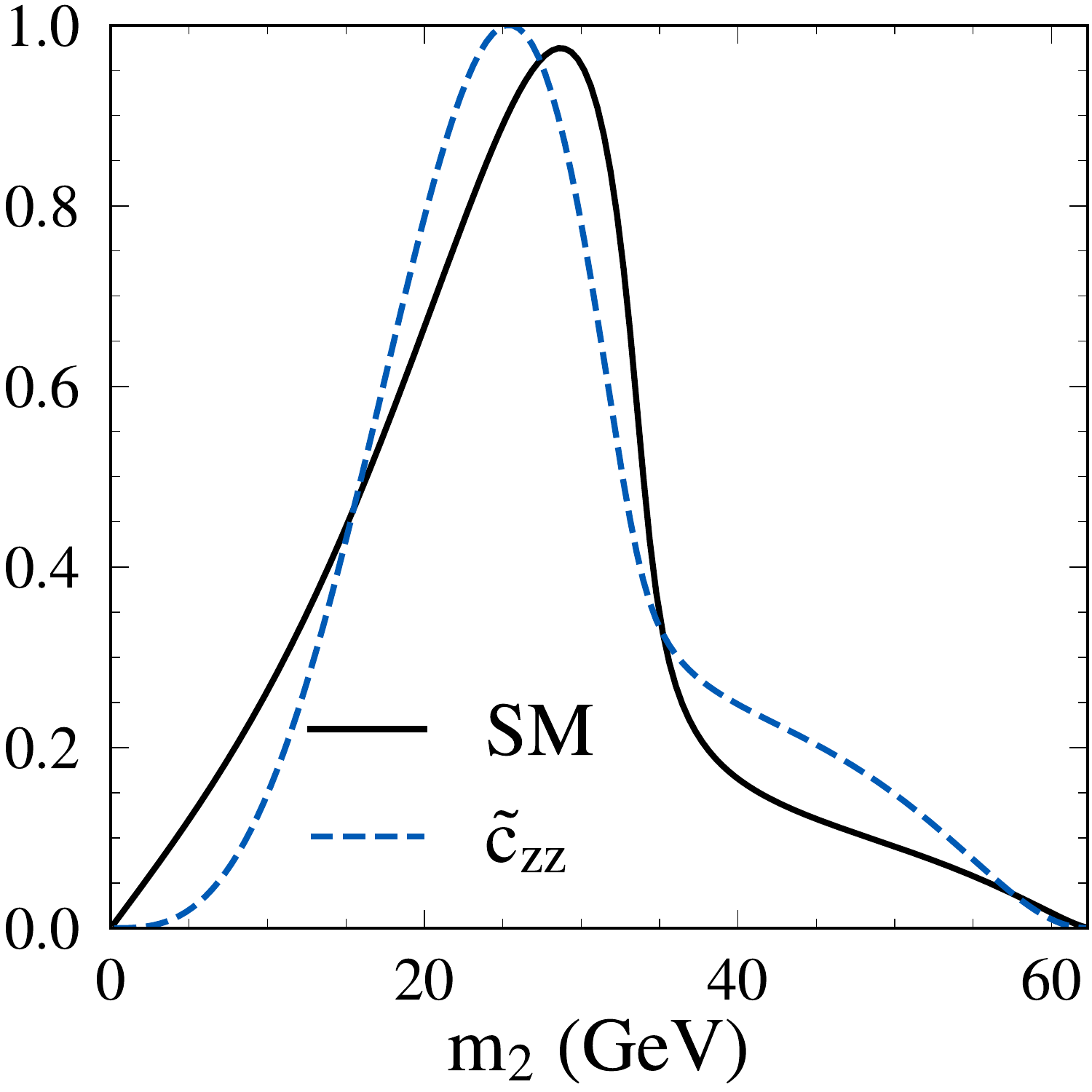}
	\includegraphics[width=0.3\textwidth]{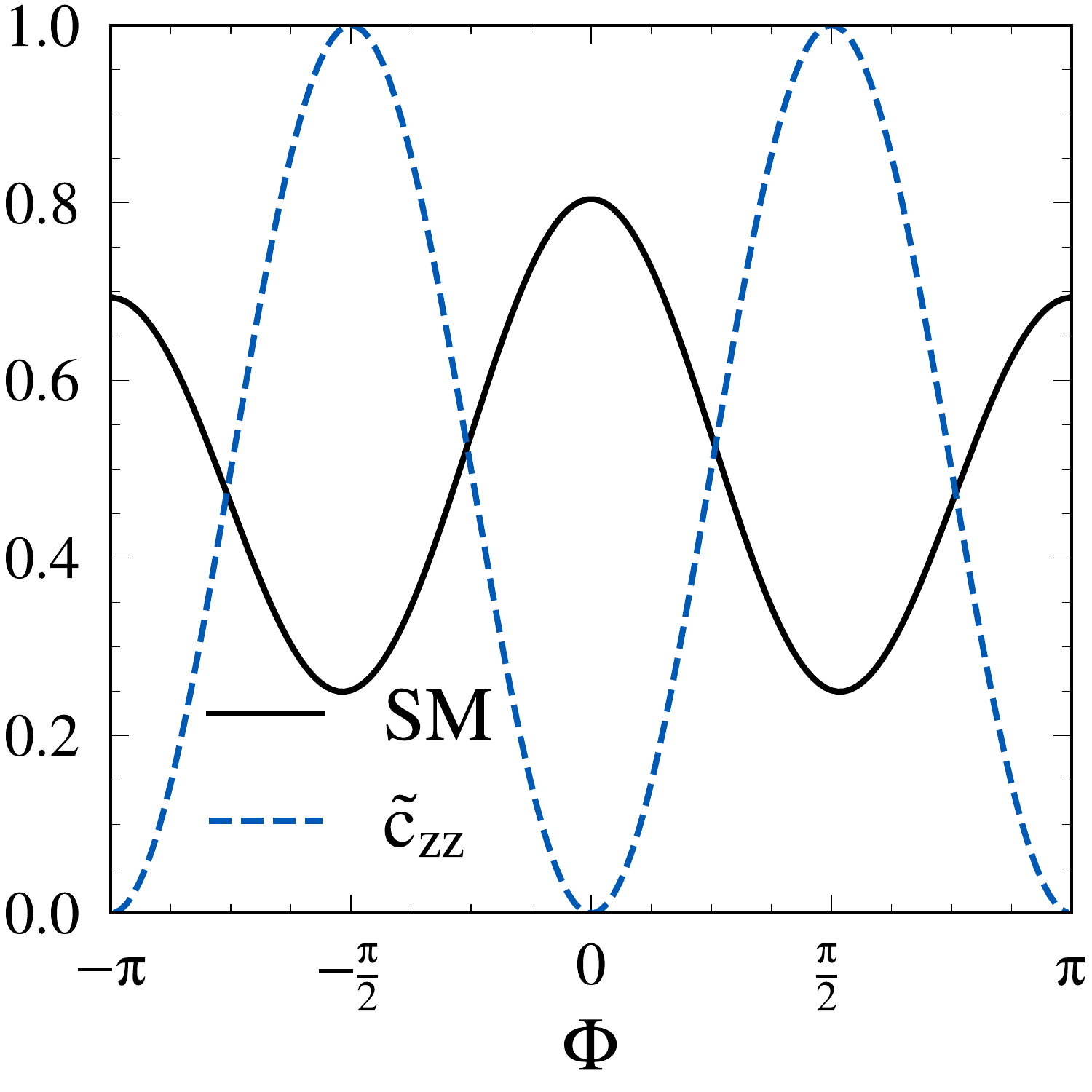}
	\caption{
Distributions of \( m_2 \) (left) and \( \Phi \) (right) in \( H \to ZZ \to 2e2\mu \) decay 
are shown for both the SM and the scenario driven by the \( \tilde{c}_{zz} \) coupling, as discussed in the text.
}
    \label{fig:observables_m2}
\end{figure}

\begin{figure}[h!]
    \captionsetup{justification=centerlast}
    \centering
    \includegraphics[width=0.3\textwidth]{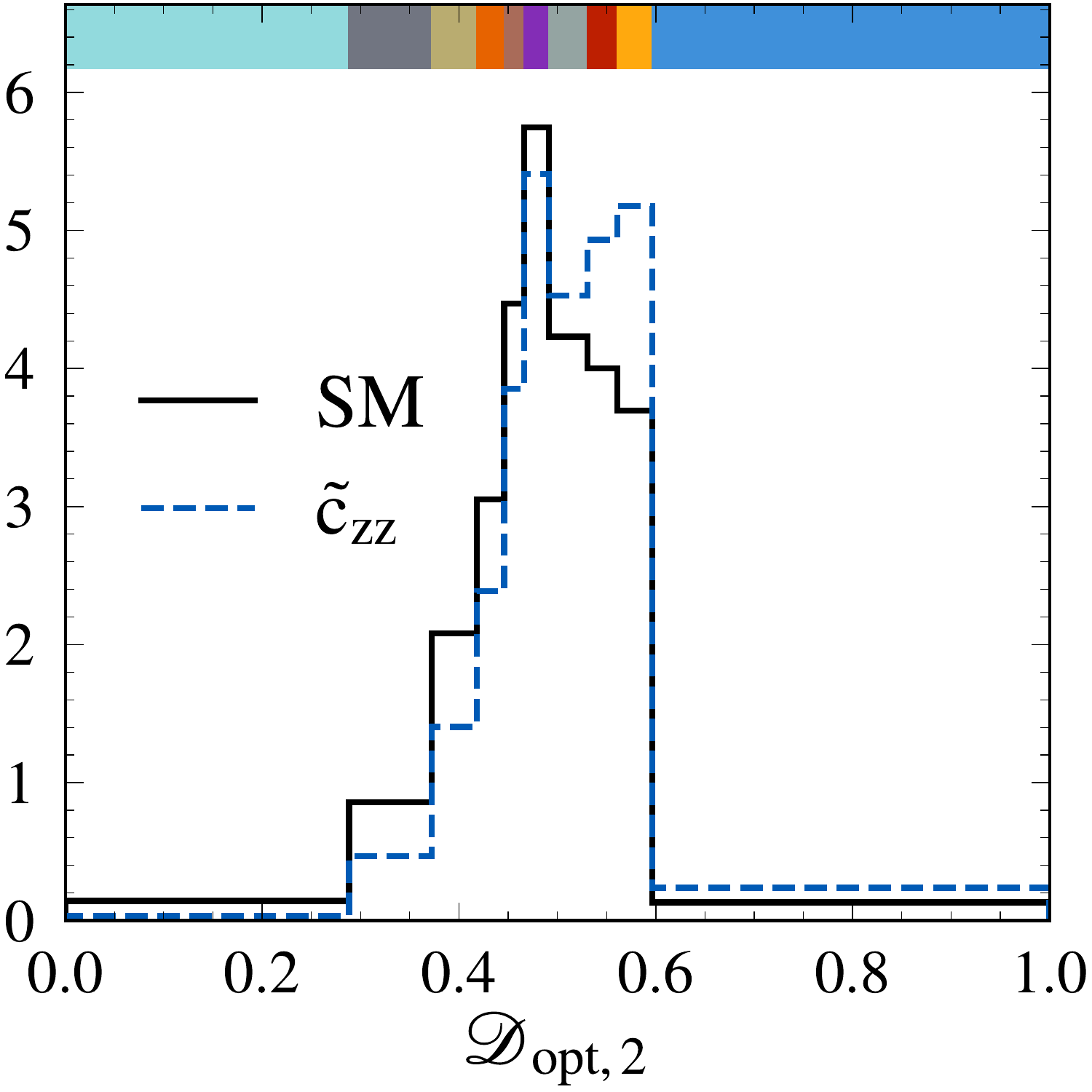}
    \includegraphics[width=0.3\textwidth]{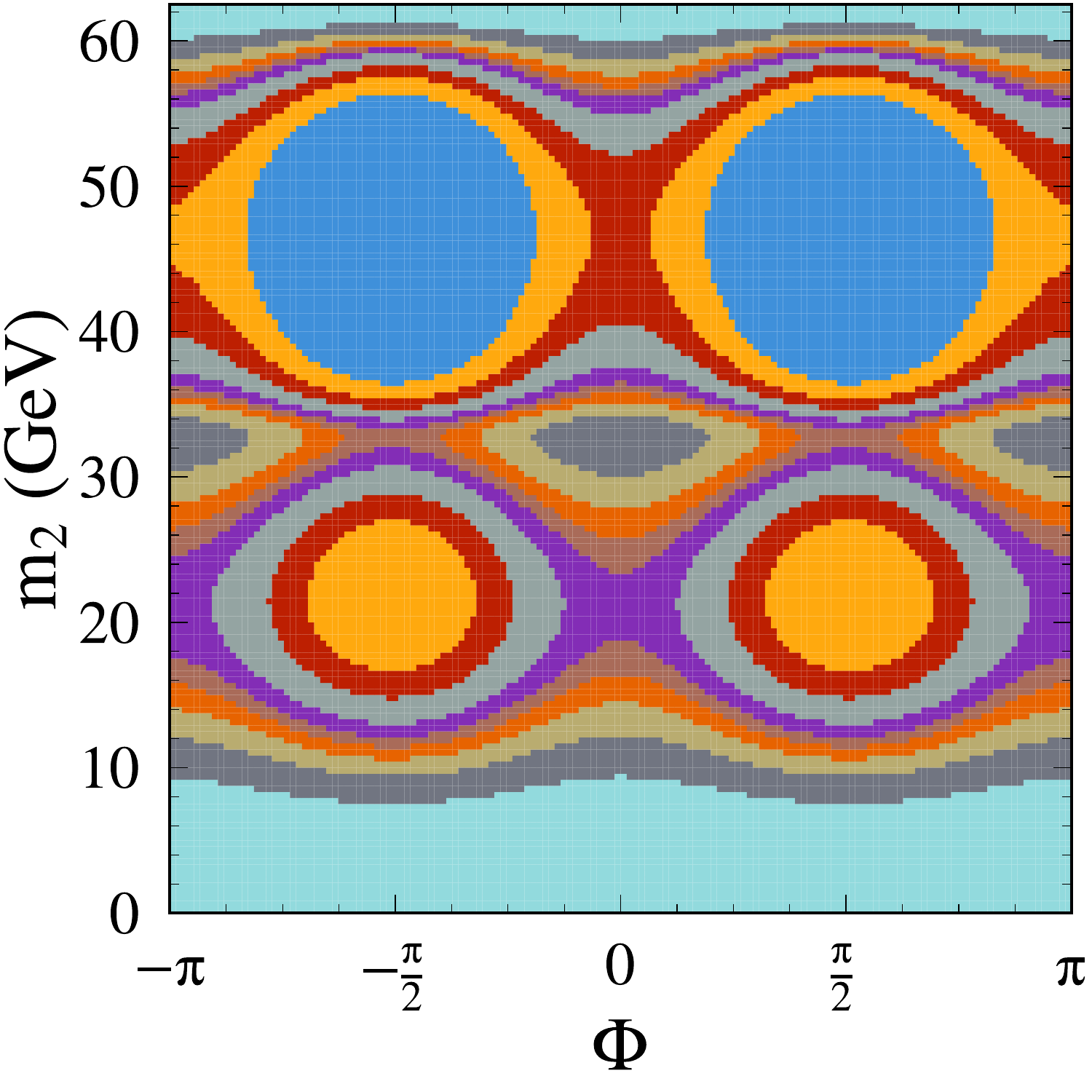}
	\caption{
Left: Distributions of \( {\cal D}_\mathrm{opt,2} \) in \( H \to ZZ \to 2e2\mu \) decay
are shown for both the SM and the scenario driven by the \( \tilde{c}_{zz} \) coupling.
The observable range is divided into 10 bins, selected to maximize the separation between the two scenarios, 
with bin color scheme at the top matching the right-hand plot.
Right: The \( (m_2, \Phi) \) plane, composed of \( 150 \times 150 \) bins, is divided into ten two-dimensional 
regions, each color-coded according to the ten bins shown in the left-hand plot.
    }
    \label{fig:observables_D}
\end{figure}

To illustrate the construction of the ROC curve, we use two observables, \( m_2 \) and \( \Phi \), 
defined in the \( H \to ZZ \to 2e2\mu \) decay process depicted in Fig.~\ref{fig:process} (right), 
along with the optimized observable \( {\cal D}_\mathrm{opt,2} \), computed according to Eq.~(\ref{eq:optimized2}) 
using a binned version of the analytical probability distribution as a function of \( m_2 \) and \( \Phi \) 
from Ref.~\cite{Bolognesi:2012mm}.
Here, \( m_2 \) refers to the smaller invariant mass of the two reconstructed \( Z \) boson candidates
and  \( \Phi \) is the angle between the decay planes of the two \( Z \) bosons. 
To simplify the visualization, we integrate over all other observables to exclude them.
Figure~\ref{fig:observables_m2} shows the distributions of \( m_2 \) and \( \Phi \).
Figure~\ref{fig:observables_D} presents the distribution of \( {\cal D}_\mathrm{opt,2} \), 
along with illustration of its correlation with \( m_2 \) and \( \Phi \).
An analytically generated toy model is presented, based on the calculations from Ref.~\cite{Bolognesi:2012mm}.

The optimal observable \( {\cal D}_\mathrm{opt,2} \) is constructed such that higher values 
correspond to a greater likelihood of model \( \theta_1 \), and therefore it can be used as a classifier score 
\( s(x) \) in Eqs.~(\ref{fig:roc_TPR}) and~(\ref{fig:roc_FPR}) directly. However, the observables  \( m_2 \) and \( \Phi \) 
are not ordered accordingly and thus cannot be directly used as classifier scores.
However, a simple transformation with the ratio of probabilities, such as \( s(x) = {\cal P}_1(m_2) / {\cal P}_0(m_2) \), 
can be applied instead. Alternatively, when using discrete binning of the observables, the bins can be reordered 
based on the increasing value of the same ratio, such as \( {\cal P}_1(m_2) / {\cal P}_0(m_2) \).

\begin{figure}[t!]
    \captionsetup{justification=centerlast}
    \centering
	\includegraphics[width=0.3\textwidth]{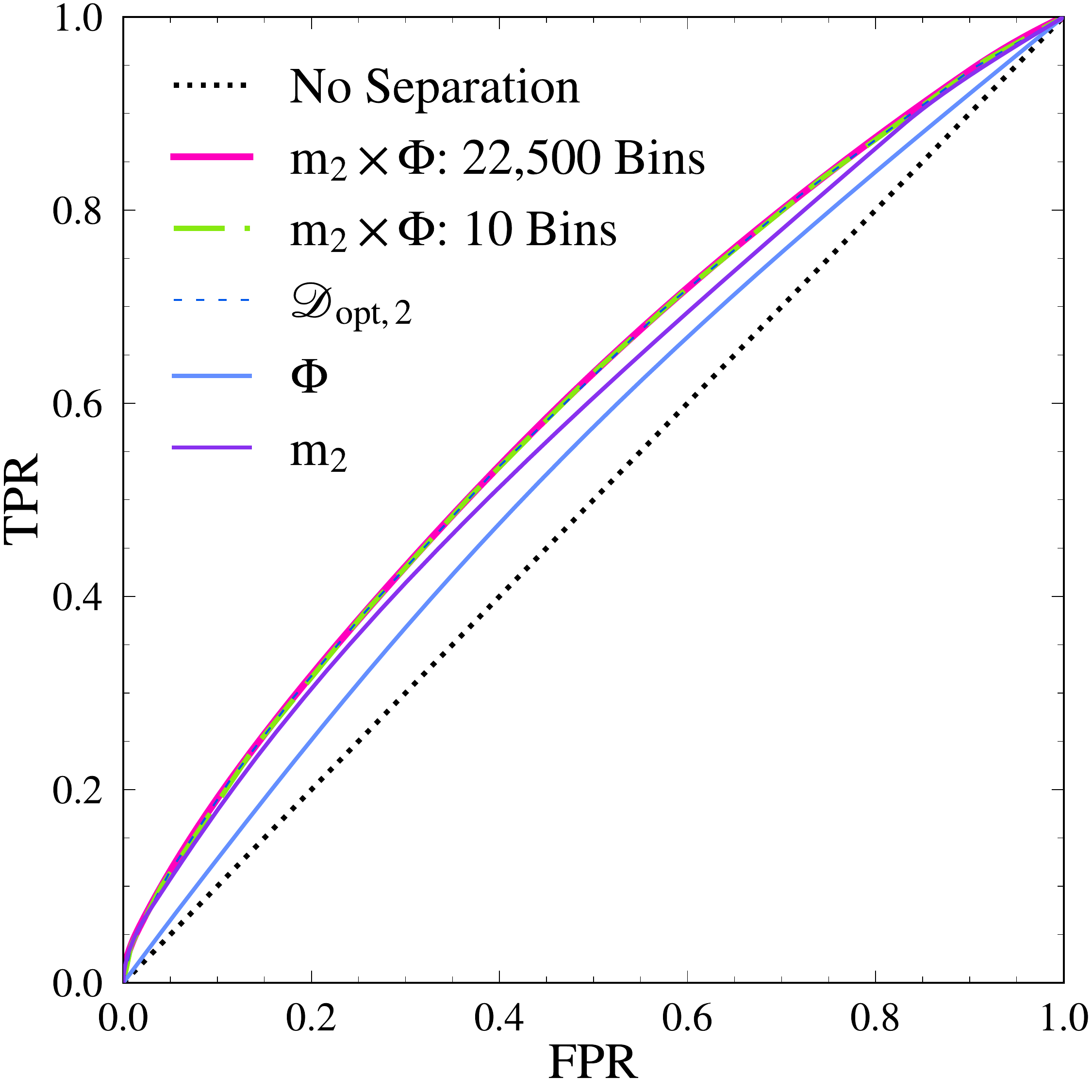}
	\includegraphics[width=0.3\textwidth]{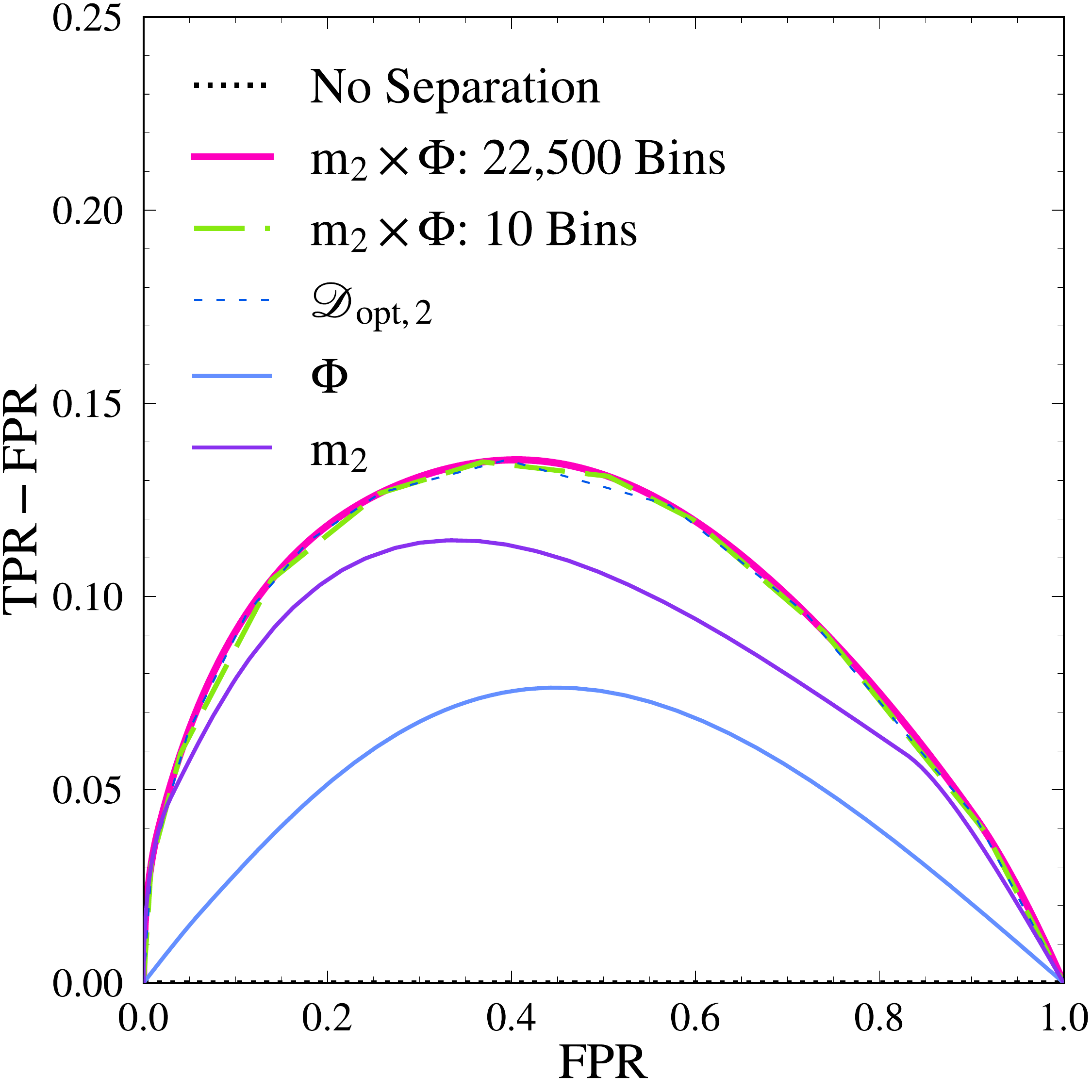}
	\caption{
The ROC curves using the observables \( m_2 \), \( \Phi \), \( {\cal D}_\mathrm{opt,2} \),
and the input ($22\,500$ bins) and the output (10 bins) of the bin-merging procedure 
in \( H \to ZZ \to 2e2\mu \) decay for the scenario driven by the \( \tilde{c}_{zz} \) 
coupling (``positive") and the SM (``negative"). 
The dotted line represents the case where there is no separation. 
The right plot is a zoomed-in version of the left, showing deviations from the no-separation diagonal.
}
    \label{fig:roc_example}
\end{figure}

Figure~\ref{fig:roc_example} presents the ROC curves for the three observables employed as classifier 
scores \( s(x) \), used to distinguish the $\theta_1 = \tilde{c}_{zz}$ model (treated as the ``positive" sample) 
from the SM baseline $\theta_0$ (treated as the ``negative" sample).
The area under the ROC curve (AUC) reflects the degree of separation between two models
and serves as a useful metric for evaluating the performance of the binary classification algorithm~\cite{Bradley_ROC}.
Since \( {\cal D}_\mathrm{opt,2} \) captures information from both observables, \( m_2 \) and \( \Phi \), 
constituting the complete information in our case, it achieves the highest AUC and therefore outperforms 
both input observables.
As explained later, the \( {\cal D}_\mathrm{opt,2} \) observable shown in Fig.~\ref{fig:observables_m2} is divided 
into 10 optimal bins, and the ROC curve presented in Fig.~\ref{fig:roc_example} is calculated based on these 10 steps.

\subsection{Minimal-Loss Merging}

\noindent An optimal data analysis would make use of the full unbinned information
based on the set of observables $\vec{x}_\mathrm{reco}^\mathrm{\,full}$ defined in Sec.~\ref{sect:input_obs}, 
which, in the context of the toy model considered here, corresponds to $(m_2, \Phi)$.
This is equivalent to having access to the full two-dimensional $(m_2, \Phi)$ probability distribution 
corresponding to Fig.~\ref{fig:observables_D} (right).
In most cases, such a continuous description is not feasible. Due to practical constraints, such as the limited 
number of events available from simulations of the processes under realistic detector conditions or from control 
samples in the data, only a finite number of discrete bins can be employed for reliable parameterization, 
typically ranging between 10 and 100.

Therefore, dimensionality reduction, referring to both the number of observables and the number of bins
used to represent them, is a key challenge in practical data analysis applications.
Fortunately, achieving near-optimal performance does not require a continuous description of the probability 
distribution, or equivalently, an extremely large number of bins.
This is demonstrated in Figs.~\ref{fig:observables_D} and~\ref{fig:roc_example}, where the 10-bin representation 
of the \( {\cal D}_\mathrm{opt,2} \) observable achieves nearly the same AUC as the full $(m_2, \Phi)$ distribution 
discretized into $150 \times 150 = 22\,500$ bins. 
In Sec.~\ref{sect:input_4l}, we will demonstrate how the ROC curve metric relates 
to the precision of constraints on the parameters of interest.

Figures~\ref{fig:observables_D} and~\ref{fig:roc_example} indicate that calculating the optimal 
discriminant, whether through matrix-element methods or machine-learning techniques, is an effective approach 
to dimensionality reduction. These methods effectively map the multi-dimensional space of observables 
$\vec{x}_\mathrm{reco}^\mathrm{\,full}$ onto classifier scores $s(x)$ that are optimal for the objectives of the analysis.
This motivates the exploration of an alternative approach to dimensionality reduction, which we outline next.
A detailed description of the full method and the metric used is provided in Sec.~\ref{sect:input_general}.
A simplified overview is presented here. 

\begin{figure}[t!]
    \captionsetup{justification=centerlast}
    \centering
    \includegraphics[width=0.3\textwidth]{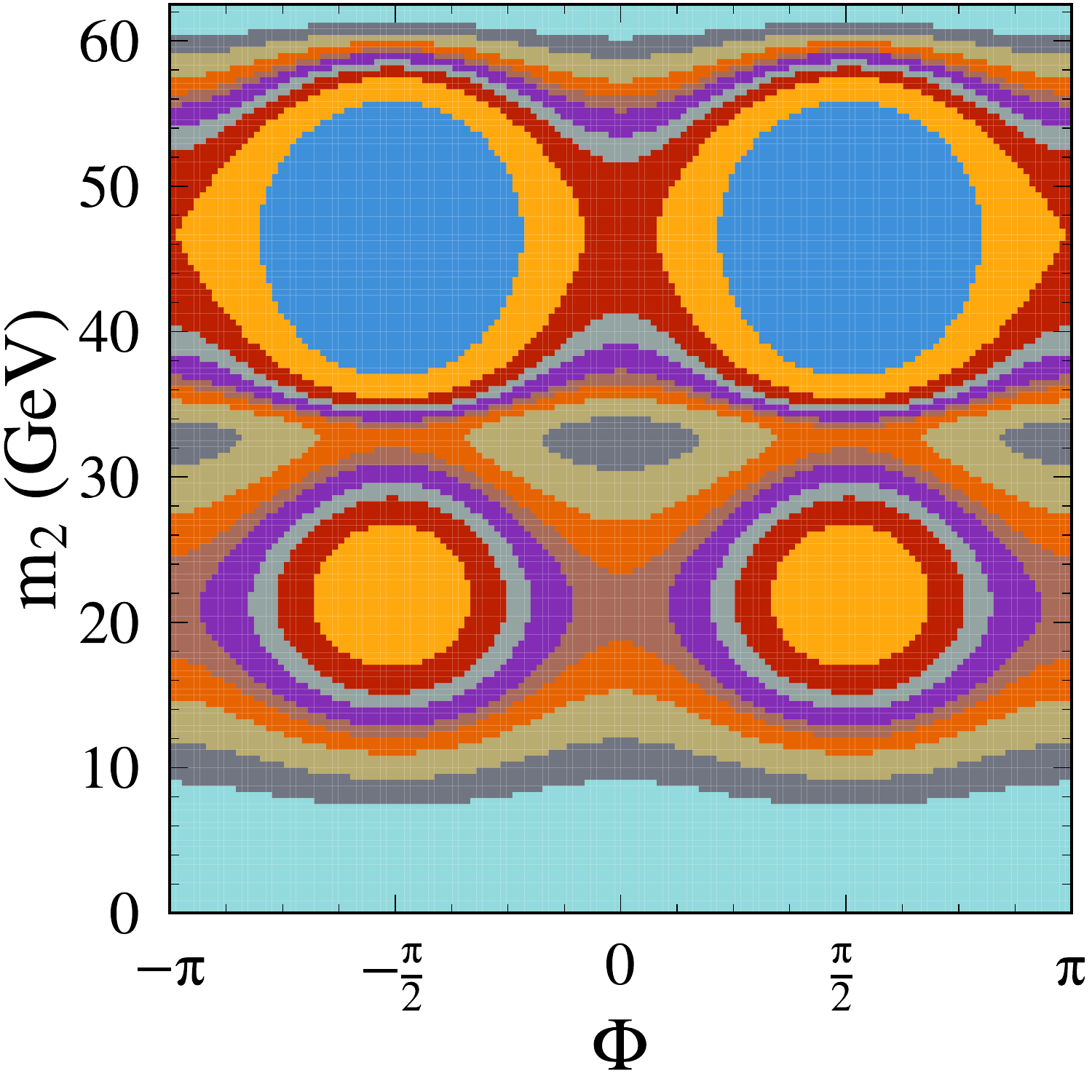}
    \includegraphics[width=0.3\textwidth]{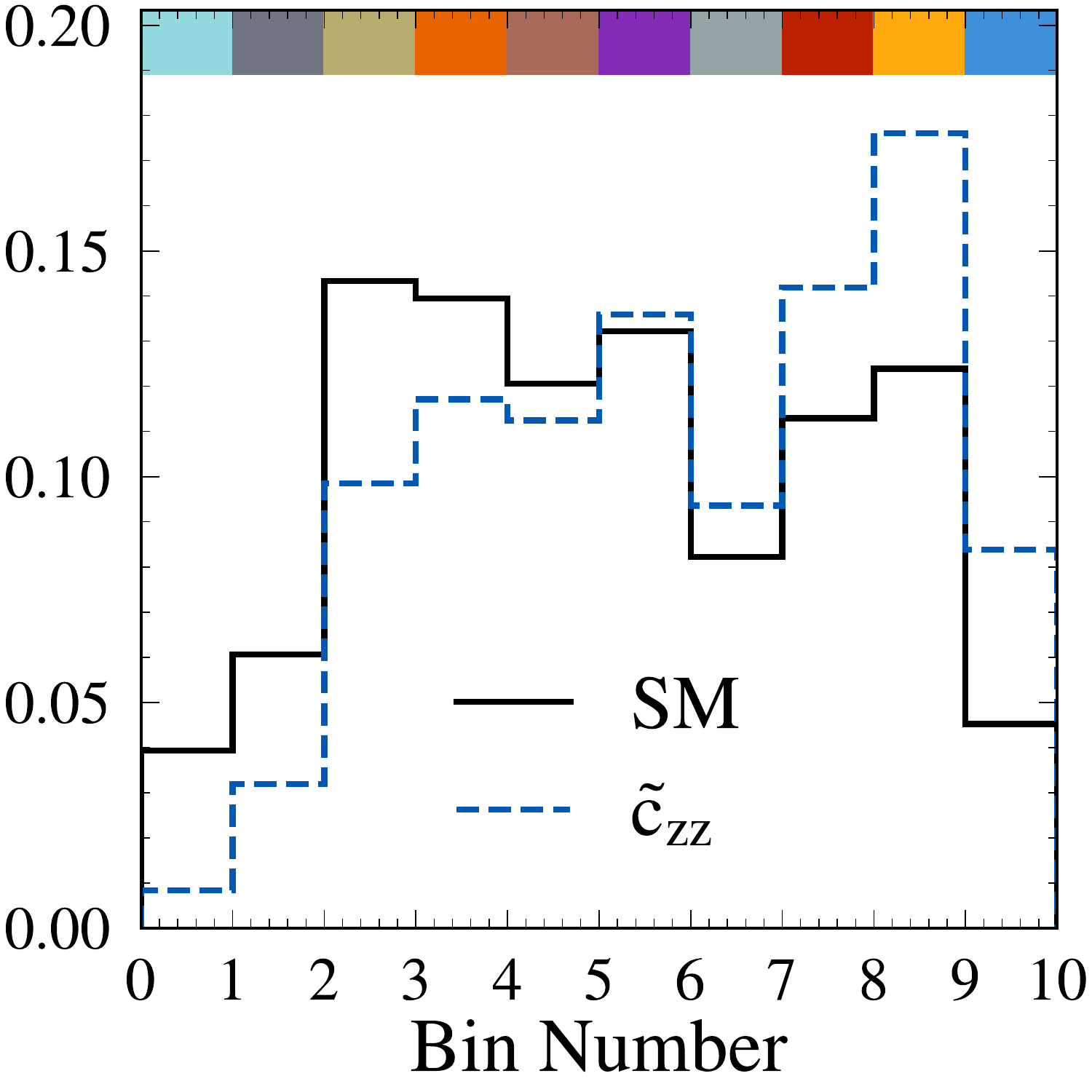}
    \caption{
Left: The \( (m_2, \Phi) \) plane, composed of \( 150 \times 150 \) bins, is divided into ten two-dimensional color-coded regions 
corresponding to the ten categories defined by the Minimal-Loss Merging procedure in the decay \( H \to ZZ \to 2e 2\mu \). 
Right: Distributions of the ten bins defined according to the two-dimensional regions in the left plot, with the corresponding 
color code at the top. The SM and the scenario driven by the \( \tilde{c}_{zz} \) coupling are shown.
    }
    \label{fig:Tiling}
\end{figure}

Beginning with the $22\,500$ bins of the $(m_2, \Phi)$ distribution shown in Fig.~\ref{fig:observables_D} (right), 
we can merge the bins in a manner that preserves a maximal achievable ROC curve metric, such as the AUC.
The procedure is performed iteratively, merging at each step a pair of bins, which need not be adjacent,
that maximizes the ROC curve metric, or equivalently, minimizes the loss.
As an example, the process can be repeated $22\,490$ times, resulting in a total of 10 bins. 
As described, this process may initially appear highly inefficient. 
However, we will discuss improvements in Sec.~\ref{sect:input_general}.

The result of this bin-merging procedure, referred to as Minimal-Loss Merging and
introduced in detail in Sec.~\ref{sect:input_milomerge}, is shown in Fig.~\ref{fig:Tiling}. 
A comparison between Figs.~\ref{fig:Tiling} and~\ref{fig:observables_D} shows that the binning derived 
from the analytic distribution closely resembles the one obtained by optimizing the ROC curve score, 
both visually and in terms of performance shown in Fig.~\ref{fig:roc_example}. 
The performance and appearance are not expected to be identical, due to approximations inherent 
in the use of finite binning. Although in this specific application the procedure offers no practical advantage, 
given the availability of an analytical description, it nevertheless demonstrates the effectiveness 
of this simple method in achieving the desired outcome in scenarios where direct analytical approaches 
are not readily available.

It is remarkable that the 10 bins produced by the Minimal-Loss Merging procedure achieve nearly 
the same performance as the original $22\,500$ bins, which retain the full information about the process,
as demonstrated in Fig.~\ref{fig:roc_example}. 
This example highlights the effectiveness of dimensionality reduction, as analyzing the data using 
10 bins is clearly much simpler than working with the original $22\,500$ bins.
Although this is a relatively simple toy model example, we will later demonstrate 
the strength of this approach in the more complex context of a full EFT analysis.
To achieve this, in the following section we extend this approach to address cases involving 
negative probabilities and multiple classifier scores.
This approach can be generalized to achieve optimal binning of any arbitrary distribution 
into any desired number of bins, provided that suitable training samples are available to calculate the ROC scores. 


\section{Generalization of binary classification}
\label{sect:input_general}

\noindent 
In this section, we review the shortcomings of the traditional ROC curve and AUC for EFT analyses.
We put forward a series of arguments advocating for the creation of a new metric, different from the AUC, 
along with a novel ROC curve approach that addresses the identified limitations while staying computationally 
feasible with current technology. The new metric and the new ROC curve approach have been integrated 
into a bin-merging framework, Minimal-Loss Merging, first introduced in Sec.~\ref{sect:input_binary}.

\subsection{Limitations of using the classical ROC curve approach in EFT}

\noindent 
When applied to Eq.~(\ref{eq:probreco}), Sec.~\ref{sect:input_binary} focused on the addition of the term proportional 
to $\theta_k^2$ to the SM. In this context, the classical ROC curve approach was well-suited.
However, in the context of EFT, the focus shifts to the term linear in $\theta_k$ in Eq.~(\ref{eq:probreco}), 
which introduces challenges unique to EFT analyses. This linear term, representing the interference between 
the BSM and SM contributions, can take both positive and negative values. 
While the full probability, incorporating all terms in Eq.~(\ref{eq:probreco}), remains positive, 
the sample associated with the interference term may include both positive and negative ``events.'' 
This feature is often essential for capturing the key information encoded in the interference effects. 
However, classical techniques, whether in machine learning algorithms or performance metrics, 
are typically not designed to accommodate such negative inputs.

\begin{figure}[t!]
    \captionsetup{justification=centerlast}
    \centering
    \includegraphics[width=0.32\textwidth]{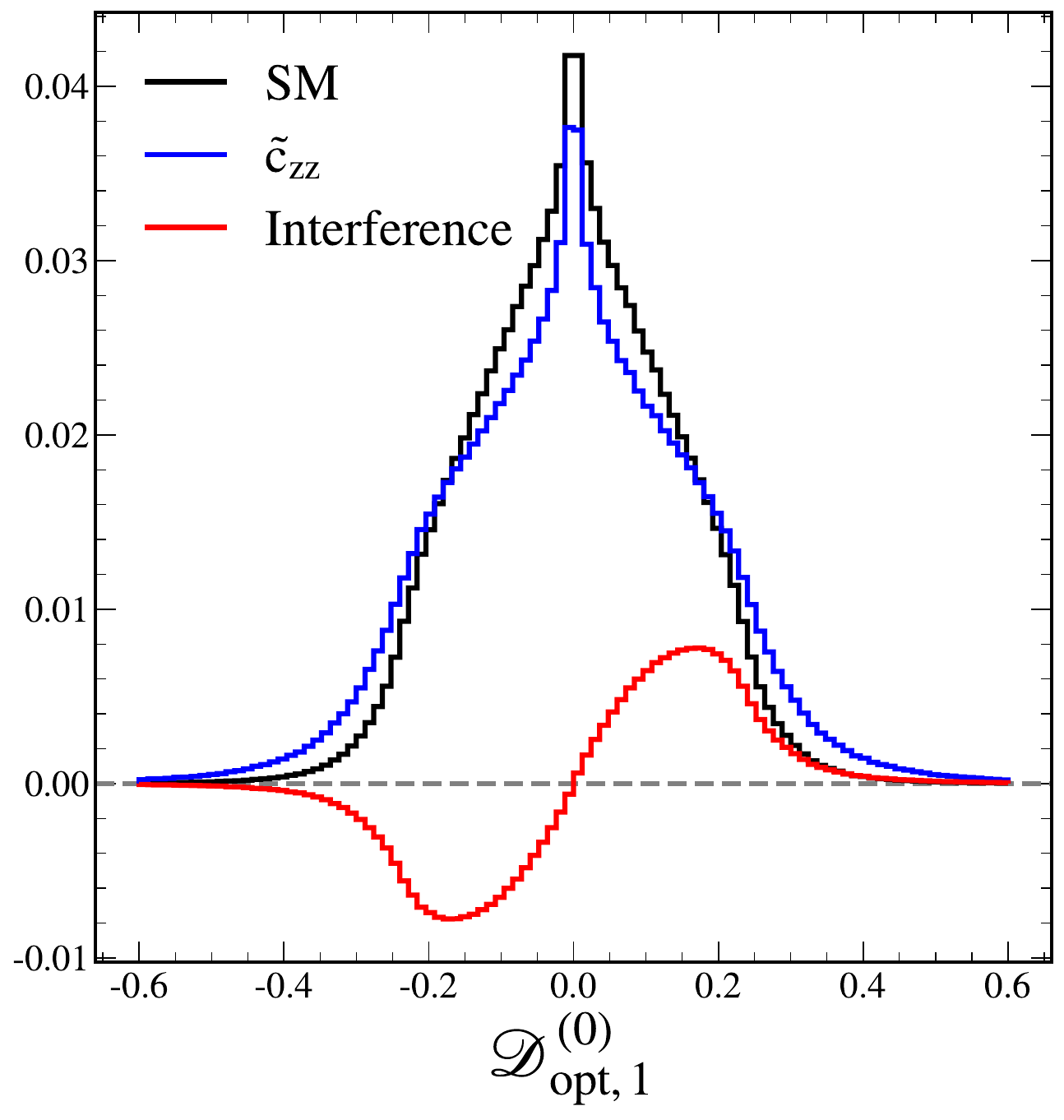}
    \includegraphics[width=0.32\textwidth]{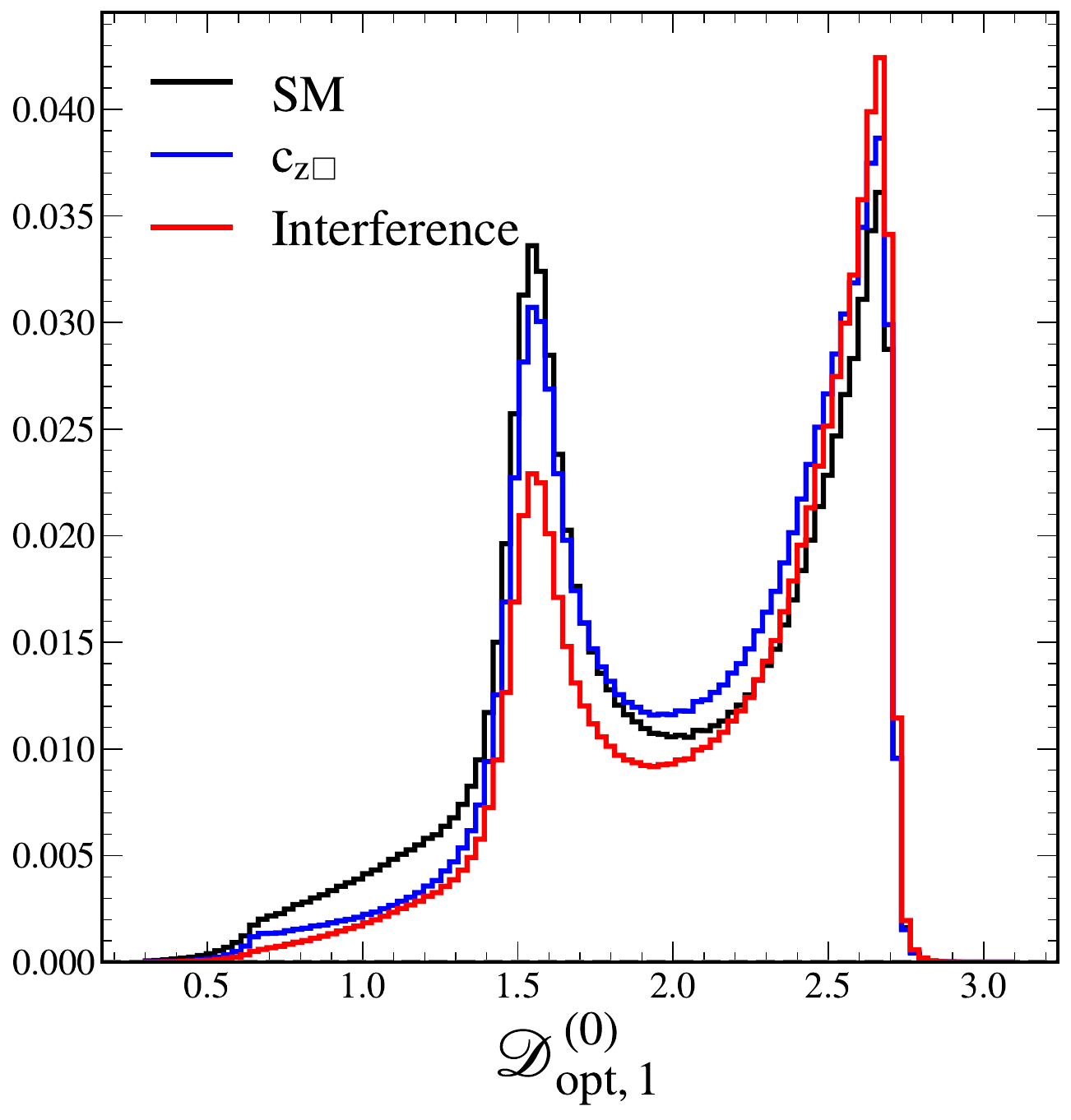}
    \includegraphics[width=0.32\textwidth]{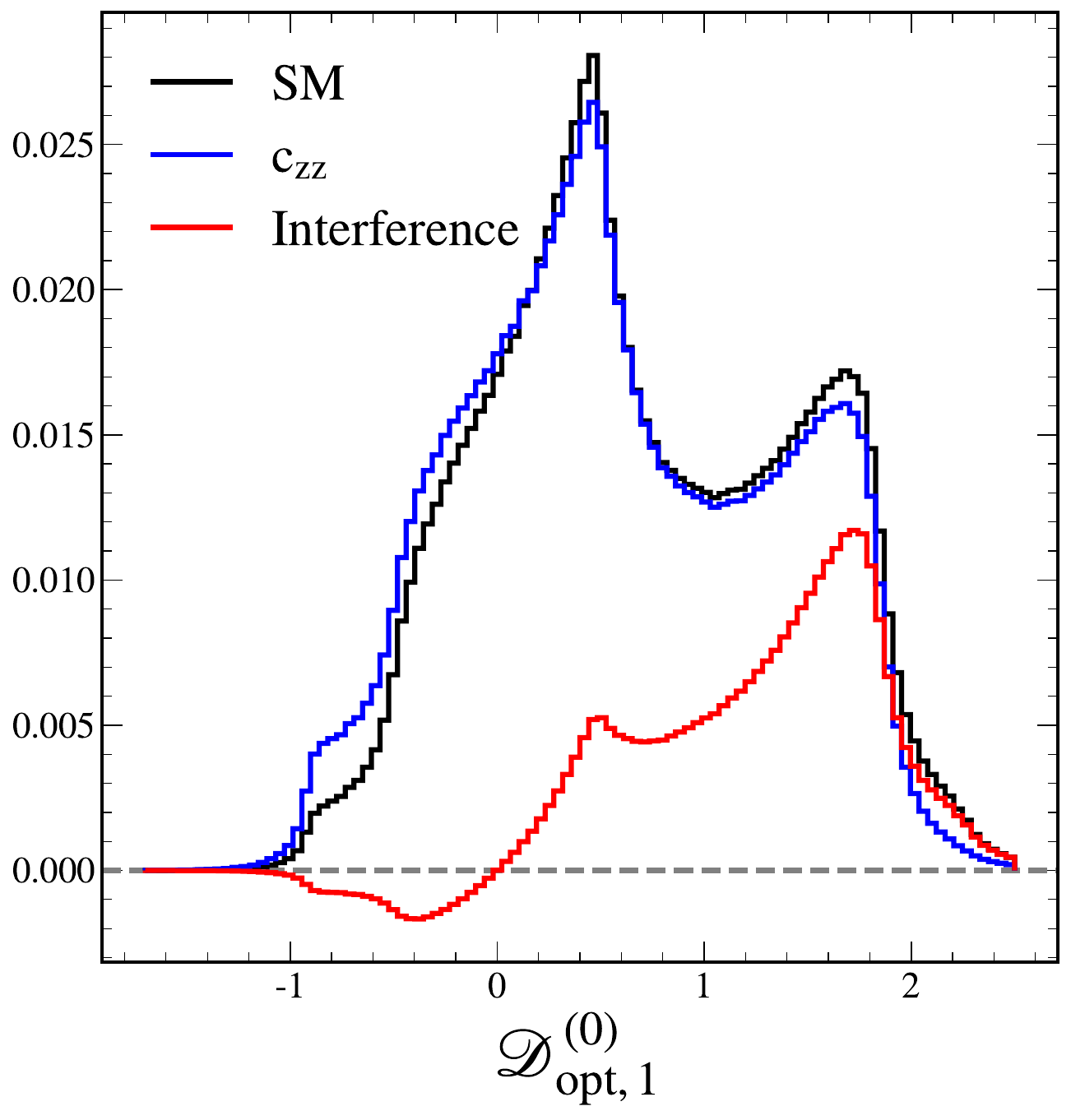}
	\caption{
Distributions of ${\cal D}_\mathrm{opt,1}^{(0)}$ for the process $H \to (Z/\gamma^*)(Z/\gamma^*) \to 2e2\mu$ 
are shown for the SM (black) and BSM (blue) scenarios, as well as for the interference (red) between the SM and BSM 
contributions driven by the couplings \( \theta_1 = \tilde{c}_{zz} \) (left), ${c}_{z\Box}$ (middle), and ${c}_{zz}$ (right). 
The BSM coupling strength is enhanced beyond the EFT validity range to make all contributions clearly visible.
The discriminant is computed using Eq.~(\ref{eq:optimized1}) with \( \alpha = 0 \).
}
    \label{fig:discr1_3cases}
\end{figure}

\begin{figure}[t!]
    \captionsetup{justification=centerlast}
    \centering
    \includegraphics[width=0.32\textwidth]{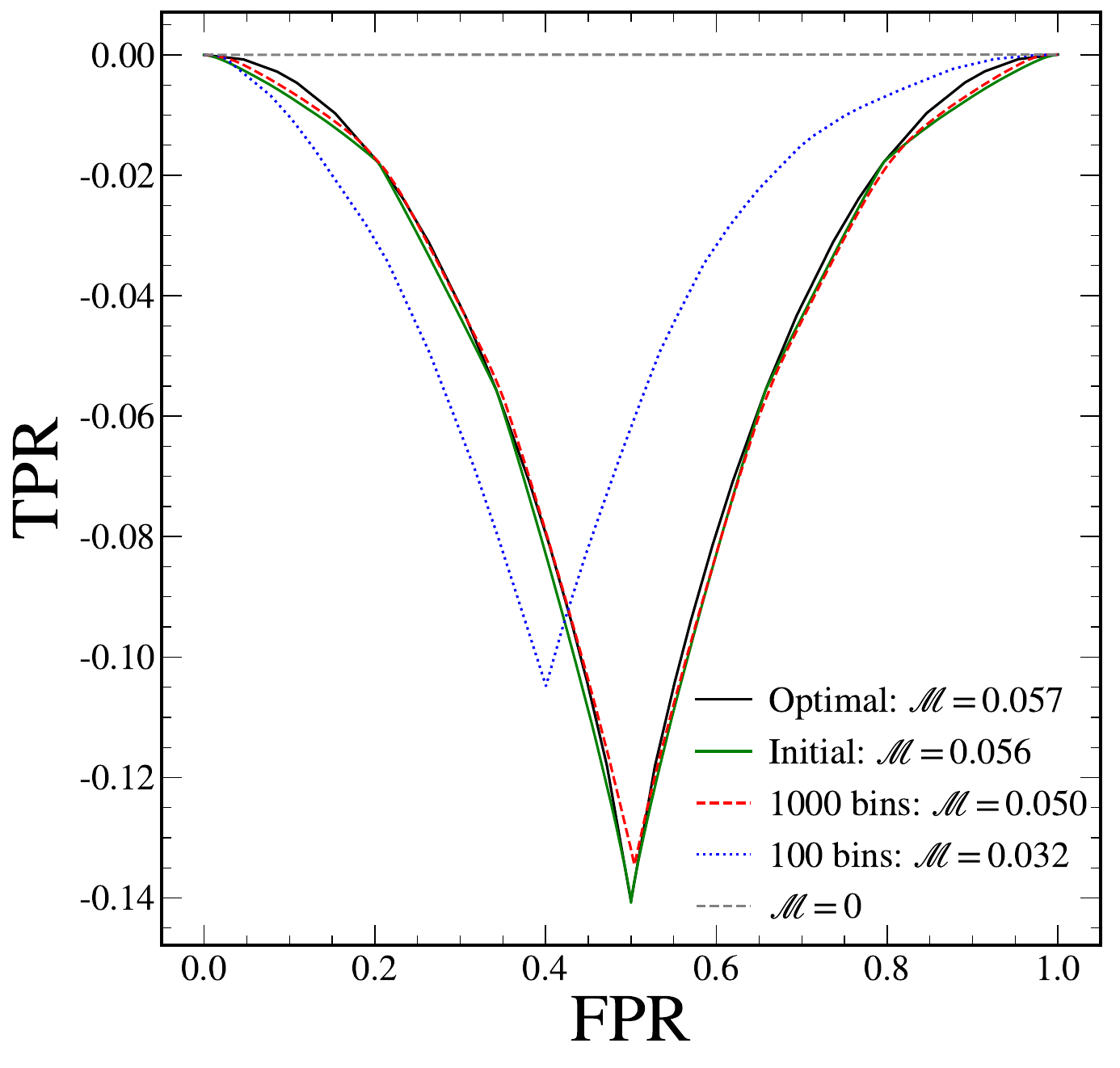}
    \includegraphics[width=0.32\textwidth]{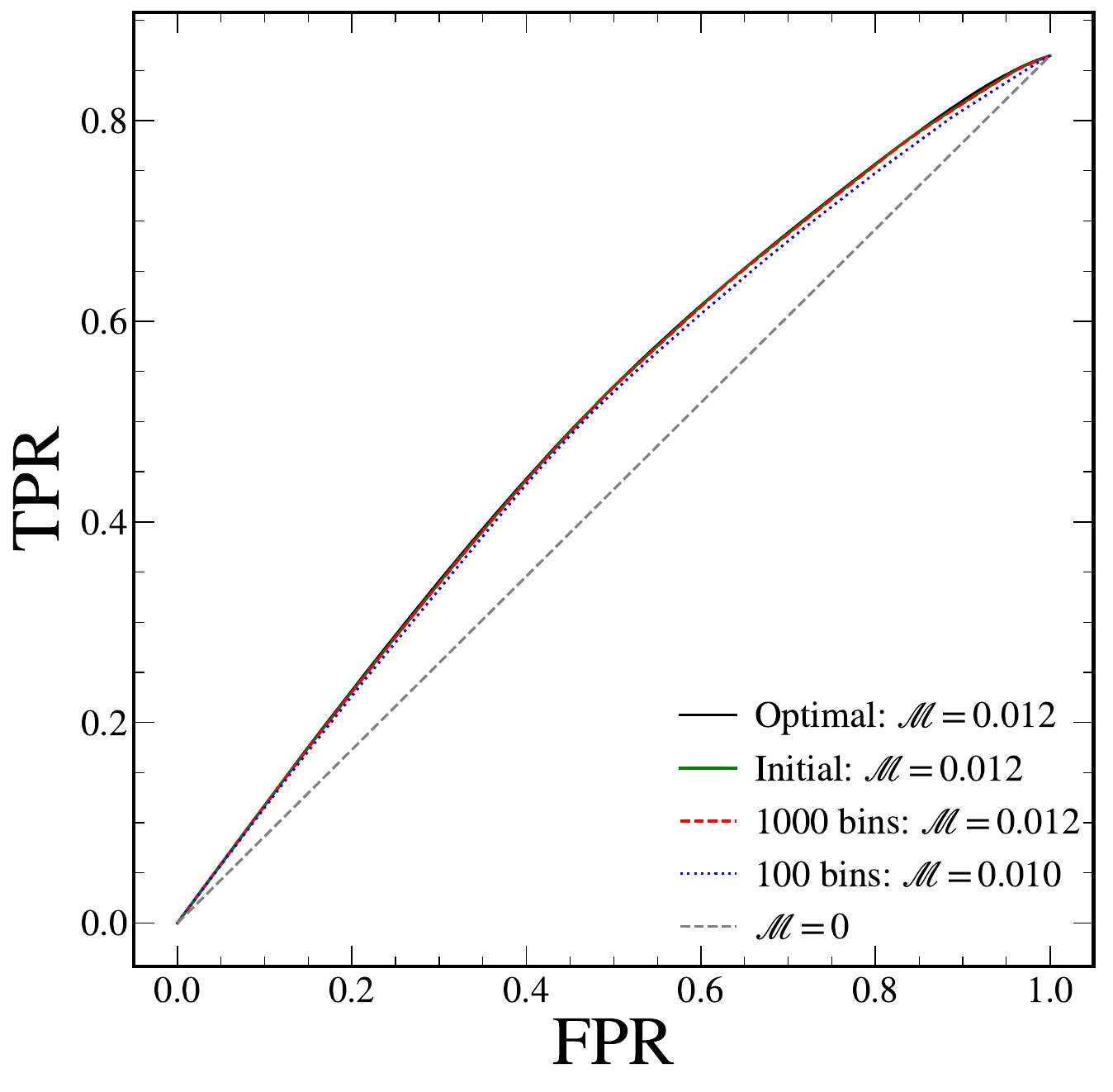}
    \includegraphics[width=0.32\textwidth]{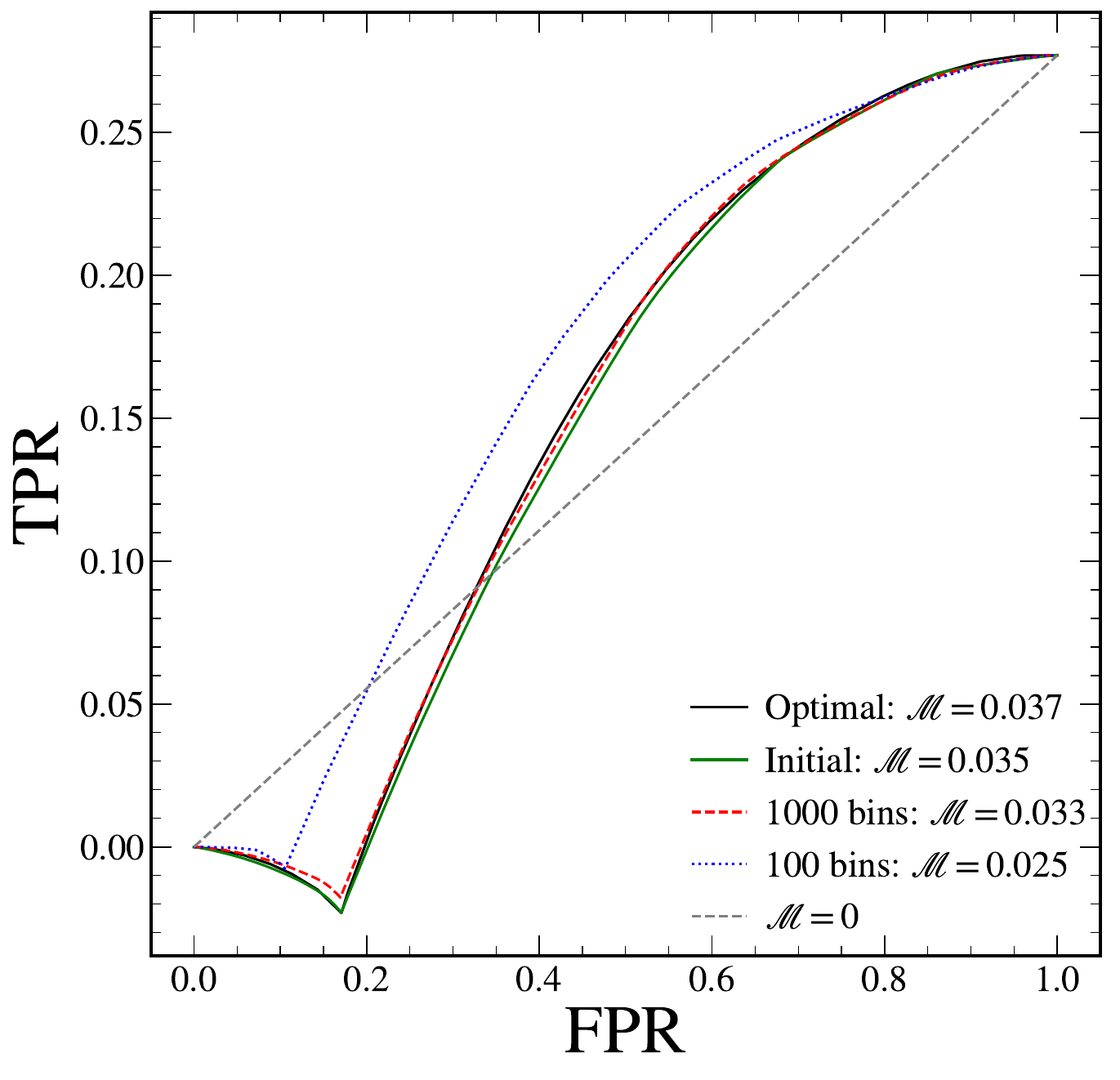}
	\caption{
The modified ROC curves based on the optimal observable ${\cal D}_{\mathrm{opt},1}^{(0)}$ (solid black) 
for the process $H \to (Z/\gamma^*)(Z/\gamma^*) \to 2e2\mu$  
are shown for the ``negative'' and ``positive'' samples, illustrated in Fig.~\ref{fig:discr1_3cases}, 
corresponding to the SM contribution and its interference with the BSM term for each choice of couplings
\( \theta_1 = \tilde{c}_{zz} \) (left), ${c}_{z\Box}$ (middle), and ${c}_{zz}$ (right). 
The performance metric ${\cal M}$ is defined in Sec.~\ref{sect:LoCmetric}
and the three scenarios considered under three different binning configurations 
(shown in green, red, and blue for the Initial, 1000, and 100 binning configurations) are discussed in Sec.~\ref{sect:input_4l}. 
The diagonal line with ${\cal M} = 0$ represents the case with no separation.
}
    \label{fig:roc_3cases}
\end{figure}

To demonstrate this, we now present the complete EFT treatment of the process $H\to (Z/\gamma^*)(Z/\gamma^*)\to 2e2\mu$, 
focusing on three illustrative examples of the relevant couplings: $\theta_1=\tilde{c}_{zz}$, ${c}_{z\Box}$, 
and ${c}_{zz}$, defined in Sec.~\ref{sect:input_eft}. 
In each case, we build the discriminant ${\cal D}_\mathrm{opt,1}^{(0)}$ according to Eq.~(\ref{eq:optimized1}), with \( \alpha = 0 \),
using the complete kinematic information available in the decay process illustrated in Fig.~\ref{fig:process} (right). 
Figure~\ref{fig:discr1_3cases} shows the ${\cal D}_\mathrm{opt,1}^{(0)}$ distributions for the three scenarios, 
while Fig.~\ref{fig:roc_3cases} displays the corresponding ROC curves, illustrating the distinct behavior observed in each case
and following the modified procedure discussed below. 

In the ROC curve calculations, the interference between the SM and BSM contributions serves as the ``positive" sample 
and the SM as the ``negative" sample, as defined in application to Eqs.~(\ref{fig:roc_TPR}) and~(\ref{fig:roc_FPR}).
It is crucial to emphasize that, for a proper analysis of the ROC curves, the observable distributions need to be arranged 
in ascending order based on the ratio, such as \( {\cal P}_{01} / {\cal P}_0 \) in this context, where \( {\cal P}_{01} \) and 
\( {\cal P}_0 \) represent the interference and SM probability distributions, respectively, as shown directly in Fig.~\ref{fig:discr1_3cases}.
This corresponds to ordering the line segments (for discrete bins of observables) 
of the ROC curve in descending order of their gradient.
The optimal observable \( {\cal D}_\mathrm{opt,1}^{(0)} \) naturally follows this ordering without requiring any rearrangement.
However, for arbitrary observables, this may not be the case, making the ordering step necessary.
This ordering corresponds to the same issue discussed in the context of \( {\cal D}_\mathrm{opt,2} \),
$m_2$, and $\Phi$ in Sec.~\ref{sect:input_binary}. 

For the \( CP \)-odd coupling \( \theta_1 = \tilde{c}_{zz} \), the interference term evaluates to zero when integrated 
over the entire range of observables, with an example shown in Fig.~\ref{fig:discr1_3cases} (left).
This presents an immediate issue in the definition of TPR in Eq.~(\ref{fig:roc_TPR}), when $|B| = 0$. 
Therefore, the classical ROC curve is no longer applicable.
While the definition of the FPR remains unchanged with respect to Eq.~(\ref{fig:roc_FPR}), 
a modification of the TPR definition, given in Eq.~(\ref{fig:roc_TPR}), is required:
\begin{align}
    \text{TPR}_h(t) = \frac{1}{|A|} \sum_{x \in B_h} \mathbb{I}[s(x) > t], 
    \label{fig:roc_TPR_2}
\end{align}
where all distributions are now normalized by the size of the ``negative" sample, 
$|A|$, which corresponds to the SM distribution and always carries positive probabilities. 
We permit multiple BSM hypotheses $h$ in the definition of the ``positive'' sample $B_h$, 
with each hypothesis giving rise to its own distinct ROC curve, $\text{TPR}_h(t)$ vs $\text{FPR}(t)$.
To enable a consistent approach, we introduce a convention for defining the size of the BSM coupling 
$\theta_h$ such that the pure BSM sample generated by coupling $\theta_h$ has size $|A|$. 
Although this convention extends beyond the strict domain of EFT validity, 
it serves to define the sample normalization used in the evaluation of the ROC curve.
We adopt this convention, along with Eq.~(\ref{fig:roc_TPR_2}), for the ROC curve definition incorporating 
interference contributions, throughout the remainder of this paper and including Fig.~\ref{fig:roc_3cases}. 

It is important to emphasize a key distinction of the new approach to normalizaiton in Eq.~(\ref{fig:roc_TPR_2}).
Unlike the classical ROC curve, where all distributions are normalized to unity and only the shape matters, 
the absolute size of the observable distribution now plays a crucial role. 
In particular, normalizing the interference distribution would be incorrect, 
as its magnitude relative to the SM is physically meaningful.

The observable \( \mathcal{D}_\mathrm{opt,1}^{(0)}\), shown in Fig.~\ref{fig:discr1_3cases} (left), 
is the most optimal for probing the interference term between the SM and $\tilde{c}_{zz}$.
It can be observed that the ROC curve for this observable
in Fig.~\ref{fig:roc_3cases} (left) exhibits a shape very different from that of a classical ROC curve, 
with all \( y \) values being negative. The characteristic ``kink" in the middle corresponds to the point where the negative probability 
values end and the positive values begin. 
This behavior is not exclusive to the $CP$-odd couplings. Certain BSM $CP$-even couplings, such as \( \theta_1 = {c}_{z\Box} \), 
do not result in negative interference with SM, as illustrated in Fig.~\ref{fig:discr1_3cases}, and their corresponding ROC curves,
such as in Fig.~\ref{fig:roc_3cases}~(middle), display the classical shape. 
Nonetheless, even the formally equivalent case \( \theta_1 = -{c}_{z\Box} \) 
would lead to entirely negative interference, which highlights the need for a more general definition capable of handling such scenarios. 

Other $CP$-even couplings, like \( \theta_1 = {c}_{zz} \), generate some negative interference, although the total integral over the 
full range remains positive. The resulting ROC curve, shown in Fig.~\ref{fig:roc_3cases}~(right), exhibits characteristics intermediate 
between the two cases described above and shown in Fig.~\ref{fig:roc_3cases}~(left and middle). 
The presence of negative probabilities in the interference sample undermines 
the classical ROC curve approach, causing the interference to evaluate to zero after the integral over a range
of an observable. As a result, AUC cannot be used as a reliable performance metric.
For instance, in Fig.~\ref{fig:roc_3cases}~(right), the no-separation scenario diagonal line and the optimal observable 
curve can exhibit identical AUC values, yet their performance is clearly quite different.
Therefore, we conclude that, in the presence of interference, which is a fundamental 
aspect of Quantum Mechanics, the choice of metric must be carefully re-evaluated.

\subsection{The LoC metric}
\label{sect:LoCmetric}

\noindent 
We propose two conditions that the ROC metric must satisfy in order to be considered a viable option. 
For clarity of presentation, we considered the binned form of the distributions. 
In the case of a multidimensional distribution, the histogram bins are transformed into a one-dimensional 
ordered distribution, noting that the specific order should not impact the results.
The unbinned case can be interpreted as the limiting case where the number of bins becomes very large.
The proper metric must remain invariant under the following two symmetry operations:
\begin{enumerate}
    \item exchange of bins in the histogram;
    \item exchange of the labels of the hypotheses.
\end{enumerate}
The AUC metric does not satisfy these symmetry conditions. 
The length of the curve (LoC), on the other hand, remains invariant under these symmetry operations.
As we will see, this metric exhibits robust behavior in the presence of negative probabilities.
We define LoC as follows:
\begin{equation}
    \mathrm{LoC}=\sum_{i}\sqrt{\left(H^{(A)}_i\right)^2+\left(H^{(B)}_i\right)^2}
    \label{eq:LOC}
\end{equation}
where the sum runs over all bins $i$, $H^{(A)}_i$ denotes the bin value from the ``positive'' sample, 
and $H^{(B)}_i$ denotes the bin value from the ``negative'' sample.
This corresponds to evaluating the ROC curve in Eqs.~(\ref{fig:roc_FPR}) and~(\ref{fig:roc_TPR_2}) at discrete thresholds $t$ defined by bin boundaries.
Continuous distributions are the limit of infinitely many bins.

For the purpose of numerical comparison, we define the quantity \({\cal M} = \mathrm{LoC} - \mathrm{LoC}_\mathrm{min}\), 
where \(\mathrm{LoC}_\mathrm{min}\) denotes the minimal possible value of the LoC metric. 
This ensures that \({\cal M} = 0\) corresponds to the LoC lower bound, which is the distance between the origin and the endpoint of the ROC curve. 
The coordinates $(x,y)$ of the endpoint are the integrals of the two samples normalized by the ``negative'' sample size $|A|$,
as discussed in application to Eqs.~(\ref{fig:roc_FPR}) and~(\ref{fig:roc_TPR_2}).
This LoC lower bound corresponds to the scenario in which the entire phase space is treated as a single bin,
meaning there is no separation between samples $B$ and $A$.

We can examine the properties of the new metric LoC in application to the classical ROC curve 
with all positive probabilities shown in Fig.~\ref{fig:roc_example}. 
The LoC is bounded between $[\sqrt{2}, 2]$. 
It reaches the upper bound in the ideal case of perfect separation (lines along the $y$- and $x$-axes), 
and the lower bound  $\mathrm{LoC}_\mathrm{min} = \sqrt{2}$ in the case of no separation (the diagonal line).
These properties resemble those of the AUC.
However, AUC requires a specific ordering of bins and hypothesis labels, whereas LoC has no such requirement,
as demonstrated in Eq.~(\ref{eq:LOC}).

The robust features of the new metric become even more evident when applied to the ROC curve with negative probabilities, 
as illustrated in Fig.~\ref{fig:roc_3cases} (left) or (right). 
The LoC upper bound, corresponding to perfect separation, is represented by a sequence of vertical and horizontal lines. 
These reflect bins that contain exclusively negative entries from sample~$A$, exclusively entries from sample~$B$ 
(always positive in application to SM), and exclusively positive entries from sample~$A$.
As noted above, the LoC reaches its lower bound \(\mathrm{LoC}_\mathrm{min}\) in the absence of separation.
For example, when the interference term evaluates to zero when integrated over the entire range, 
as shown in Fig.~\ref{fig:roc_3cases} (left), the  \(\mathrm{LoC}_\mathrm{min}=1.\) 
Unlike AUC, the LoC metric naturally supports negative probabilities.

\subsection{The LOC curve for multiple EFT hypotheses}

Up to this point, we have focused solely on comparing a single ``positive'' sample (BSM) with a single ``negative'' sample (SM).
However, within the EFT framework, multiple BSM hypotheses arise, 
each of which must be distinguished from the SM, and also from one another.
The classical ROC curve can only assess the performance of a cut-based analysis between 
two hypotheses at a time, with the AUC being defined for each pair individually.
In contrast, the LoC metric can be readily extended to multiple hypotheses as follows:
\begin{equation}
    \mathrm{LoC}=\sum_{i}\sqrt{\sum_{h}\left(H^{(h)}_i\right)^2}\,,
    \label{eq:LOC_2}
\end{equation}
where the sum is taken over all bins $i$, and $H^{(h)}_i$ represents the bin value from the sample associated with hypothesis~$h$,
including the SM.
The normalization of each sample follows our definition as applied in Eqs.~(\ref{fig:roc_FPR}) and~(\ref{fig:roc_TPR_2}), 
but with multiple TPR$_h(t)$ now defined, one for each BSM sample. 

\begin{figure}[t!]
    \captionsetup{justification=centerlast}
    \centering
    \includegraphics[width=0.55\textwidth]{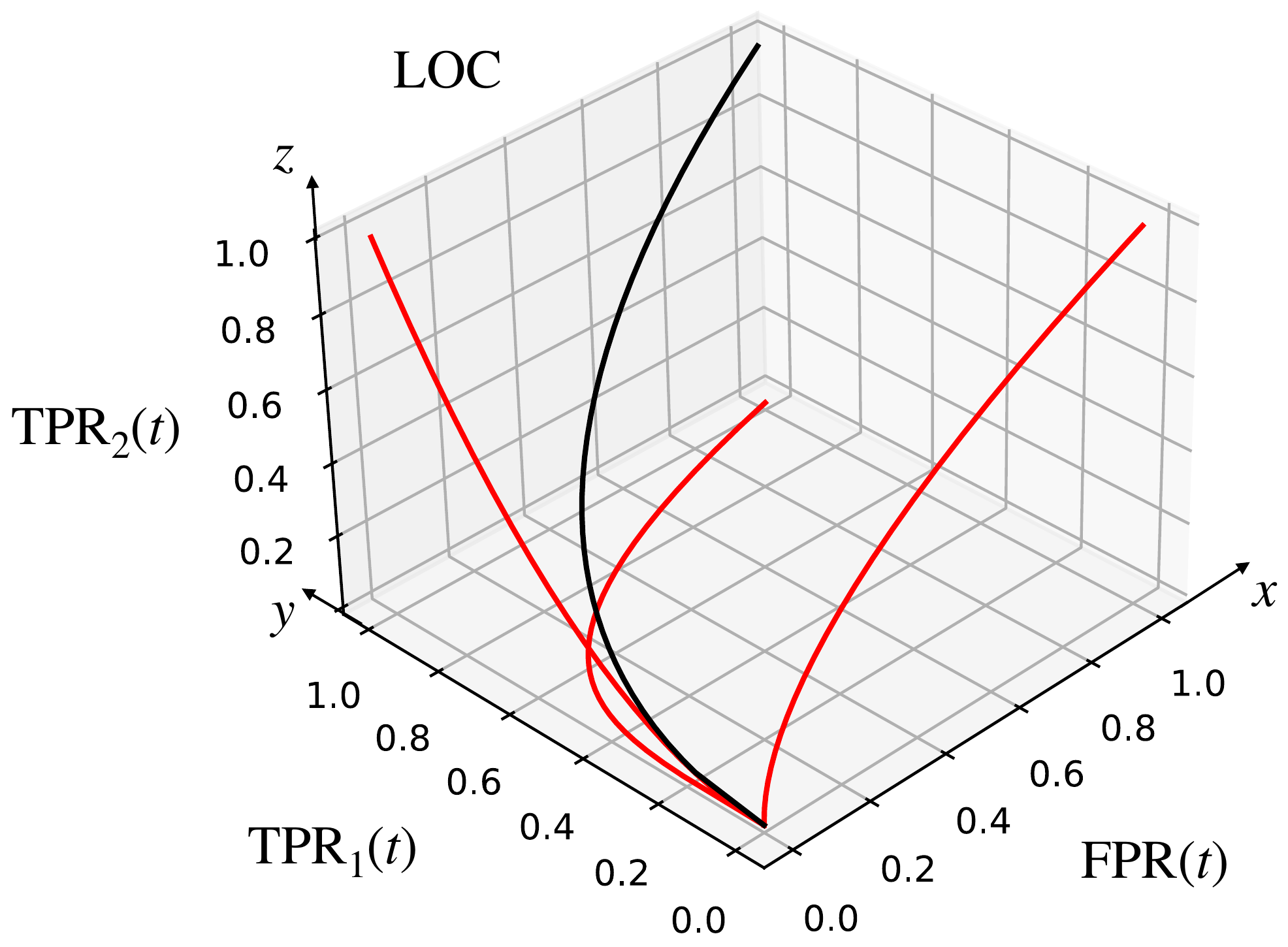}
	\caption{
	Illustration of the LOC curve in 3D space, featuring two TPR$_h(t)$ dimensions and one FPR$(t)$. 
	The complete LOC curve is depicted in black, while the individual curve projections are shown in red.}
    \label{fig:roc_space}
\end{figure}

This motivates the introduction of a new Likelihood Operating Characteristic (LOC) analysis
in high-dimensional hypothesis spaces, which incorporates negative probabilities in the likelihood parameterization, 
introduces a novel LoC metric to achieve optimal performance under these extended conditions, 
and accomplishes everything in an input-invariant manner.
The LOC framework is built on the ROC approach, but it offers a significantly wider range of applications.
An illustration of the LOC curve in 3D space is shown in Fig.~\ref{fig:roc_space}.

In Fig.~\ref{fig:roc_space}, each planar projection of the curve clearly resembles a standard ROC curve 
between the two corresponding hypotheses. However, these projected curves do not represent classical ROC curves, 
as the bin ordering determined by the threshold $t$ sampling is generally suboptimal for any given hypothesis $h$, 
and cannot be simultaneously optimal for all hypotheses. As a result, the relationship between $\mathrm{TPR}_h(t)$ 
and $\mathrm{FPR}(t)$ does not follow the correct ranking required for a proper ROC curve.
In contrast, the LoC metric within the LOC framework, defined in Eq.~(\ref{eq:LOC_2}),
is invariant to both the ordering of bins and the ordering of hypotheses. 
It is capable of handling an arbitrary number of hypotheses in a multidimensional space, 
accommodating negative probabilities as encountered in EFT, and supporting any bin ordering, 
none of which are feasible within the framework of classical ROC analysis.

\subsection{The MiLoMerge framework for optimal binning}
\label{sect:input_milomerge}

\noindent
In Sec.~\ref{sect:input_binary}, we introduced the Minimal-Loss Merging procedure using a toy model, 
which led to the results shown in Fig.~\ref{fig:Tiling}.
At that point, we lacked the necessary background to fully characterize the metric used for optimization.
We can now clarify that the LoC metric defined in Eq.~(\ref{eq:LOC_2}) was used to achieve this result.
It is also important to discuss the computational complexity associated with this approach.
This Minimal-Loss Merging procedure, along with the LOC framework,
has undergone significant optimization in an accompanying Python package, {\tt MiLoMerge},
available from Ref.~\cite{mela}, which has been employed to obtain the results presented in this paper.

In a simple case, such as the one considered in the toy model in Sec.~\ref{sect:input_binary}, 
the AUC metric could have been used as well.
A bin-merging algorithm must re-sort the bins at each trial step to compute the AUC,
to make a decision which pair of bins to merge. 
This means that the computational complexity of that algorithm scales as $\mathbb{O}\sim N!$, 
where $N$ is the number of bins to be merged. 
Even with a relatively small number of bins $150\times150$ merged in the study shown in Fig.~\ref{fig:Tiling}, 
the computational complexity becomes excessive.

The new LoC metric is completely order-invariant, on account of addition being commutative. 
This means that each bin pair requires a merging calculation only once and no sorting step is required to do so, 
which brings the computational complexity down to $\mathbb{O}\sim N^2$, which is a drastic improvement 
and can allow for optimizations starting at a far greater initial bin density in phase-space than the AUC-based technique.

To achieve optimal computational performance, a bin-merging score $\mathbb{D}_{ij}$ has been derived,
and the Minimal-Loss Merging procedure has been formulated. 
The $\mathbb{D}_{ij}$ value is calculated for each pair of bins $i$ and $j$ as
\begin{equation}
    \mathbb{D}_{ij}=\sum_{h}\sum_{h^\prime>h}
  w_{hh^\prime}  \left(H^{(h)}_iH^{(h^\prime)}_j-H^{(h)}_jH^{(h^\prime)}_i\right)^2 \,,
    \label{eq:Dscore}
\end{equation}
where the two sums are computed over all pairs of different hypotheses $h$ and $h^\prime$, including the SM,
$H^{(h)}_i$ represents the $i$-th bin value from the sample associated with hypothesis $h$,
and the weight $w_{hh^\prime}$ is set to 1 by default, but it can be modified by the analyst as described below.
The $\mathbb{D}_{ij}$ value needs to be computed only once for each pair of bins in the initial unrolled histogram with $N$ bins.
Sorting is performed once by $\mathbb{D}_{ij}$ value, with recomputation limited to bins affected by merging only. 
The bin pairs which have the lowest $\mathbb{D}_{ij}$ value should be merged first. 
This procedure is performed iteratively, merging at each step the pair of bins with the lowest $\mathbb{D}_{ij}$ value, 
which need not be adjacent.
The $\mathbb{D}_{ij}$ metric ensures that the procedure maximizes LoC, or equivalently, minimizes the loss. 

The $\mathbb{D}_{ij}$ score defines the unique and optimal sequence of bin merging in the procedure, 
starting from $N$ bins down to $2$ bins. The desired number of bins is determined by the objectives of the analyst.
In cases where two lowest $\mathbb{D}_{ij}$ values are equal, the order of operations is pseudo-random and 
has no impact on the LoC. 
With each bin-merging step, the LoC value is inevitably reduced, leading to a less optimal performance,
unless $\mathbb{D}_{ij}=0$ and the LoC value remains the same.
However, as demonstrated in the toy example in Fig.~\ref{fig:roc_example}, even reduction from 22\,500 down 
to 10 bins may lead to a negligible loss of performance. 

In the procedure, the merged bins are not required to be adjacent. 
However, as a variation of this method, one could enforce merging of only adjacent bins, 
which further reduces computational complexity by requiring consideration of a significantly smaller number of bin pairs $i$ and $j$. 
This feature can be useful when determining the optimal representation of a one-dimensional distribution 
with a finite number of bins, as demonstrated in the bin selection process shown in Fig.~\ref{fig:observables_D}\,(left). 

The weight $w_{hh^\prime}$ introduced in Eq.~(\ref{eq:Dscore}) provides additional flexibility in the optimization process. 
For instance, in EFT optimization, one may aim to enhance the separation between each BSM model and the SM, 
without prioritizing the separation among BSM models themselves. 
In such cases, $w_{hh^\prime} = 1$ when $h$ corresponds to the SM hypothesis and $h^\prime$ to a BSM model, 
and $w_{hh^\prime} = 0$ otherwise.
A further refinement of this approach involves assigning greater or lesser importance to a particular BSM model 
$h^\prime$ in the optimization process. In such cases, the corresponding weights $w_{hh^\prime}$ can be 
increased or decreased to reflect the desired emphasis. 
One example of such a weight adjustment is to normalize the individual LOC curve projections for each hypothesis pair
to a common metric value, effectively setting $w_{hh^\prime} = 1/\text{LoC}_{hh^\prime}$, where $\text{LoC}_{hh^\prime}$
represents the metric used in the LOC curve projection to distinguish between the two models $h$ and $h^\prime$.
Choosing $w_{hh^\prime} \ne \mathrm{const}$ departs from the straightforward graphical interpretation in Fig.~\ref{fig:roc_space} 
and the simple metric definition in Eq.~(\ref{eq:LOC_2}), but offers a practical means of prioritizing specific information.

While an illustration of the Minimal-Loss Merging procedure was presented with a simple toy model in Sec.~\ref{sect:input_binary}, 
with the corresponding results shown in Fig.~\ref{fig:Tiling}, a complete implementation of the concept requires 
the full LOC framework applicable to the general case of multiple hypotheses, potentially involving negative 
probabilities. Such a comprehensive analysis is provided in Sec.~\ref{sect:input_4l}.


\section{Generalized classification for EFT analysis of the $H\to4\ell$ process at the LHC}
\label{sect:input_4l}

\noindent 
With the full set of tools in hand, such as 
{\tt JHUGen} for simulation of BSM effects in LHC events,  
{\tt MELA} for calculation of optimal observables and re-weighting, and 
{\tt MiLoMerge} for optimal binning, 
we are now prepared to illustrate the concept of a complete EFT analysis with an example 
the process $H \to (Z/\gamma^*)(Z/\gamma^*) \to 2e\,2\mu,\ 4e,\ \text{and}\ 4\mu$, as produced at the LHC.
Although the $4e$ and $4\mu$ channels were previously omitted to avoid complications arising from 
the permutation of identical leptons, they are straightforwardly included in the complete analysis.
However, due to the substantial differences in their kinematic distributions~\cite{Davis:2021tiv}, 
all optimization procedures are performed separately from the $2e\,2\mu$ channel.

We have established the complete analysis workflow for the process $gg\to H \to (Z/\gamma^*)(Z/\gamma^*) \to 4\ell$, 
in accordance with the schematic representation shown in Fig.~\ref{fig:templates},
following realistic LHC data analysis examples~\cite{CMS:2014nkk,CMS:2021nnc}. 
A total of 130 million $gg \to H \to (Z/\gamma^*)(Z/\gamma^*) \to 4\ell$ events were generated using {\tt JHUGen}, 
evenly distributed across 13 different models featuring alternative BSM hypotheses, 
as well as their interferences with the SM. Of these, 70\% of the events were allocated for bin-merging training, 
while 30\% were set aside for testing through a simulated data analysis with thousands of realistic toy experiments. 
To maximize statistics, each sample was re-weighted using the {\tt MELA} package to represent each target model.
Detector effects were incorporated through selection requirements on the final-state leptons, 
namely \(p_T^\ell > 5\) GeV and \(|\eta^\ell| < 2.5\). 
For the purposes of this simplified analysis, non-uniform detection efficiency and track momentum 
smearing were neglected, as these effects are not crucial for demonstrating the approaches in this study.
The background process were also neglected, as those only complicate somewhat the analysis chain, 
but could be effectively suppressed with the dedicated ${\cal D}_{\mathrm{opt},2}$ observable in this
already very clean channel. 
The final inference of the EFT parameters was conducted using a maximum likelihood fit~\cite{CMS:2024onh}, 
utilizing a probability parameterization as specified in Eq.~(\ref{eq:probreco}) and templates of observables 
based on the binning choices derived following the procedures in Sec.~\ref{sect:input_general}.

Our target is to characterize the full decay process using 100 bins of observables, while keeping 
the performance nearly optimal. 
A successful analysis would serve as a proof of principle that information about such a complex process,
characterized by five kinematic observables, as defined in Secs.~\ref{sect:input_intro} and~\ref{sect:input_obs}, 
and eight parameters of interest, as specified in Eq.~(\ref{eq:massbasis}),
can be captured with sufficient accuracy using a reasonably small number of bins.
Moreover, it provides the technical means to achieve these results.
For comparison, a simple five-dimensional histogram with only six bins along each dimension would 
have two orders of magnitude more bins. 

\begin{figure}[b!]
    \captionsetup{justification=centerlast}
    \centering
    \includegraphics[width=0.27\textwidth]{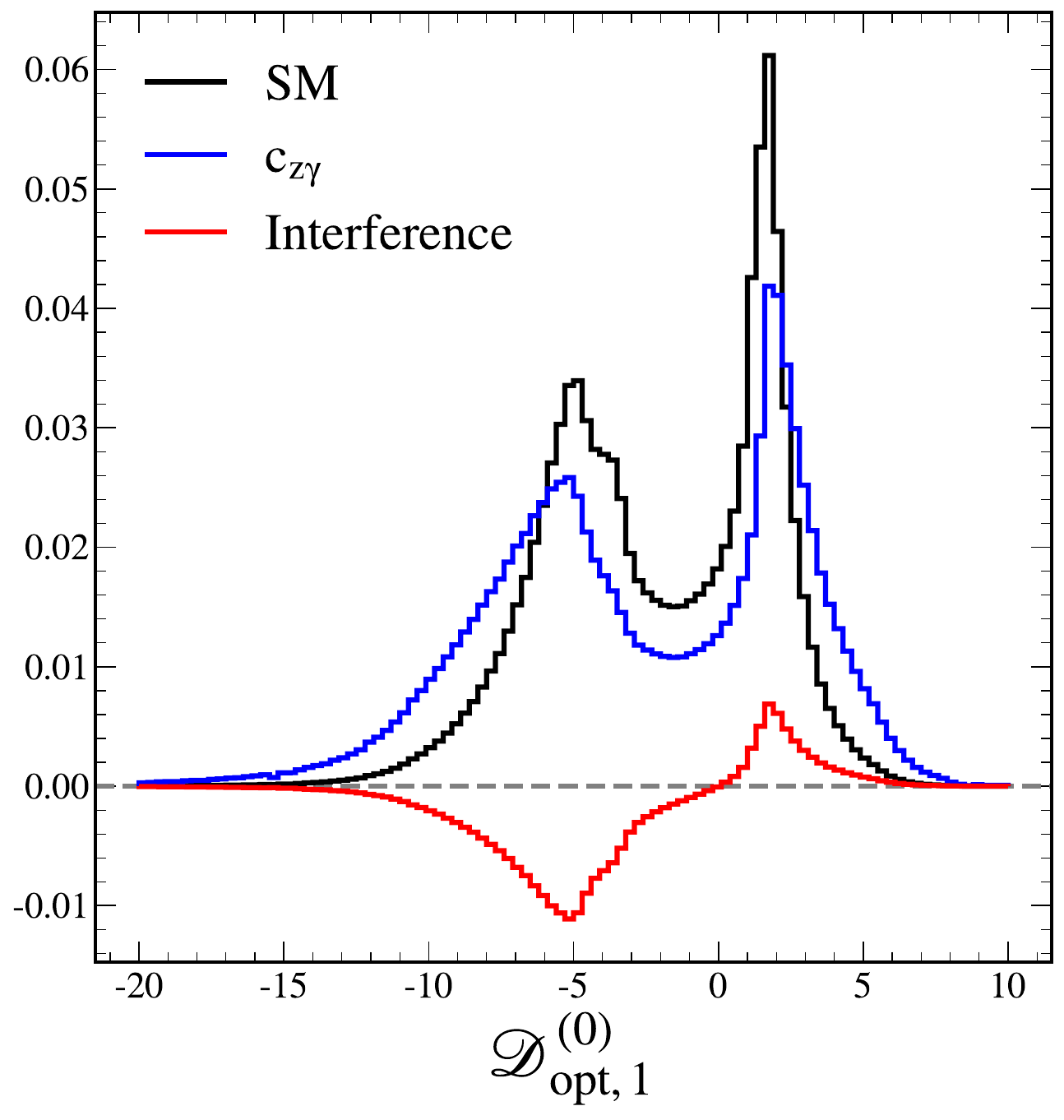}
    \includegraphics[width=0.3\textwidth]{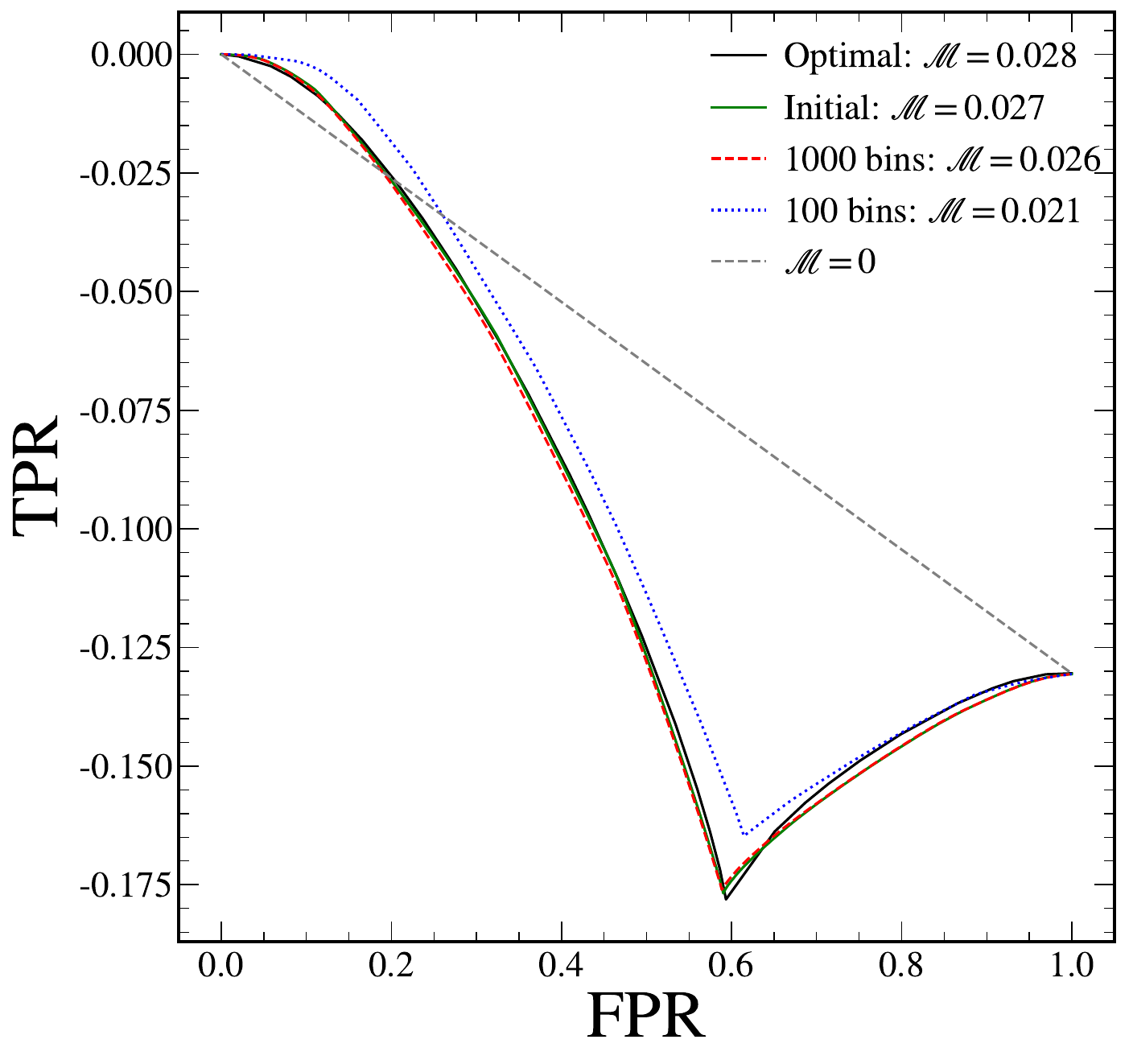} \\
    \includegraphics[width=0.27\textwidth]{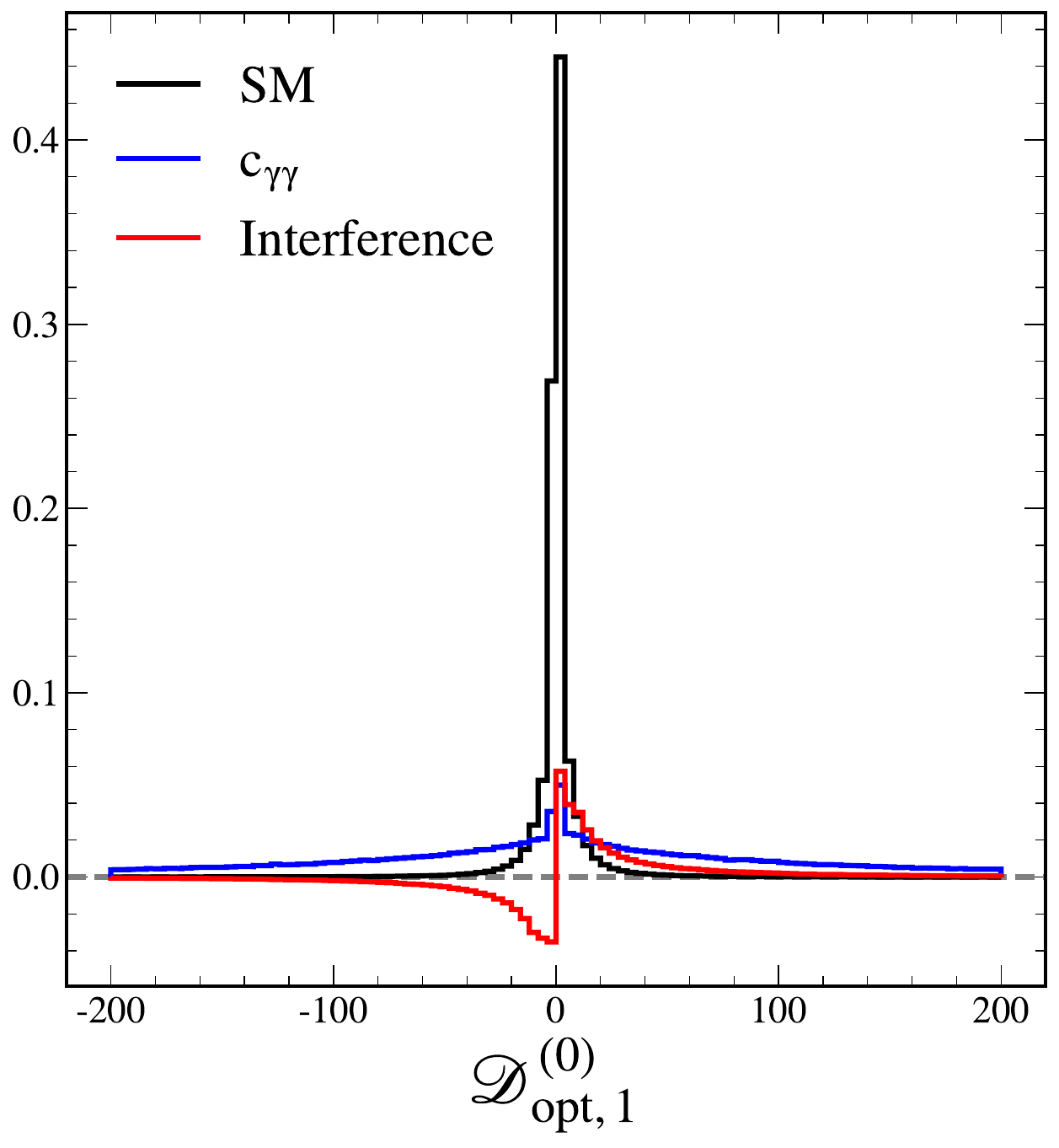} 
    \includegraphics[width=0.3\textwidth]{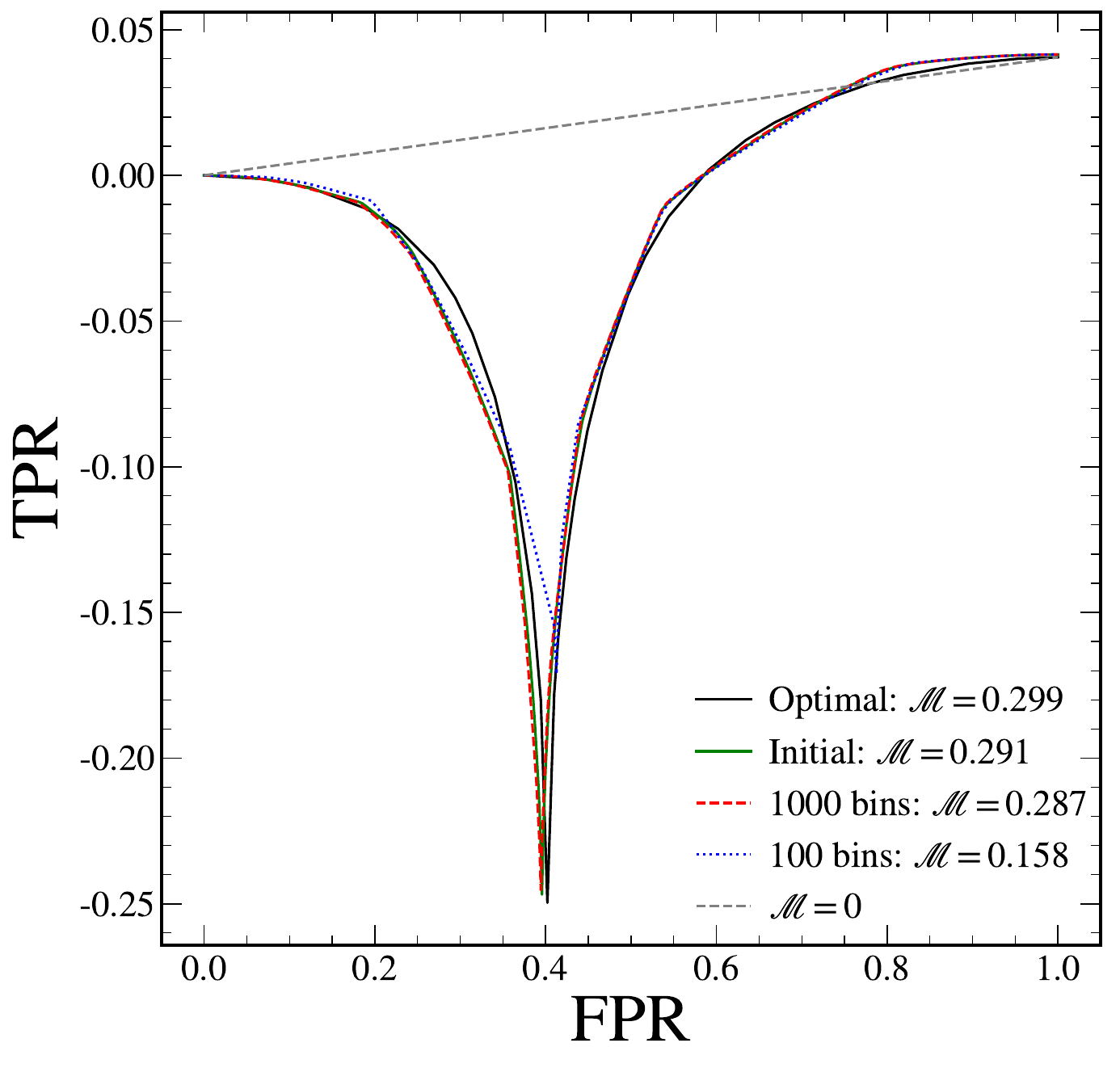} \\
    \includegraphics[width=0.27\textwidth]{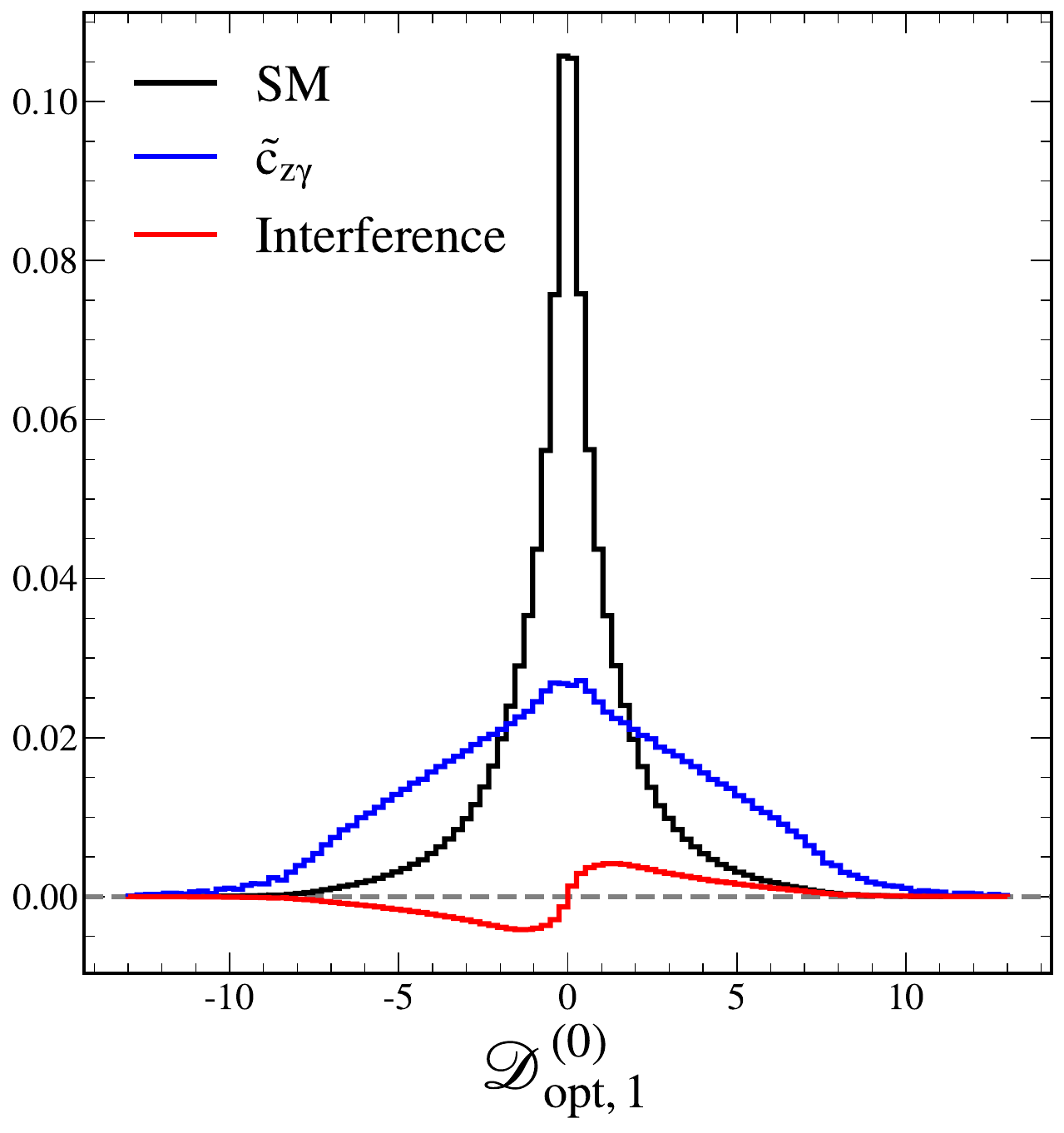}
    \includegraphics[width=0.3\textwidth]{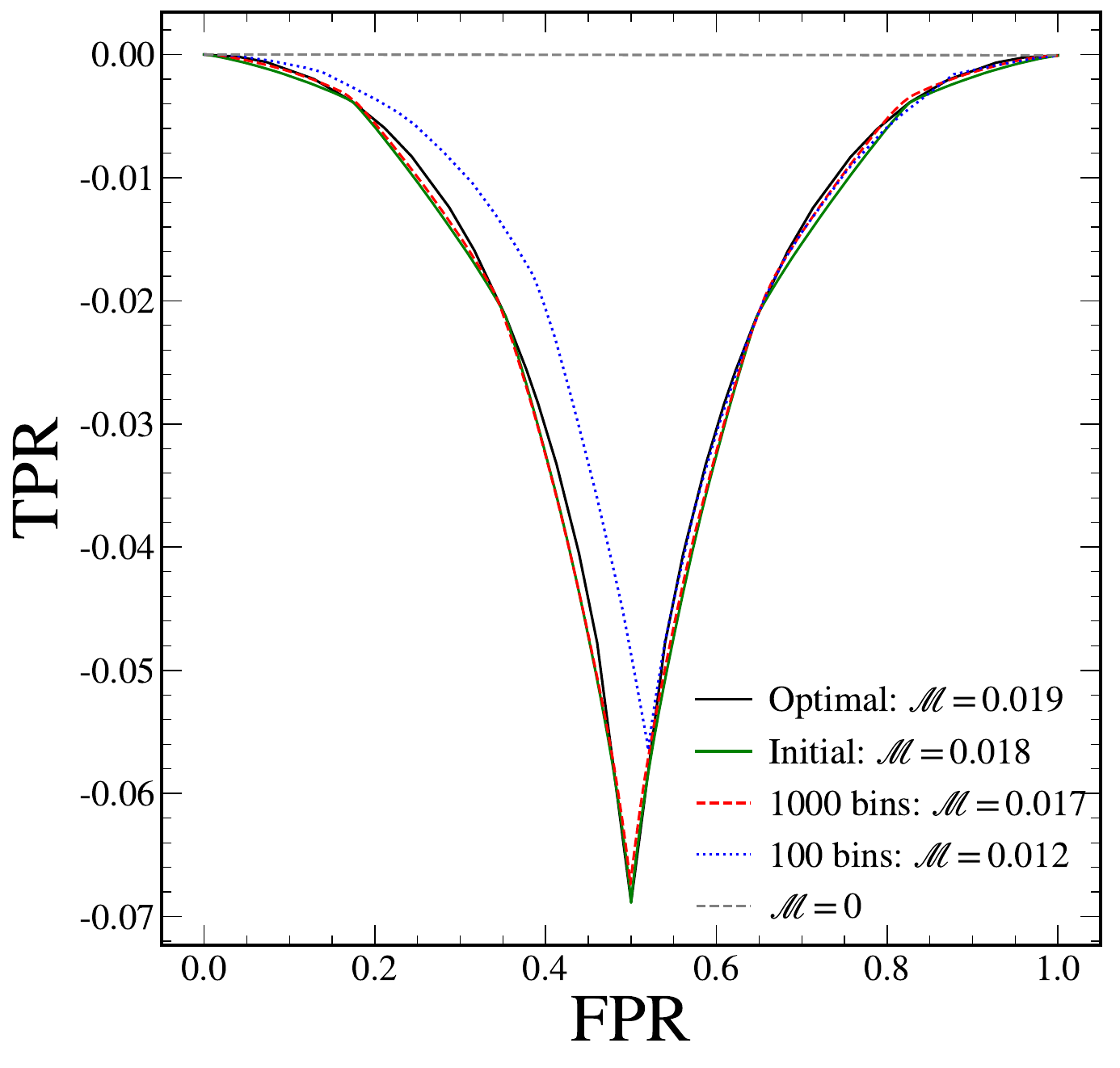} \\
    \includegraphics[width=0.27\textwidth]{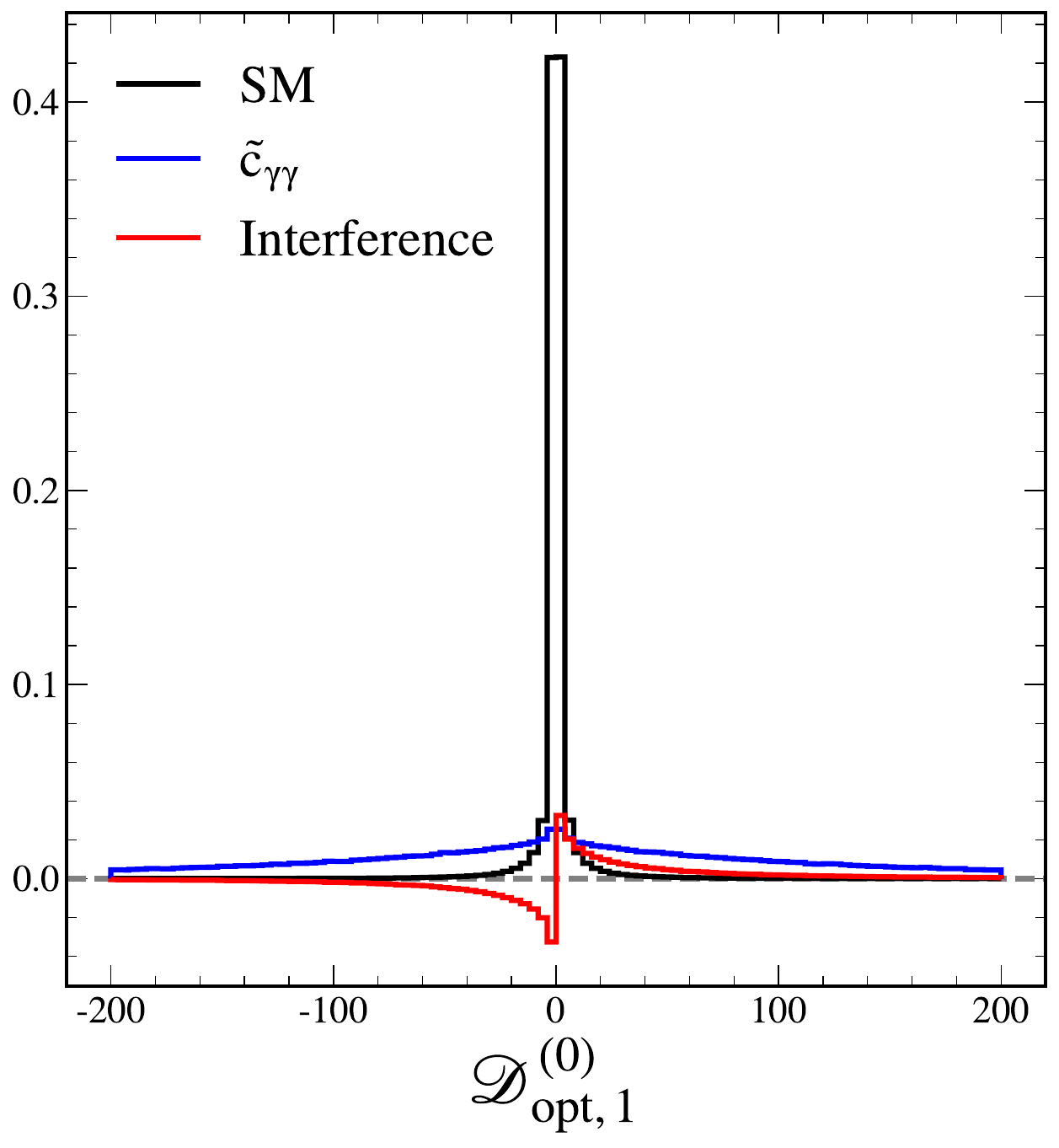} 
    \includegraphics[width=0.3\textwidth]{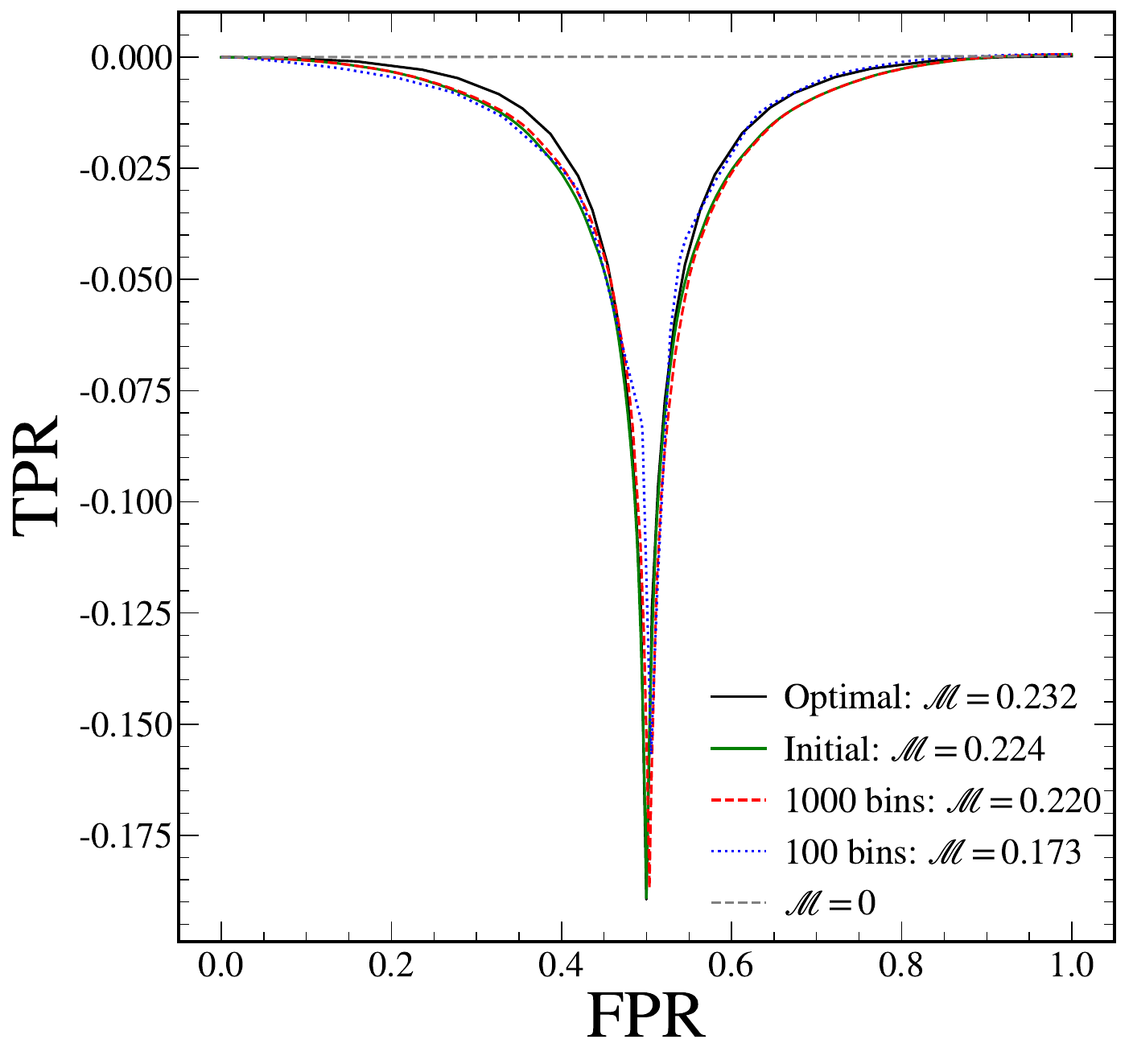} 
	\caption{
Left: Distributions of ${\cal D}_\mathrm{opt,1}^{(0)}$ for the process $H \to (Z/\gamma^*)(Z/\gamma^*) \to 2e2\mu$, 
obtained using the same approach as in Fig.~\ref{fig:discr1_3cases}, but shown for the four remaining couplings
$c_{z\gamma}, c_{\gamma\gamma}, \tilde{c}_{z\gamma},$ and $\tilde{c}_{\gamma\gamma}$.
Right: The corresponding LOC curves obtained using the same approach as in Fig.~\ref{fig:roc_3cases}.
}
    \label{fig:roc_4cases}
\end{figure}

Historically, EFT analyses of this final state at the LHC have employed optimal observables, 
such as those defined in Eqs.~(\ref{eq:optimized2}) and~(\ref{eq:optimized1})~\cite{CMS:2014nkk,CMS:2021nnc}. 
However, the number of such discriminants increases rapidly with the number of parameters 
being considered, and no prior study has attempted to simultaneously target all eight parameters 
of interest using optimal discriminants specific to each.
At the same time, this decay process can be fully described by five observables:
$m_1$, $m_2$, $\cos\Theta_1$, $\cos\Theta_2$, and $\Phi$.
Therefore, a five-dimensional histogram with sufficiently fine binning could be constructed, 
allowing the bins to be merged in the most optimal manner to retain sensitivity to the eight
parameters of interest. 

However, because extracting information sensitive to quantum mechanical interference is complex, 
performing an initial calculation of observables optimized for such interference allows us to begin with 
fewer initial bins, even if the number of observables is slightly larger.
With eight couplings, we define one overall signal strength 
parameter and seven ratios defined in Eq.~(\ref{eq:coupleratio}): $c'_{z\Box}, c'_{zz}, c'_{z\gamma}, 
c'_{\gamma\gamma}, \tilde{c}'_{zz}, \tilde{c}'_{z\gamma},$ and $\tilde{c}'_{\gamma\gamma}$.
This implies that there are seven optimal observables ${\cal D}_{\mathrm{opt},1}^{(0)}$,
with one corresponding to each coupling. A total of 27\,648 initial bins are used before merging, 
corresponding to 4, 4, 4, 6, 6, 6, and 2 bins in each dimension, respectively.
The number of bins in each dimension was tested and gradually reduced while maintaining high performance.


\begin{figure*}[b!]
    \captionsetup{justification=centerlast}
    \centering
    \includegraphics[width=0.42\textwidth]{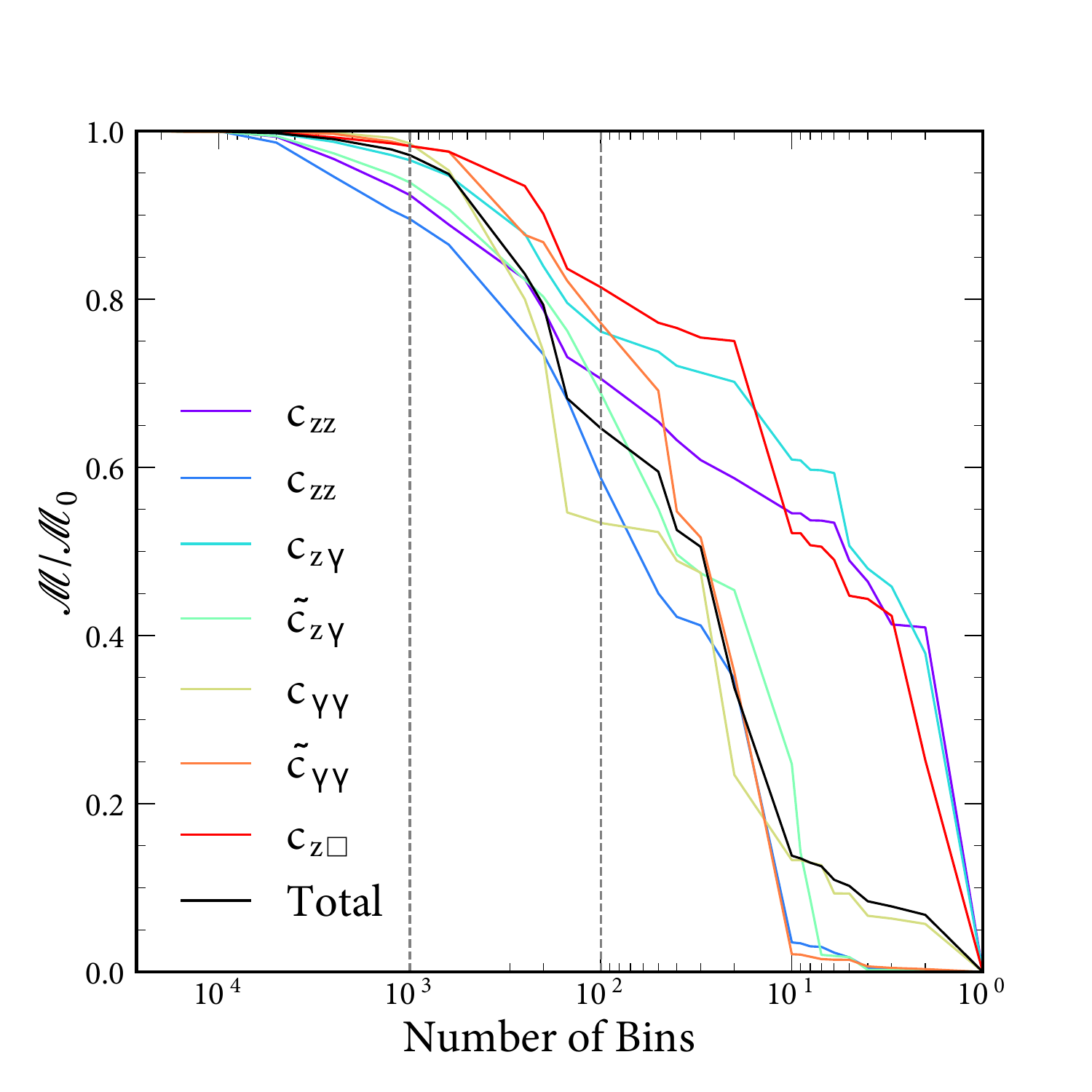}
    \includegraphics[width=0.42\textwidth]{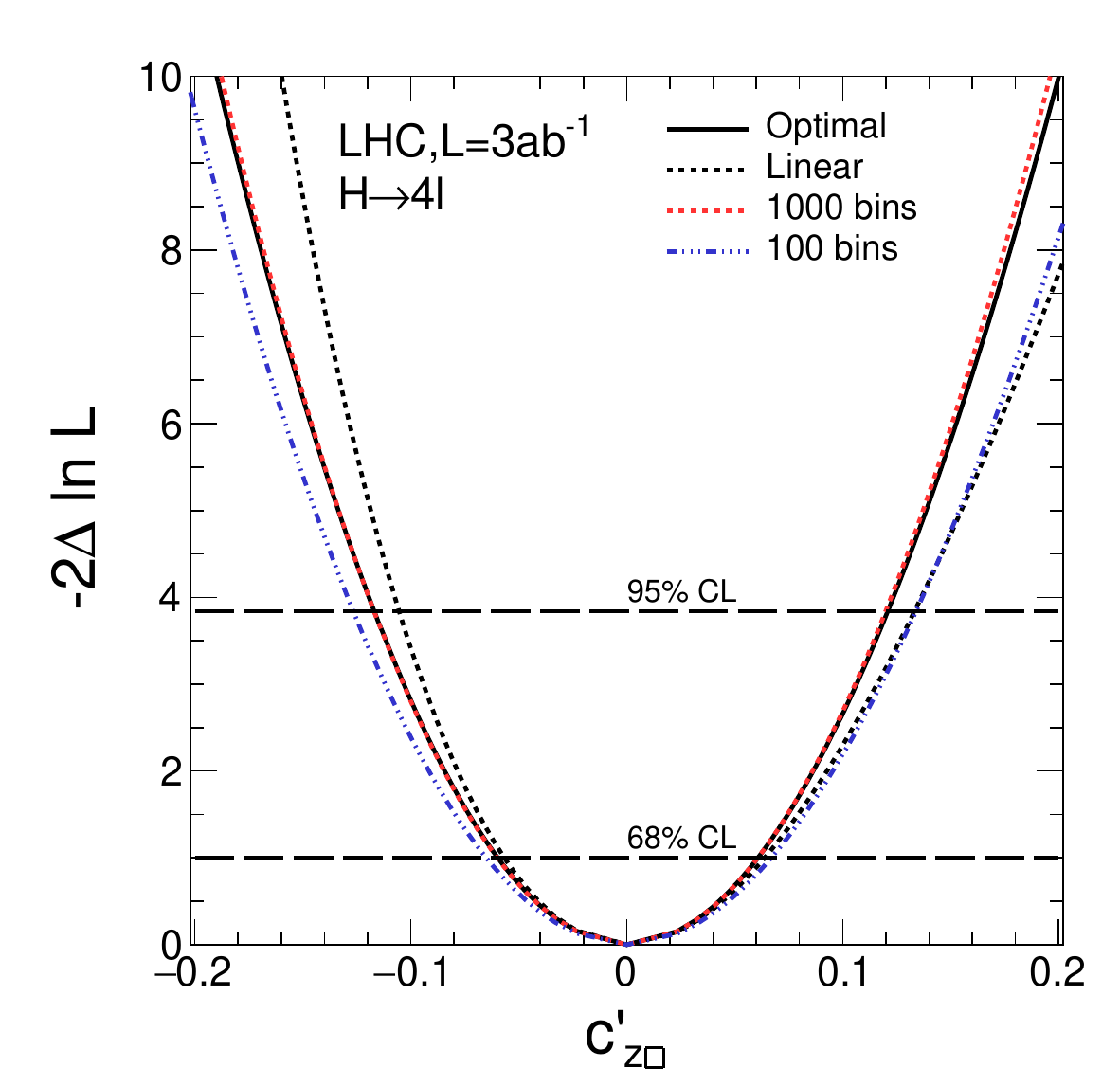}
	\caption{
The projected performance of the $H \to (Z/\gamma^*)(Z/\gamma^*) \to 4\ell$ analysis at the HL-LHC 
with an integrated luminosity of $3~\mathrm{ab}^{-1}$.
Left: The LOC performance metric normalized to its initial value, ${\cal M}/{\cal M}_0$, 
shown as a function of the number of bins for each of the seven couplings individually (colored lines)
and for all couplings combined (black line).
Right: Likelihood scans of $c^{\prime}_{z\Box}$ using 1\,000 (red) and 100 (blue) bins for the merged templates 
of the observables indicated by the vertical dashed lines in the left plot.
The performance obtained using the observables ${\cal D}_\mathrm{opt,1}^{(0)}$, 
which are optimal for the coupling of interest, is shown by the solid black line, while the dashed line shows 
the performance when only the linear terms are included in the parameterization.
    }
    \label{fig:roc_summary}  
\end{figure*}

The distributions of the seven observables are shown in Figs.~\ref{fig:discr1_3cases} and ~\ref{fig:roc_4cases}, 
while the corresponding LOC curves with the indicated metric  ${\cal M}$ for several cases are presented
in Figs.~\ref{fig:roc_3cases} and ~\ref{fig:roc_4cases}. 
In each case, the metric is given for the initial 27\,648, 1\,000, and 100 binning configurations, as 
well as for the the observables ${\cal D}_\mathrm{opt,1}^{(0)}$, which are optimal for the coupling of interest.
As indicated by the ${\cal M}$ values, the performance obtained with a large initial binning 
and with 1\,000 bins is nearly identical, and is also comparable to the optimal performance 
achievable with dedicated discriminants. 
The performance begins to degrade when using 100 bins, but still retains relatively large values, 
which may be acceptable for a practical data analysis.

A more detailed presentation of the performance as a function of the number of bins is shown in Fig.~\ref{fig:roc_summary} (left),
displaying only the $2e2\mu$ final state, while the $4e$ and $4\mu$ optimizations are qualitatively very similar.
As expected, the performance gradually decreases as the number of bins is reduced. 
However, it is remarkable that even with 1\,000 bins, the overall performance remains above 95\% according to the ${\cal M}$ metric. 
This performance is averaged over the seven parameters. Any further reduction in the number of bins 
is a matter of judgement for a given analysis, as it is limited purely by the practical constraints 
of the available modeling samples required to populate the bins.
Importantly, we now have the tools to reduce the bin size to any desired value in an optimal way,
along with a metric that enables rapid evaluation of performance.

While the LoC metric ${\cal M}$ allows for a rapid evaluation of performance, the ultimate performance 
in a data analysis can be assessed through a full likelihood scan of the parameters of interest. 
While the initial 27\,648 binning is not practical for a maximum likelihood fit, the three other configurations 
with the 1\,000 and 100 bins, along with the optimal discriminants for each coupling, have been implemented
in this analysis. 
The projected constraints on $c^{\prime}_{z\Box}$ at the HL-LHC with $3~\mathrm{ab}^{-1}$ of integrated luminosity 
are shown in Fig.~\ref{fig:roc_summary} (right).  The performance with 1\,000 bins is nearly identical 
to that with the optimal discriminants, with only a slight degradation observed for 100 bins.
Also shown is the expected performance considering only the linear terms in the coupling expansion of 
Eq.~(\ref{eq:probreco}), namely ${\cal P}_{0}$ and ${\cal P}_{0k}$. The small difference between including 
or omitting the quadratic terms indicates that they are relatively unimportant, implying operation near full EFT validity. 
These differences are, however, similar in magnitude to the modest performance loss with 100 bins.

\begin{figure*}[!t]
\centering
\includegraphics[width=0.32\textwidth]{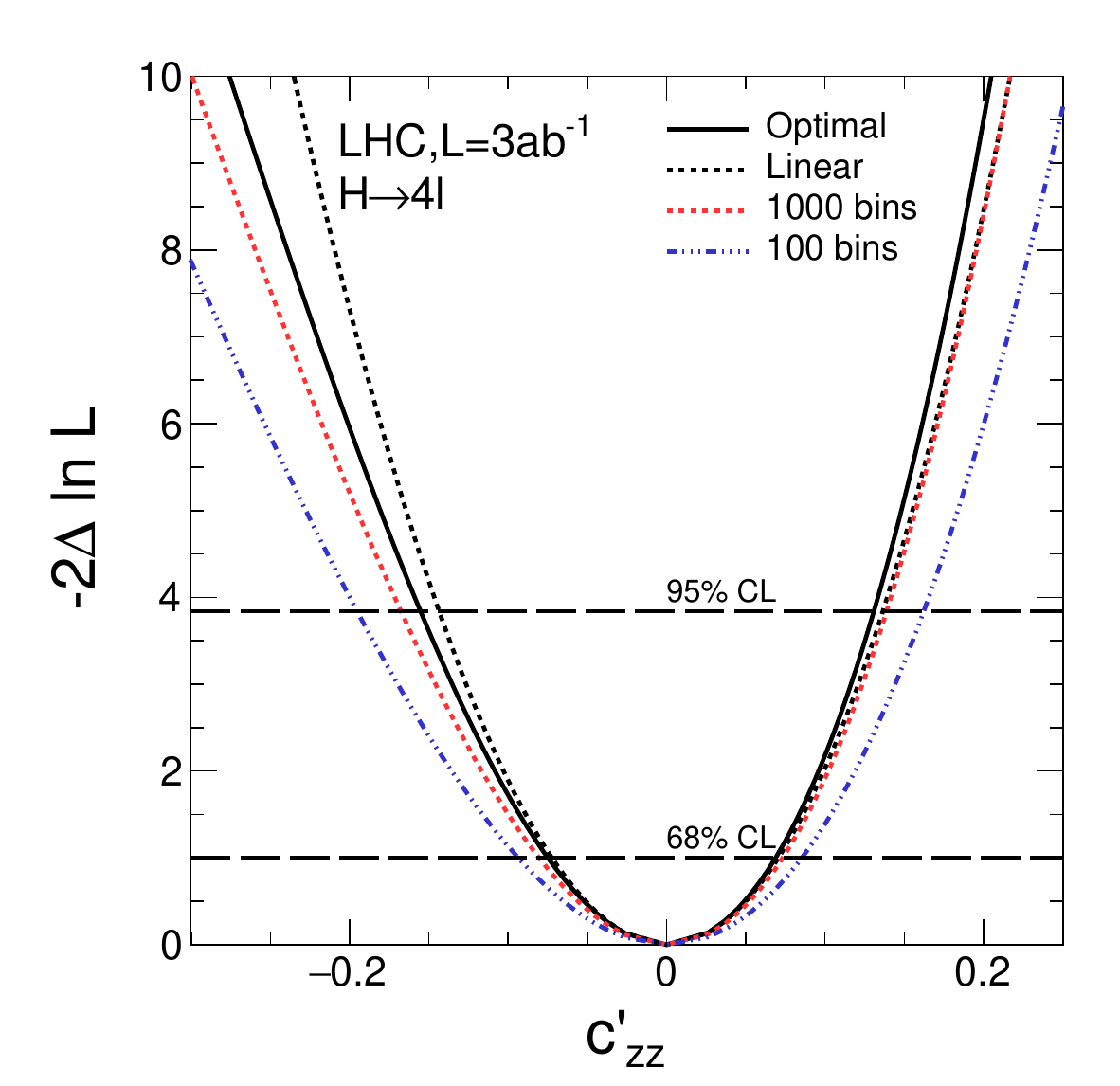}
\includegraphics[width=0.32\textwidth]{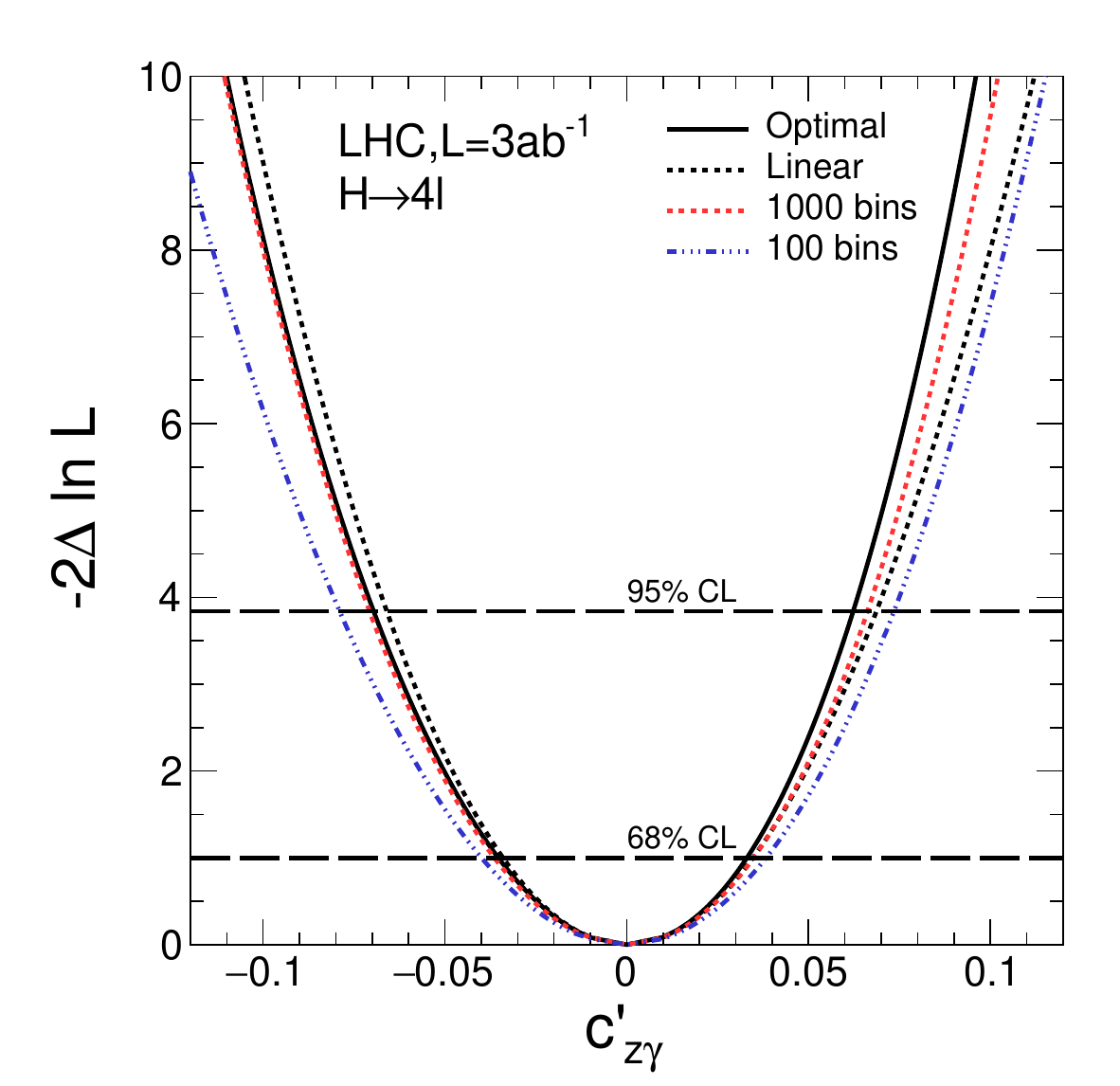}
\includegraphics[width=0.32\textwidth]{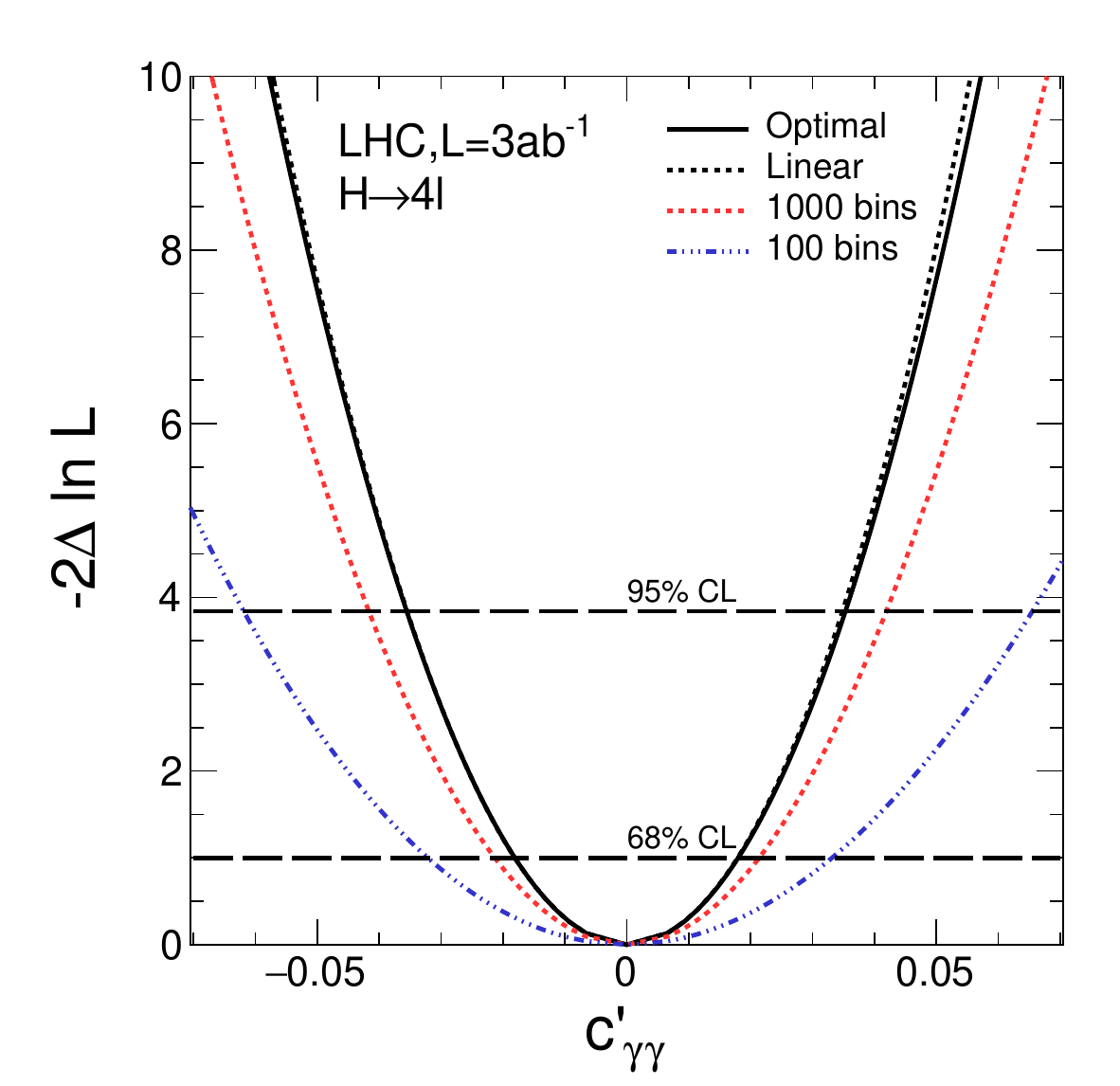}
\includegraphics[width=0.32\textwidth]{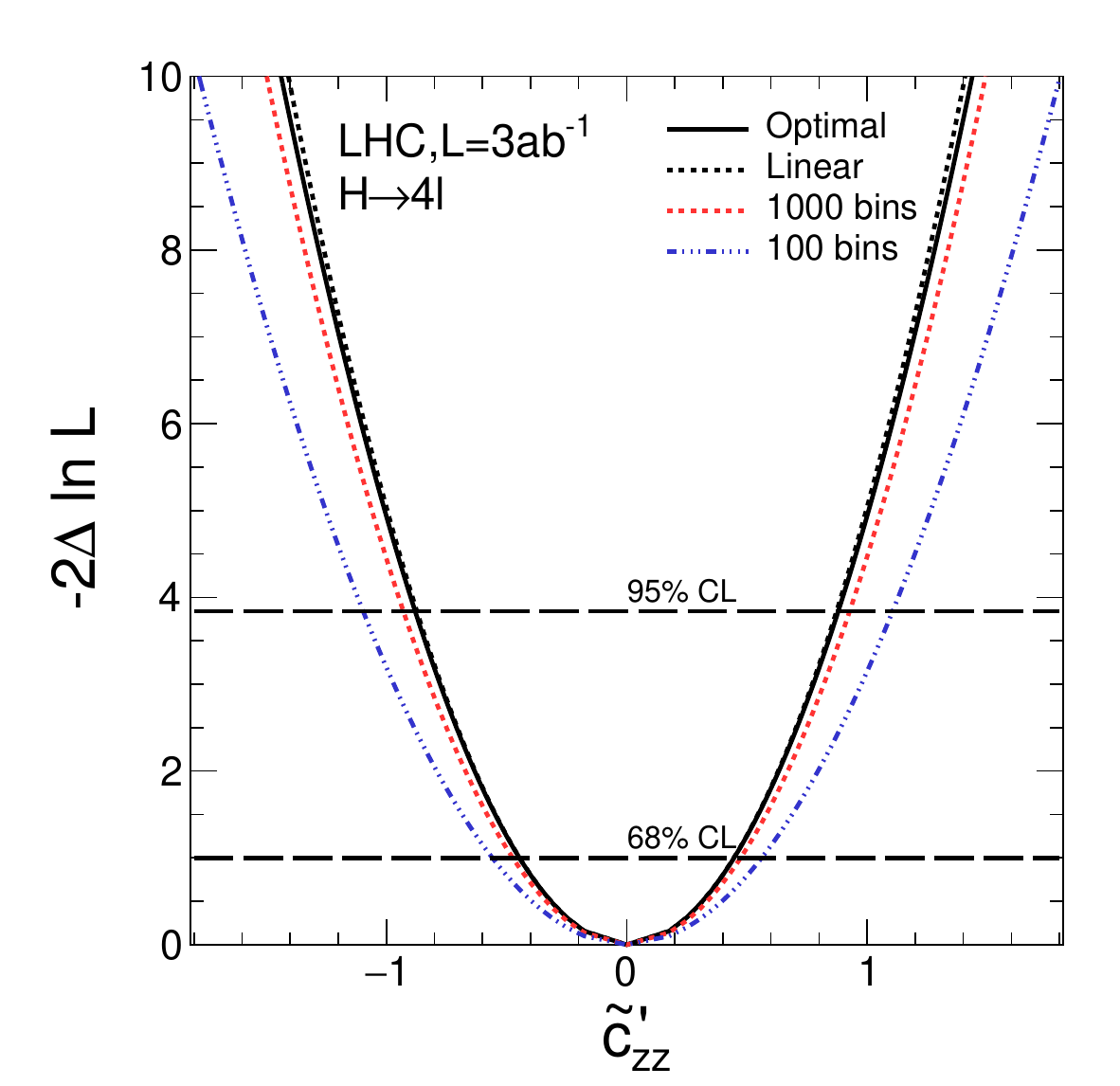}
\includegraphics[width=0.32\textwidth]{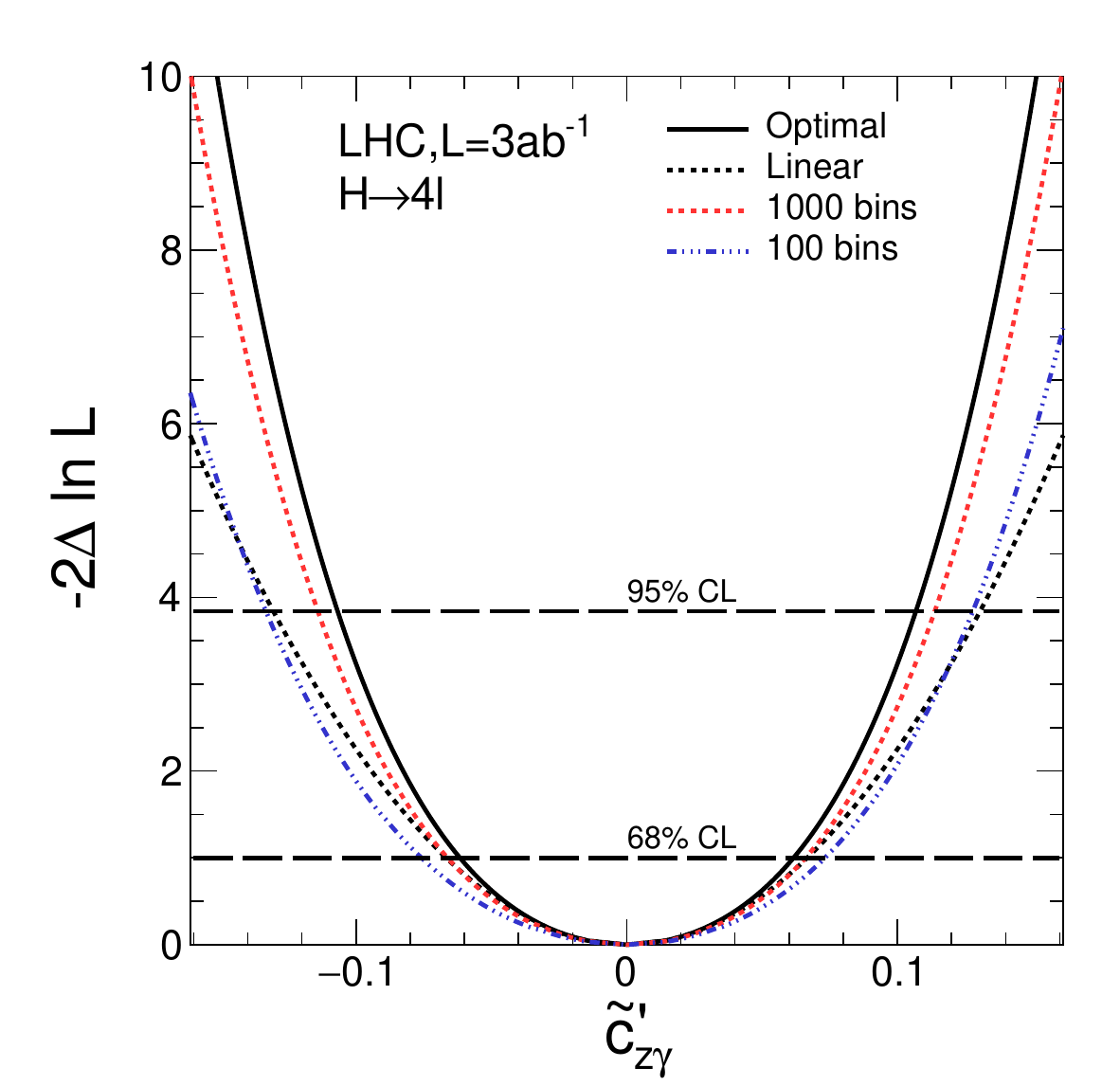}
\includegraphics[width=0.32\textwidth]{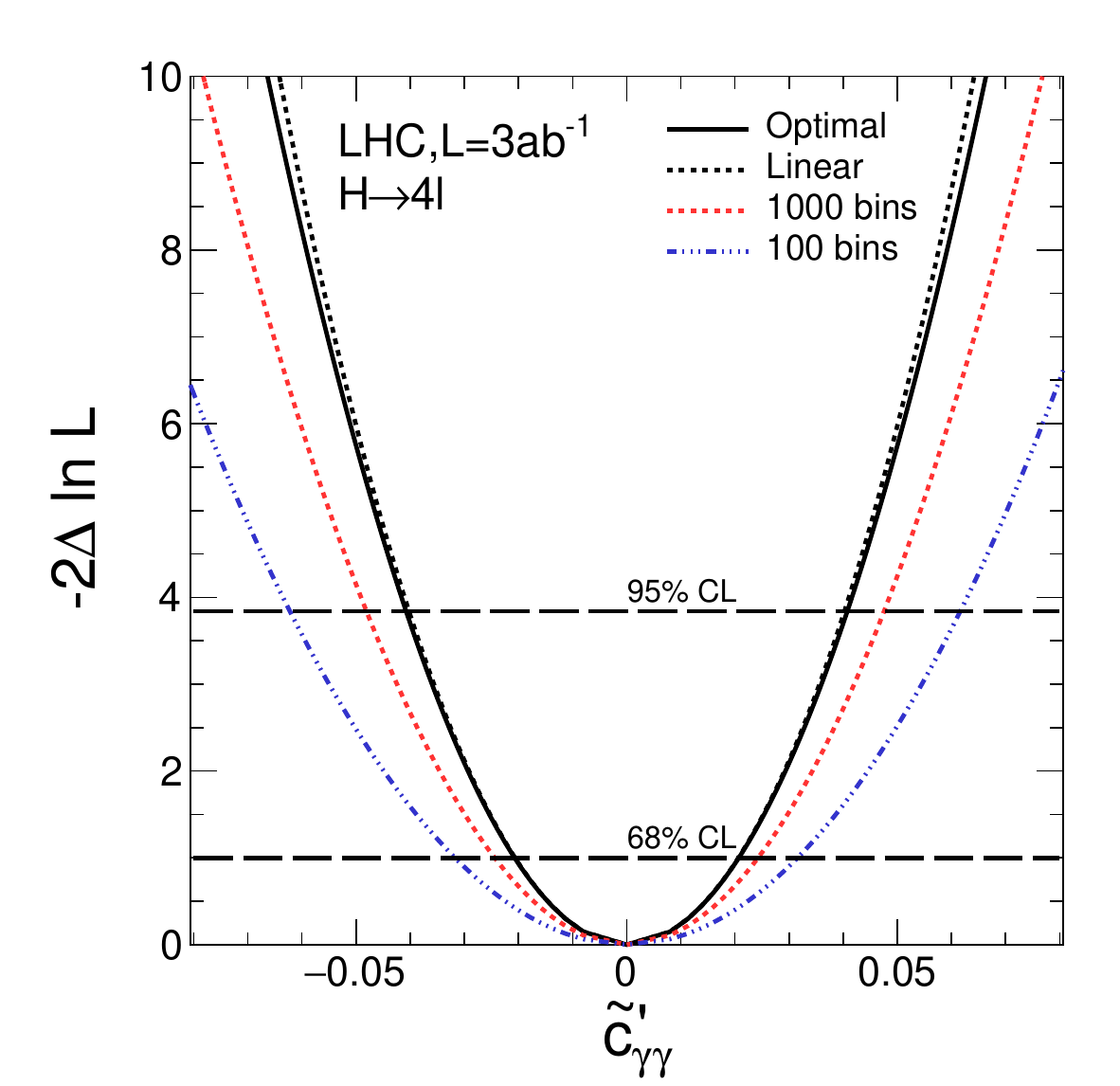}
\caption{
The projected constraints on 
$c^{\prime}_{zz}$, $c^{\prime}_{z\gamma}$,  $c^{\prime}_{\gamma\gamma}$,
$\tilde{c}^{\prime}_{zz}$, $\tilde{c}^{\prime}_{z\gamma}$, and $\tilde{c}^{\prime}_{\gamma\gamma}$
following the notation of Fig.~\ref{fig:roc_summary} (right). 
}
\label{fig:result_fa2-couplings}
\end{figure*}

A similar pattern is observed for the six other couplings shown in Fig.~\ref{fig:result_fa2-couplings}. 
It is important to note that, for illustration purposes, the analysis is performed one coupling at a time, 
while all other couplings are fixed to their SM values.
Floating all couplings simultaneously, as required in a proper EFT analysis, is fully possible and indeed 
the goal of this project: to develop bin-merged observables optimized for simultaneous analysis of all couplings. 
However, examining one coupling at a time allows performance to be illustrated in a carefully controlled setting.
This approach is necessary because the optimal performance shown in each likelihood scan 
in Figs.~\ref{fig:roc_summary} and~\ref{fig:result_fa2-couplings} is achievable for each coupling individually 
using the discriminants optimized specifically for that coupling. Seven sets of optimal observables are used 
as a reference in each fit. In contrast, the bin-merged observables are optimized for all couplings simultaneously, 
which is an important distinction and the reason this procedure was developed. 
Once the performance is verified for each coupling individually, the bin-merged observables can be employed 
in a proper EFT analysis with a simultaneous fit of all couplings.

Based on the study of the $H \to (Z/\gamma^*)(Z/\gamma^*) \to 4\ell$ channel presented here, 
we conclude that it is entirely feasible to develop a fully optimal EFT analysis of a channel targeting 
approximately eight operators using around 1\,000 bins, provided that simulation or control samples 
allow proper population of the templates.
Alternatively, with only a modest loss in performance, an analysis can be performed using approximately 100 bins. 
Reducing the number of bins to around 10, however, would significantly affect the precision.

\section{A machine-learning approach to the $VH$ process at the LHC}
\label{sect:input_vh}

\noindent 
The application of the techniques described in this work extends beyond clean, fully reconstructed channels at LHC, 
such as the $H \to 4\ell$ channel examined in the preceding section.
One key difference may be the absence of well-defined matrix-element-based observables in other channels. 
In such cases, machine-learning approaches, as discussed in Sec.~\ref{sect:input_ml}, can provide effective 
approximations to matrix-element concepts.
In all cases, the guiding principle remains unchanged. To illustrate the practical applications and potential 
limitations within the EFT framework, we consider the $q\bar{q} \to (Z/\gamma^*) \to H(Z/\gamma^*) \to H(\ell^+\ell^-)$ 
process at the LHC, introduced in Sec.~\ref{sect:input_intro} and previously discussed in Sec.~\ref{sect:input_ml}. 
Although this channel also allows for the evaluation of matrix elements with full reconstruction~\cite{Anderson:2013afp}, 
we leverage this fact to establish a well-defined reference for machine-learning approaches.
Among other aspects, we provide additional details regarding Figs.~\ref{fig:MELA_d0m} and~\ref{fig:MLA_d0m}, 
along with the conclusions that follow.

In Sect.~\ref{sect:input_ml}, we discussed the discriminants generated using both machine learning and 
the analytically described matrix elements. At that time, we did not possess the technical capability to assess 
the performance of the discriminants, as the presence of negative probabilities due to interference hindered 
our ability to utilize classical ROC scores. With the introduction of the LOC approach in Sec.~\ref{sect:input_general}, 
we are now prepared to conduct more in-depth studies and comparisons of the techniques.

In every scenario, our guiding principle for identifying optimal observables is provided by Eqs.~(\ref{eq:optimized2}) 
and~(\ref{eq:optimized1}). For the MLA, the aim is to train the machine to approximate these analytical methods, 
as they have been demonstrated to deliver optimal performance.
In most cases, we employ a Boosted Decision Tree (BDT) based on the XGBOOST algorithm~\cite{Chen_2016}, 
using five kinematic observables as input: $\vec{x} = (m_{VH}, m_{\ell\ell}, \cos\Theta_1, \cos\Theta_2, \Phi)$, 
as defined in Fig.~\ref{fig:process} (left) and Ref.~\cite{Bolognesi:2012mm}. 
However, when training the $\mathcal{D}_{\text{opt,1}}^{(0)}$ observable, XGBOOST does not permit the use 
of negative weights, prompting us to utilize the TMVA BDT~\cite{TMVA}. 
The sample preparation adhered to the logic outlined in the diagram presented in Fig.~\ref{fig:templates}. 
A total of ten million SM $q\bar{q} \to (Z/\gamma^*) \to H(Z/\gamma^*) \to H(\ell^+\ell^-)$ events were 
generated using {\tt JHUGen}, with 64\% of the events allocated for training, 16\% for training validation, 
and 20\% for testing. The BSM model was obtained through the {\tt MELA} re-weighting.
For illustrative purposes, we focus solely on the $\tilde{c}_{zz}$ coupling that generates the BSM contributions.

Below, we evaluate the performance of three observables: $\mathcal{D}_{\text{opt,2}}$, 
$\mathcal{D}_{\text{opt,1}}^{(1)}$, and $\mathcal{D}_{\text{opt,1}}^{(0)}$, 
comparing the direct analytical MELA calculations presented in Fig.~\ref{fig:MELA_d0m} 
with the MLA approximation trained by a machine, as illustrated in Fig.~\ref{fig:MLA_d0m}.
We employ the LoC metric ${\cal M}$, defined in Sec.~\ref{sect:input_general}, to assess the performance, 
where the capacity to handle negative probabilities arising from interference is crucial.
For illustrative purposes, we also present the performance in terms of ${\cal M}$ using the single representative
kinematic observables $m_{\ell\ell}$ or $\Phi$. 

\begin{figure}[b!]
  \begin{center}
    \captionsetup{justification=centerlast}
    \includegraphics[width=0.45\textwidth]{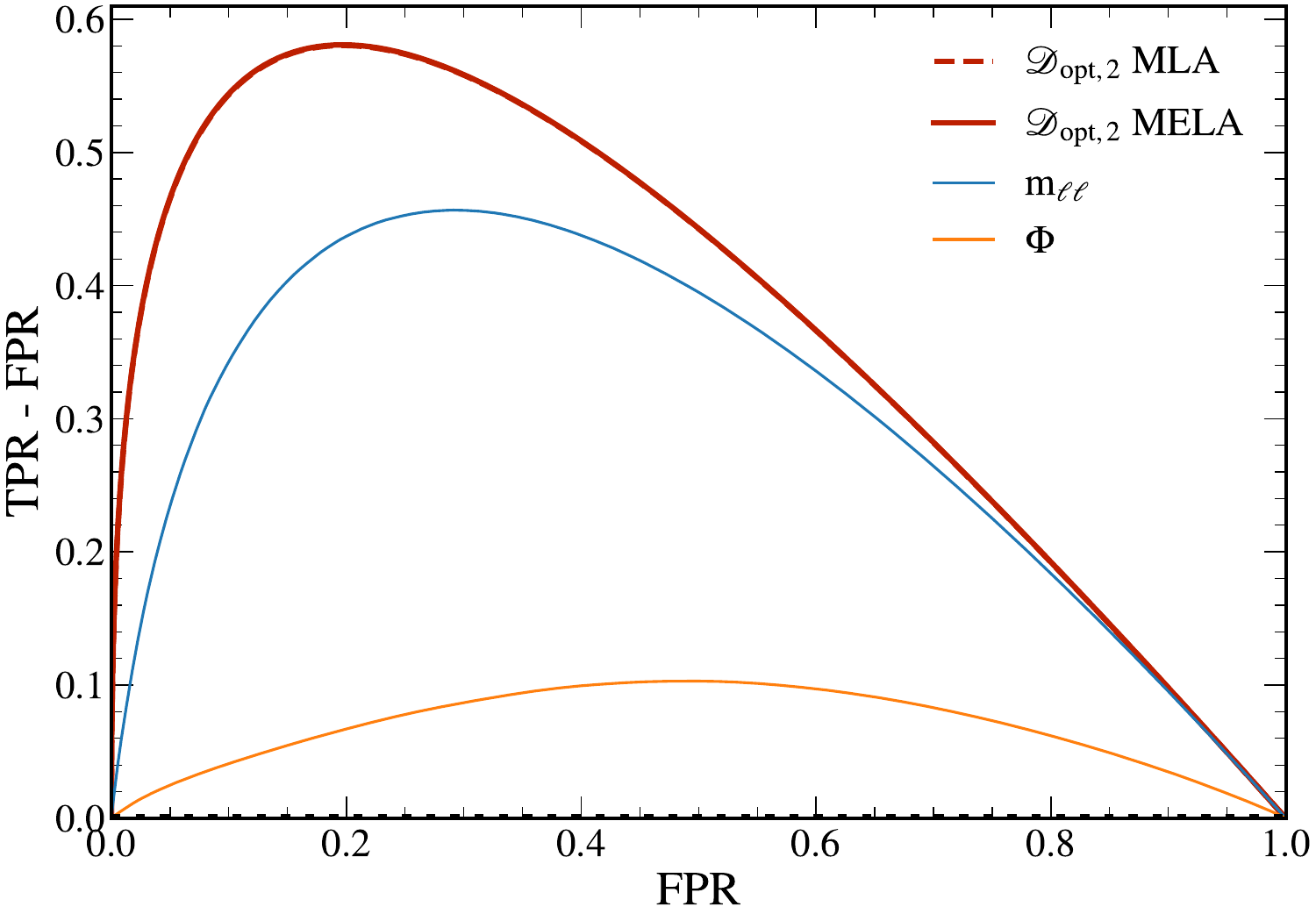}
    \includegraphics[width=0.45\textwidth]{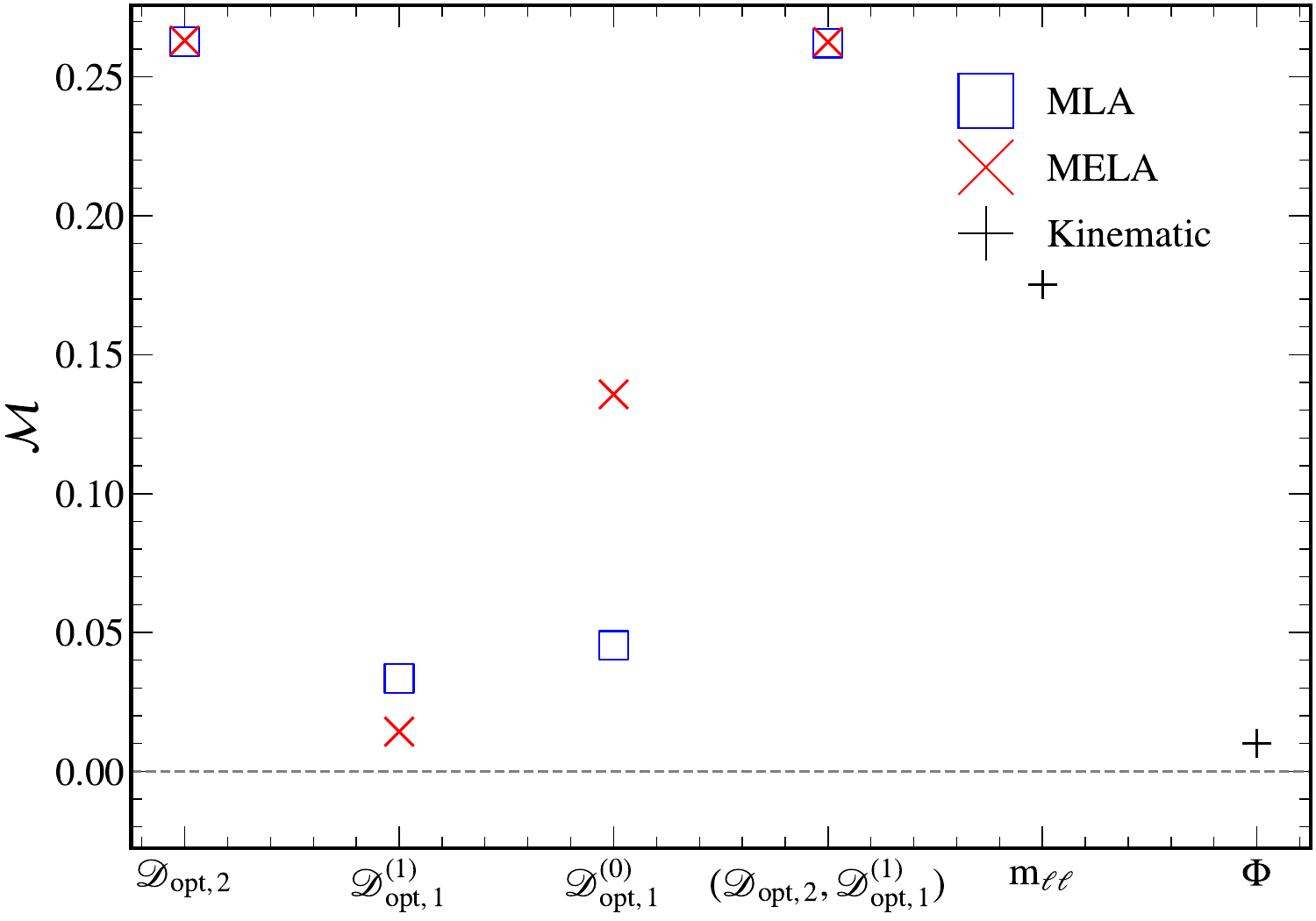}
    \caption{
    Left: The LOC curves evaluating the separation between the two alternative hypotheses, SM and $\theta_1=\tilde{c}_{zz}$, 
    for four observables: $\mathcal{D}_{\text{opt,2}}$ computed using both MLA and MELA, as well as $m_{\ell\ell}$ and $\Phi$.
    Right: A summary of the LoC metric ${\cal M}$ derived from the LOC construction on the left, displayed for four discriminant 
    configurations, utilizing both MLA (blue box) and MELA (red cross), along with $m_{\ell\ell}$ and $\Phi$ (black cross).
     }
    \label{fig:d0m}
  \end{center}
\end{figure}

From Fig.~\ref{fig:d0m}, it is clear that the $\mathcal{D}_{\text{opt,2}}$ discriminant demonstrates 
the same performance regardless of whether it is calculated using MLA or MELA techniques.
The distributions used to derive the LOC curves are presented in the left panels of 
Figs.~\ref{fig:MELA_d0m} and~\ref{fig:MLA_d0m}.
Neither the $\mathcal{D}_{\text{opt,1}}^{(1)}$ nor the $\mathcal{D}_{\text{opt,1}}^{(0)}$ discriminants 
are designed for distinguishing the pure models tested in Fig.~\ref{fig:d0m}, and individual kinematic
observables, such as $m_{\ell\ell}$ and $\Phi$, lack sufficient information. Therefore, none of these
observables can match the performance of $\mathcal{D}_{\text{opt,2}}$.
The evaluation of the two-dimensional discriminant $(\mathcal{D}_{\text{opt,2}}, \mathcal{D}_{\text{opt,1}}^{(1)})$ 
 is a trivial test in this case. However, it demonstrates that no new information pertinent to the measurement 
can be added to the already optimal observable $\mathcal{D}_{\text{opt,2}}$.

\begin{figure}[b!]
  \begin{center}
    \captionsetup{justification=centerlast}
    \includegraphics[width=0.45\textwidth]{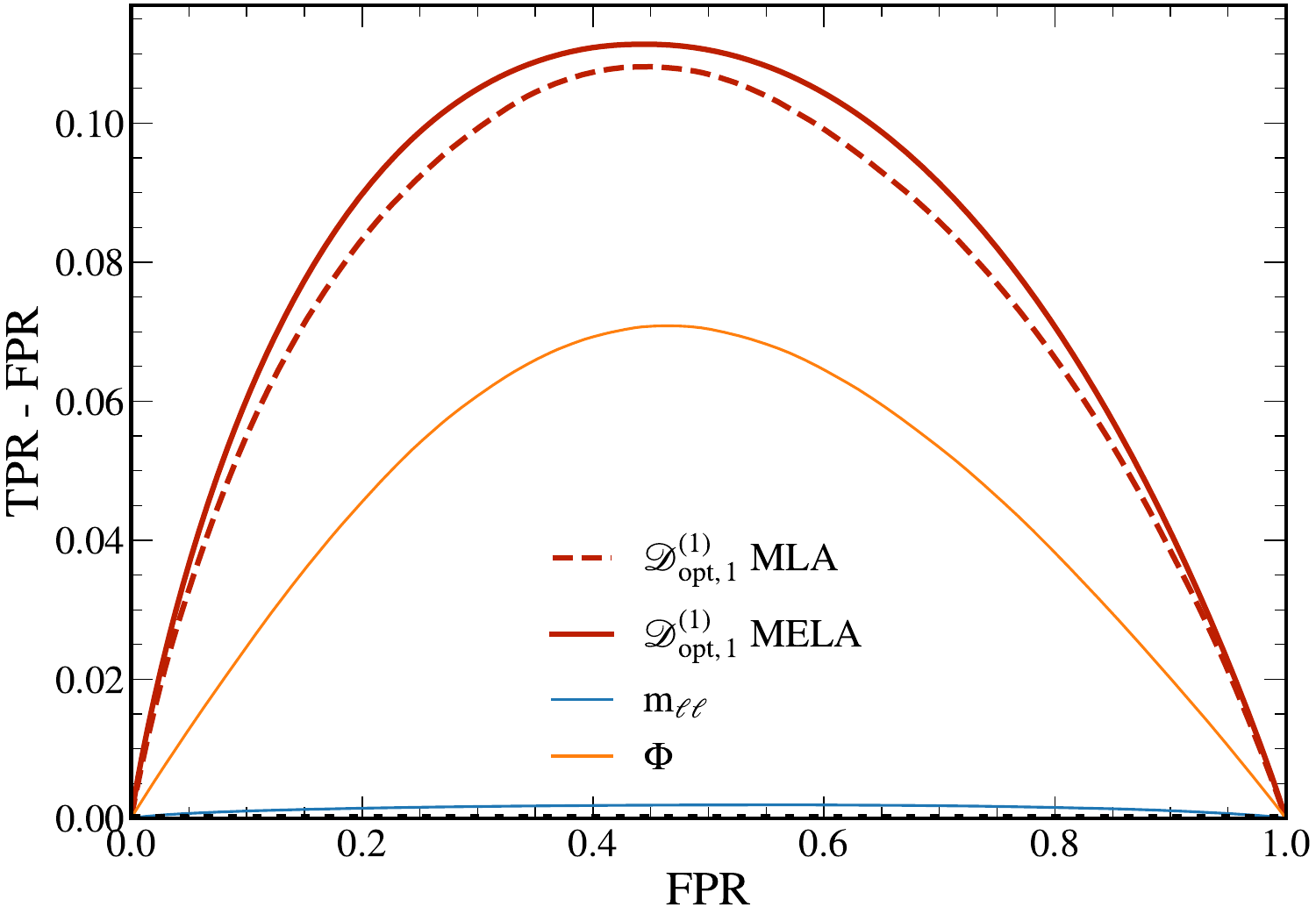}
    \includegraphics[width=0.45\textwidth]{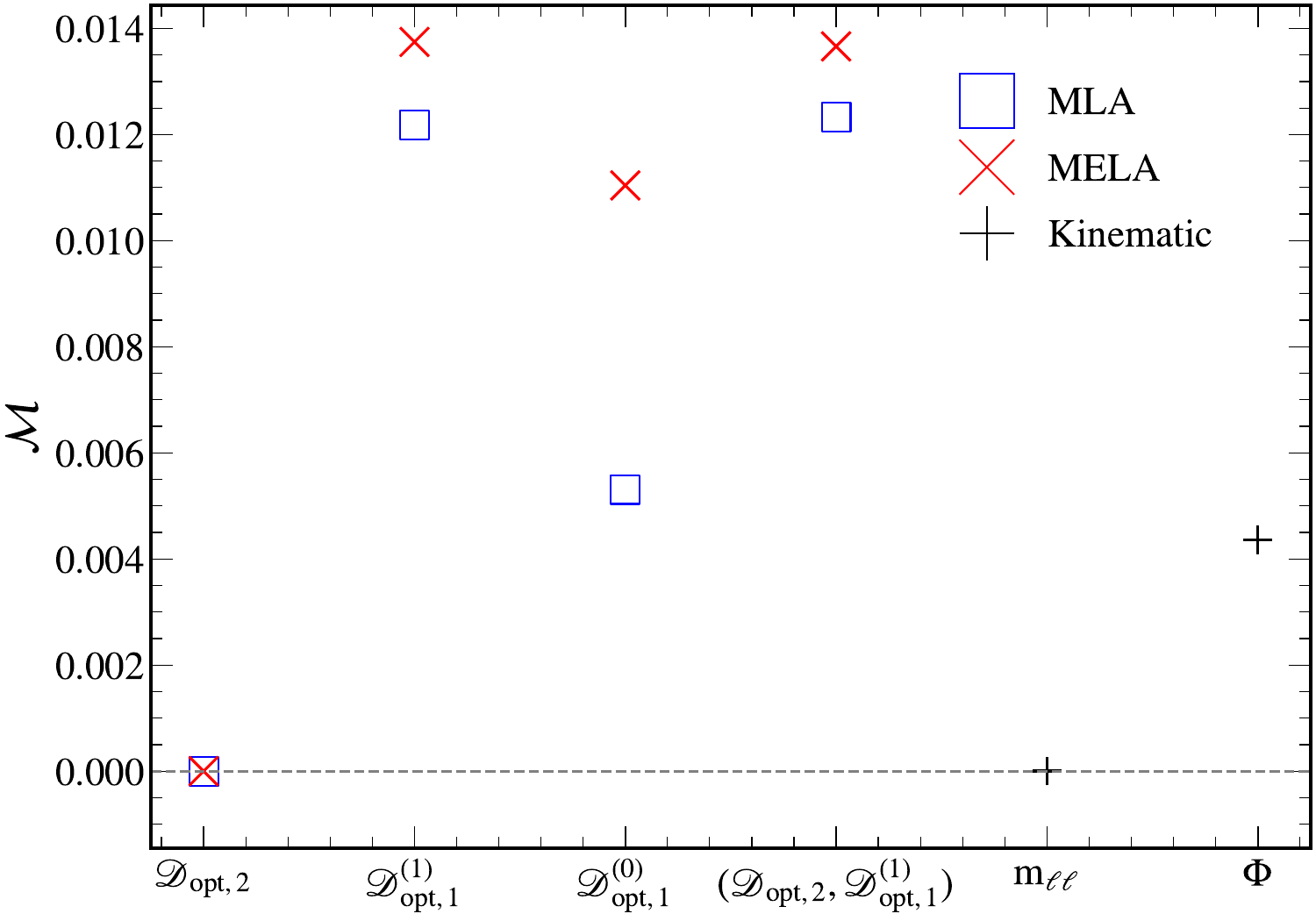}
    \caption{
     Left: The LOC curves evaluating the separation between the two alternative hypotheses, 
     which differ in sign of the SM interference with the BSM contribution with 
     $\theta_1=+\sqrt{\sigma_0/\sigma_1}\tilde{c}_{zz}$ and $\theta_1=-\sqrt{\sigma_0/\sigma_1}\tilde{c}_{zz}$,
    for four observables: $\mathcal{D}_{\text{opt,1}}^{(1)}$ computed using both MLA and MELA, as well as $m_{\ell\ell}$ and $\Phi$.
    Right: A summary of the LoC metric ${\cal M}$ derived from the LOC construction on the left, displayed for four discriminant 
    configurations, utilizing both MLA (blue box) and MELA (red cross), along with $m_{\ell\ell}$ and $\Phi$ (black cross).
    }
    \label{fig:dCP_5050}
  \end{center}
\end{figure}

\begin{figure}[t!]
  \begin{center}
    \captionsetup{justification=centerlast}
    \includegraphics[width=0.45\textwidth]{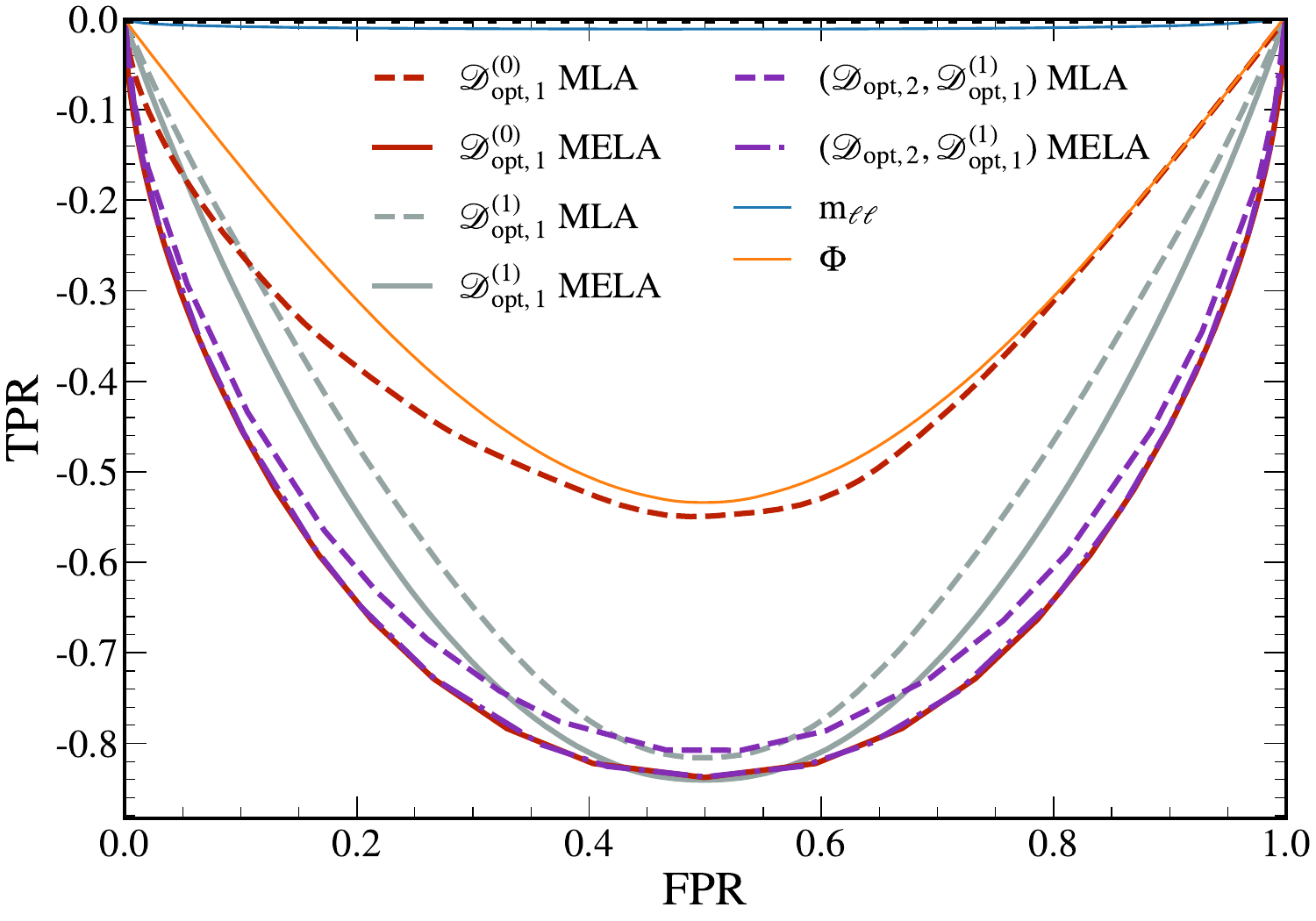}
    \includegraphics[width=0.45\textwidth]{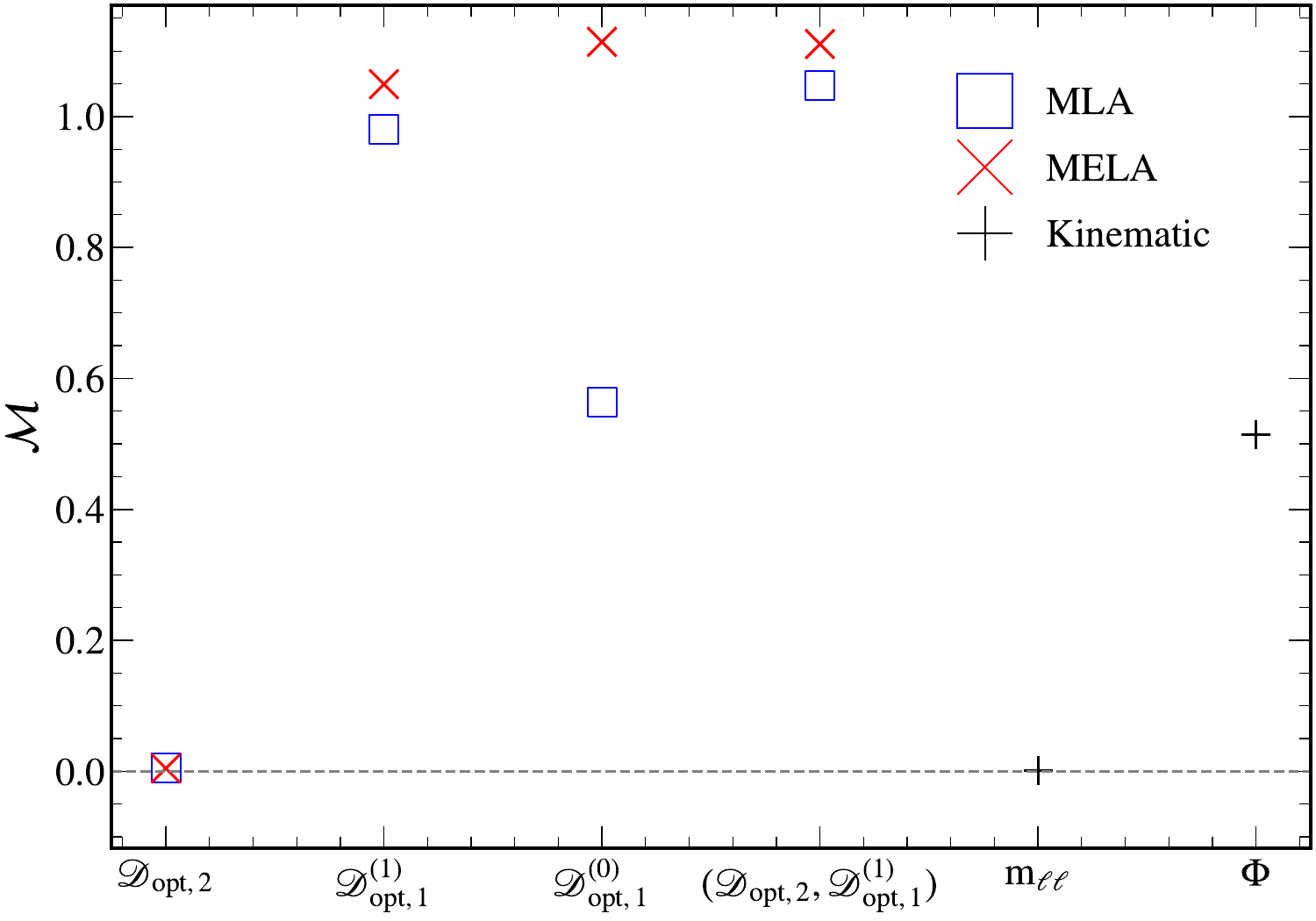}
    \caption{
    Left: The LOC curves evaluating the separation between the two alternative hypotheses, the SM and the pure interference 
    from the SM and the $\tilde{c}_{zz}$ contribution, for eight observables: $\mathcal{D}_{\text{opt,1}}^{(0)}$, 
    $\mathcal{D}_{\text{opt,1}}^{(1)}$, the two-dimensional discriminant $(\mathcal{D}_{\text{opt,2}},\mathcal{D}_{\text{opt,1}}^{(1)})$, 
    all computed using both MLA and MELA, as well as $m_{\ell\ell}$ and $\Phi$. 
    Right: A summary of the LoC metric ${\cal M}$ derived from the LOC construction on the left, displayed for four discriminant 
    configurations, utilizing both MLA (blue box) and MELA (red cross), along with $m_{\ell\ell}$ and $\Phi$ (black cross).
    }
    \label{fig:dCP_1000}
  \end{center}
\end{figure}

The validation of the $\mathcal{D}_{\text{opt,1}}^{(1)}$ discriminants in Fig.~\ref{fig:dCP_5050} reveals a similar pattern. 
These optimal discriminants perform best when evaluated on the appropriate samples corresponding to the positive and 
negative signs of interference between the SM and BSM contributions, which are depicted in the middle panels of 
Figs.~\ref{fig:MELA_d0m} and~\ref{fig:MLA_d0m}.
However, one can observe a slight degradation in the performance of the MLA discriminant compared to that of MELA.
This highlights the challenges of achieving ideal training to fully capture the quantum mechanical interference effects, 
which can be complex. Nevertheless, we believe that with enhanced training, these challenges can be addressed. 
A similar observation was made in Ref.~\cite{Gritsan:2020pib}, which involved a similar study of VBF production.
Similar to the study in Fig.~\ref{fig:d0m}, the non-optimal observables cannot match the performance of the optimal 
discriminant, while the two-dimensional discriminant $(\mathcal{D}_{\text{opt,2}}, \mathcal{D}_{\text{opt,1}}^{(1)})$ 
achieves the same performance as the optimal observable $\mathcal{D}_{\text{opt,1}}^{(1)}$.

The $\mathcal{D}_{\text{opt,1}}^{(0)}$ discriminants are validated in Fig.~\ref{fig:dCP_1000} using the LOC curve with 
the SM and pure interference distributions. As shown in the right panels of Figs.~\ref{fig:MELA_d0m} and~\ref{fig:MLA_d0m}, 
the integral of the interference distributions is zero, with both negative and positive entries. Due to the ordering of the
distributions according to the ratio of the interference and SM probabilities, the LOC curve exhibits negative TPR values. 
This illustrates a scenario where the classical ROC score cannot be used.
The observable $\mathcal{D}_{\text{opt,1}}^{(1)}$ contains significant information about the interference.
However, it is slightly less sensitive as it is intended for a different task.
It is notable that the two-dimensional discriminant $(\mathcal{D}_{\text{opt,2}}, \mathcal{D}_{\text{opt,1}}^{(1)})$ 
performs as well as $\mathcal{D}_{\text{opt,1}}^{(0)}$ when calculated using MELA, even though neither 
of the two discriminants is optimal for the task on its own. This highlights the fact that the joint distribution 
comprises all the necessary information for the task.

One notable observation is that the MLA-based $\mathcal{D}_{\text{opt,1}}^{(0)}$ discriminant performs quite poorly, 
with a score approximately two times lower than that obtained with MELA, and even lower than the $\mathcal{D}_{\text{opt,1}}^{(1)}$ 
trained with MLA. This underscores the fact that machine learning algorithms are not optimal for training on events with 
negative weights. Machine learning algorithms are generally not designed to work with negative probabilities.
Nonetheless, negative weights are a common feature in simulations of events at the LHC.
The TMVA package, originally designed for applications in particle physics, with BDT being the most robust choice, 
can handle negative-weight events treating them as signed contributions to loss functions. 
However, when the interference effects are $\mathcal{O}(100\%)$ and change sign, BDTs may no longer be valid.
Although developing the fundamental MLA approach to manage negative probabilities at $\mathcal{O}(100\%)$ 
could be a project in itself, we present a few practical alternatives.

This challenge can be addressed by considering potential solutions such as training with positive probabilities only. 
For example, one might train using scenarios such as SM versus SM+interference, SM versus SM+EFT, 
or SM+EFT versus SM-SFT, where the additions to the SM should be small.
The latter scenarios correspond to $\mathcal{D}_{\text{opt,1}}^{(\alpha)}$ with 
a small $\alpha$, as we have previously noted that as $\alpha$ approaches~0, it results in the EFT limit.
However, each of these solutions has its own limitations that do not exactly replicate the ultimate performance 
of $\mathcal{D}_{\text{opt,1}}^{(0)}$, and they are often unstable due to the smallness of the additions. 
We have already observed that the performance of $\mathcal{D}_{\text{opt,1}}^{(1)}$ shown 
in Fig.~\ref{fig:dCP_1000} is quite close to that of $\mathcal{D}_{\text{opt,1}}^{(0)}$ using MELA. 
A practical strategy to achieve full performance again is to train the two stable discriminants 
$(\mathcal{D}_{\text{opt,2}}, \mathcal{D}_{\text{opt,1}}^{(1)})$. 
Subsequently, \texttt{MiLoMerge} can be applied to produce an effective discriminant from the two-dimension
distribution with a limited number of bins, which is equivalent to $\mathcal{D}_{\text{opt,1}}^{(0)}$,
as demonstrated in this study in Fig.~\ref{fig:dCP_1000}.

The study presented here demonstrates the effectiveness of optimal observables through the concepts 
introduced in Eqs.~(\ref{eq:optimized2}) and~(\ref{eq:optimized1}). Both matrix elements and machine learning 
techniques can serve as effective methods for constructing such observables. However, when training the machine, 
it is important to be mindful of the challenges associated with extracting information about interferences, 
particularly the potential issues related to handling negative weights.
We also provide practical solutions to address these issues.


\section{Summary}
\label{sect:summary}

\noindent
We have presented a framework that integrates analytical and machine learning methods for calculating 
observables that are optimal for EFT and other applications at the LHC. This framework, 
discussed in Secs.~\ref{sect:input_obs} and~\ref{sect:input_ml}, is supported by the {\tt JHUGen} 
and {\tt MELA} packages, which facilitate simulation, re-weighting, and matrix-element 
calculations of processes involving the Higgs boson at the LHC. The fundamental concepts 
for training the pair of optimal observables behind Eqs.~(\ref{eq:optimized2}) and~(\ref{eq:optimized1})
for any parameter of interest have broader relevance in the data analysis of processes involving 
quantum-mechanical calculations, particularly in the context of EFT, discussed in Sec.~\ref{sect:input_eft}. 

The quantum effects of interference pose specific challenges in testing the observables and in 
training with machine learning techniques, particularly when the interference is substantial and changes sign.
The classical ROC score techniques for testing fails and machine learning algorithms cannot directly work 
with negative probabilities. To address the latter problem, in Sec.~\ref{sect:input_vh} we offer potential 
solutions which may rely on training with positive probabilities only.
To address the former problem, in Sec.~\ref{sect:input_general}
we introduce a new Likelihood Operating Characteristic (LOC) 
analysis in high-dimensional hypothesis spaces, which incorporates negative probabilities in the likelihood 
parameterization, introduces a novel LoC (length-of-the-curve) metric to achieve optimal performance under 
these extended conditions, and accomplishes everything in an input-invariant manner. 
The LOC framework is built on the ROC approach, but it offers a significantly wider range of applications.

Once the observables optimal for data analysis are constructed, the challenge of high dimensionality 
in the data remains. The presence of multiple observables, a large granularity of data for a given observable, 
or likely a combination of both can make practical analysis difficult.
A new Minimal-Loss Merging approach, built on the LOC framework and its associated metric, has been developed to 
address these challenges, supported by the Python {\tt MiLoMerge} package, which provides the necessary tools
and is accessible from Ref.~\cite{mela}.
With this method discussed in Secs.~\ref{sect:input_binary} and ~\ref{sect:input_general}, 
most relevant information can be effectively stored in a limited number of bins with minimal performance loss. 
This approach facilitates data analysis, data preservation, and global data combination.

The sequence of detailed Monte Carlo modeling, optimization of observables with proper testing, and optimal
dimensionality reduction of resulting data, especially when it is sensitive to a potentially large number of parameters 
of interest, ensures the highest return on investment.
All these features have been demonstrated through simulated analyses of the Higgs boson production 
and decay processes, with a detailed discussion provided in Secs.~\ref{sect:input_4l} and~\ref{sect:input_vh}.


\bigskip

\noindent
{\bf Acknowledgments}:

\noindent
This research is partially supported by the U.S. NSF under grants PHY-2012584 and PHY-2310072. 
Calculations reported in this paper were performed on the Advanced Research Computing at Hopkins (ARCH). 
We thank our colleagues involved in the JHU event generator / MELA project for their assistance 
and support throughout this initiative. 
We thank members of the LHC Higgs and EFT Working Groups for stimulating discussions.
Furthermore, we appreciate our collaboration with the members of the CMS Joint 4Lepton Study Team (CJLST), 
which has fostered engaging discussions and produced significant physics results. 


\input{paper-MiLo.bbl}

\end{document}

%% file: paper-MiLo.bbl
\providecommand{\href}[2]{#2}\begingroup\raggedright\endgroup

%% file: paper-MiLo.bbl
\begin{thebibliography}{10}%
\makeatletter
\providecommand{\hrefCMSnoop }[0]{\@secondoftwo}%
\makeatother
\providecommand{\doi}{\texttt{doi:}\begingroup \urlstyle{tt}\Url}

\bibitem{CMS:2012qbp}
\hrefCMSnoop {}{{CMS} Collaboration, ``{Observation of a New Boson at a Mass of
  125 GeV with the CMS Experiment at the LHC}'',} \textit{ Phys. Lett. B}
  \textbf{ 716} (2012) 30--61,
  \href{http://dx.doi.org/10.1016/j.physletb.2012.08.021}{\doi{10.1016/j.physletb.2012.08.021}},
  \href{http://www.arXiv.org/abs/1207.7235}{\texttt{ arXiv:1207.7235}}.

\bibitem{ATLAS:2012yve}
\hrefCMSnoop {}{{ATLAS} Collaboration, ``{Observation of a new particle in the
  search for the Standard Model Higgs boson with the ATLAS detector at the
  LHC}'',} \textit{ Phys. Lett. B} \textbf{ 716} (2012) 1--29,
  \href{http://dx.doi.org/10.1016/j.physletb.2012.08.020}{\doi{10.1016/j.physletb.2012.08.020}},
  \href{http://www.arXiv.org/abs/1207.7214}{\texttt{ arXiv:1207.7214}}.

\bibitem{Gao:2010qx}
Y.~Gao\hrefCMSnoop {}{ {et~al.}, ``{Spin determination of single-produced
  resonances at hadron colliders}'',} \textit{ Phys. Rev. D} \textbf{ 81}
  (2010) 075022,
  \href{http://dx.doi.org/10.1103/PhysRevD.81.075022}{\doi{10.1103/PhysRevD.81.075022}},
\href{http://www.arXiv.org/abs/1001.3396}{\texttt{ arXiv:1001.3396}}.

\bibitem{Bolognesi:2012mm}
S.~Bolognesi\hrefCMSnoop {}{ {et~al.}, ``{Spin and parity of a single-produced
  resonance at the LHC}'',} \textit{ Phys. Rev. D} \textbf{ 86} (2012) 095031,
  \href{http://dx.doi.org/10.1103/PhysRevD.86.095031}{\doi{10.1103/PhysRevD.86.095031}},
\href{http://www.arXiv.org/abs/1208.4018}{\texttt{ arXiv:1208.4018}}.

\bibitem{Anderson:2013afp}
I.~Anderson\hrefCMSnoop {}{ {et~al.}, ``{Constraining anomalous HVV
  interactions at proton and lepton colliders}'',} \textit{ Phys. Rev. D}
  \textbf{ 89} (2014) 035007,
  \href{http://dx.doi.org/10.1103/PhysRevD.89.035007}{\doi{10.1103/PhysRevD.89.035007}},
\href{http://www.arXiv.org/abs/1309.4819}{\texttt{ arXiv:1309.4819}}.

\bibitem{Gritsan:2016hjl}
\hrefCMSnoop {}{A.~V. Gritsan, R.~R{\"o}ntsch, M.~Schulze, and M.~Xiao,
  ``{Constraining anomalous Higgs boson couplings to the heavy flavor fermions
  using matrix element techniques}'',} \textit{ Phys. Rev. D} \textbf{ 94}
  (2016) 055023,
  \href{http://dx.doi.org/10.1103/PhysRevD.94.055023}{\doi{10.1103/PhysRevD.94.055023}},
\href{http://www.arXiv.org/abs/1606.03107}{\texttt{ arXiv:1606.03107}}.

\bibitem{Gritsan:2020pib}
A.~V. Gritsan\hrefCMSnoop {}{ {et~al.}, ``{New features in the JHU generator
  framework: constraining Higgs boson properties from on-shell and off-shell
  production}'',} \textit{ Phys. Rev. D} \textbf{ 102} (2020), no.~5, 056022,
  \href{http://dx.doi.org/10.1103/PhysRevD.102.056022}{\doi{10.1103/PhysRevD.102.056022}},
  \href{http://www.arXiv.org/abs/2002.09888}{\texttt{ arXiv:2002.09888}}.

\bibitem{Martini:2021uey}
\hrefCMSnoop {}{T.~Martini, R.-Q. Pan, M.~Schulze, and M.~Xiao, ``{Probing the
  CP structure of the top quark Yukawa coupling: Loop sensitivity versus
  on-shell sensitivity}'',} \textit{ Phys. Rev. D} \textbf{ 104} (2021) 055045,
  \href{http://dx.doi.org/10.1103/PhysRevD.104.055045}{\doi{10.1103/PhysRevD.104.055045}},
  \href{http://www.arXiv.org/abs/2104.04277}{\texttt{ arXiv:2104.04277}}.

\bibitem{Davis:2021tiv}
J.~Davis\hrefCMSnoop {}{ {et~al.}, ``{Constraining anomalous Higgs boson
  couplings to virtual photons}'',} \textit{ Phys. Rev. D} \textbf{ 105} (2022)
  096027,
  \href{http://dx.doi.org/10.1103/PhysRevD.105.096027}{\doi{10.1103/PhysRevD.105.096027}},
  \href{http://www.arXiv.org/abs/2109.13363}{\texttt{ arXiv:2109.13363}}.

\bibitem{Georgi:1993mps}
\hrefCMSnoop {}{H.~Georgi, ``Effective Field Theory'',} \textit{ Annual Review
  of Nuclear and Particle Science} \textbf{ 43} (1993) 209--252,
  \href{http://dx.doi.org/10.1146/annurev.ns.43.120193.001233}{\doi{10.1146/annurev.ns.43.120193.001233}}.

\bibitem{Weinberg:1995mt}
S.~Weinberg, ``The Quantum Theory of Fields. Volume 2: Modern Applications''.
\newblock Cambridge University Press, Cambridge, 1996.

\bibitem{Nelson:1986ki}
\hrefCMSnoop {}{C.~A. Nelson, ``{Correlation between decay planes in
  Higgs-boson decays into a W Pair (into a Z Pair)}'',} \textit{ Phys. Rev. D}
  \textbf{ 37} (1988) 1220,
\href{http://dx.doi.org/10.1103/PhysRevD.37.1220}{\doi{10.1103/PhysRevD.37.1220}}.

\bibitem{Soni:1993jc}
\hrefCMSnoop {}{A.~Soni and R.~M. Xu, ``{Probing CP violation via Higgs decays
  to four leptons}'',} \textit{ Phys. Rev. D} \textbf{ 48} (1993) 5259,
  \href{http://dx.doi.org/10.1103/PhysRevD.48.5259}{\doi{10.1103/PhysRevD.48.5259}},
\href{http://www.arXiv.org/abs/hep-ph/9301225}{\texttt{ arXiv:hep-ph/9301225}}.

\bibitem{Plehn:2001nj}
\hrefCMSnoop {}{T.~Plehn, D.~L. Rainwater, and D.~Zeppenfeld, ``{Determining
  the structure of Higgs couplings at the LHC}'',} \textit{ Phys. Rev. Lett.}
  \textbf{ 88} (2002) 051801,
  \href{http://dx.doi.org/10.1103/PhysRevLett.88.051801}{\doi{10.1103/PhysRevLett.88.051801}},
\href{http://www.arXiv.org/abs/hep-ph/0105325}{\texttt{ arXiv:hep-ph/0105325}}.

\bibitem{Choi:2002jk}
\hrefCMSnoop {}{S.~Y. Choi, D.~J. Miller, M.~M. M{\" u}hlleitner, and P.~M.
  Zerwas, ``{Identifying the Higgs spin and parity in decays to Z pairs}'',}
  \textit{ Phys. Lett. B} \textbf{ 553} (2003) 61,
  \href{http://dx.doi.org/10.1016/S0370-2693(02)03191-X}{\doi{10.1016/S0370-2693(02)03191-X}},
\href{http://www.arXiv.org/abs/hep-ph/0210077}{\texttt{ arXiv:hep-ph/0210077}}.

\bibitem{Buszello:2002uu}
\hrefCMSnoop {}{C.~P. Buszello, I.~Fleck, P.~Marquard, and J.~J. van~der Bij,
  ``{Prospective analysis of spin- and CP-sensitive variables in $H \to ZZ \to
  \ell_1^+ \ell_1^- \ell_2^+ \ell_2^-$ at the LHC}'',} \textit{ Eur. Phys. J.
  C} \textbf{ 32} (2004) 209,
  \href{http://dx.doi.org/10.1140/epjc/s2003-01392-0}{\doi{10.1140/epjc/s2003-01392-0}},
\href{http://www.arXiv.org/abs/hep-ph/0212396}{\texttt{ arXiv:hep-ph/0212396}}.

\bibitem{Hankele:2006ma}
\hrefCMSnoop {}{V.~Hankele, G.~Klamke, D.~Zeppenfeld, and T.~Figy, ``{Anomalous
  Higgs boson couplings in vector boson fusion at the CERN LHC}'',} \textit{
  Phys. Rev. D} \textbf{ 74} (2006) 095001,
  \href{http://dx.doi.org/10.1103/PhysRevD.74.095001}{\doi{10.1103/PhysRevD.74.095001}},
\href{http://www.arXiv.org/abs/hep-ph/0609075}{\texttt{ arXiv:hep-ph/0609075}}.

\bibitem{Accomando:2006ga}
\hrefCMSnoop {}{E.~Accomando {et~al.}, ``{Workshop on CP studies and
  non-standard Higgs physics}'',} (2006).
\href{http://www.arXiv.org/abs/hep-ph/0608079}{\texttt{ arXiv:hep-ph/0608079}}.

\bibitem{Godbole:2007cn}
\hrefCMSnoop {}{R.~M. Godbole, D.~J. Miller, and M.~M. M{\" u}hlleitner,
  ``{Aspects of CP violation in the HZZ coupling at the LHC}'',} \textit{ JHEP}
  \textbf{ 12} (2007) 031,
  \href{http://dx.doi.org/10.1088/1126-6708/2007/12/031}{\doi{10.1088/1126-6708/2007/12/031}},
\href{http://www.arXiv.org/abs/0708.0458}{\texttt{ arXiv:0708.0458}}.

\bibitem{Hagiwara:2009wt}
\hrefCMSnoop {}{K.~Hagiwara, Q.~Li, and K.~Mawatari, ``{Jet angular correlation
  in vector-boson fusion processes at hadron colliders}'',} \textit{ JHEP}
  \textbf{ 07} (2009) 101,
  \href{http://dx.doi.org/10.1088/1126-6708/2009/07/101}{\doi{10.1088/1126-6708/2009/07/101}},
\href{http://www.arXiv.org/abs/0905.4314}{\texttt{ arXiv:0905.4314}}.

\bibitem{DeRujula:2010ys}
A.~De~R{\' u}jula\hrefCMSnoop {}{ {et~al.}, ``{Higgs look-alikes at the
  LHC}'',} \textit{ Phys. Rev. D} \textbf{ 82} (2010) 013003,
  \href{http://dx.doi.org/10.1103/PhysRevD.82.013003}{\doi{10.1103/PhysRevD.82.013003}},
\href{http://www.arXiv.org/abs/1001.5300}{\texttt{ arXiv:1001.5300}}.

\bibitem{Christensen:2010pf}
\hrefCMSnoop {}{N.~D. Christensen, T.~Han, and Y.~Li, ``{Testing CP Violation
  in ZZH Interactions at the LHC}'',} \textit{ Phys. Lett. B} \textbf{ 693}
  (2010) 28,
  \href{http://dx.doi.org/10.1016/j.physletb.2010.08.008}{\doi{10.1016/j.physletb.2010.08.008}},
\href{http://www.arXiv.org/abs/1005.5393}{\texttt{ arXiv:1005.5393}}.

\bibitem{Ellis:2012xd}
\hrefCMSnoop {}{J.~Ellis, D.~S. Hwang, V.~Sanz, and T.~You, ``{A fast track
  towards the `Higgs' spin and parity}'',} \textit{ JHEP} \textbf{ 11} (2012)
  134,
  \href{http://dx.doi.org/10.1007/JHEP11(2012)134}{\doi{10.1007/JHEP11(2012)134}},
\href{http://www.arXiv.org/abs/1208.6002}{\texttt{ arXiv:1208.6002}}.

\bibitem{Chen:2012jy}
\hrefCMSnoop {}{Y.~Chen, N.~Tran, and R.~Vega-Morales, ``{Scrutinizing the
  Higgs signal and background in the $2e2\mu$ golden channel}'',} \textit{
  JHEP} \textbf{ 01} (2013) 182,
  \href{http://dx.doi.org/10.1007/JHEP01(2013)182}{\doi{10.1007/JHEP01(2013)182}},
\href{http://www.arXiv.org/abs/1211.1959}{\texttt{ arXiv:1211.1959}}.

\bibitem{Artoisenet:2013puc}
P.~Artoisenet\hrefCMSnoop {}{ {et~al.}, ``{A framework for Higgs
  characterisation}'',} \textit{ JHEP} \textbf{ 11} (2013) 043,
  \href{http://dx.doi.org/10.1007/JHEP11(2013)043}{\doi{10.1007/JHEP11(2013)043}},
\href{http://www.arXiv.org/abs/1306.6464}{\texttt{ arXiv:1306.6464}}.

\bibitem{Chen:2013waa}
M.~Chen\hrefCMSnoop {}{ {et~al.}, ``{Role of interference in unraveling the ZZ
  couplings of the newly discovered boson at the LHC}'',} \textit{ Phys. Rev.
  D} \textbf{ 89} (2014) 034002,
  \href{http://dx.doi.org/10.1103/PhysRevD.89.034002}{\doi{10.1103/PhysRevD.89.034002}},
\href{http://www.arXiv.org/abs/1310.1397}{\texttt{ arXiv:1310.1397}}.

\bibitem{Maltoni:2013sma}
\hrefCMSnoop {}{F.~Maltoni, K.~Mawatari, and M.~Zaro, ``{Higgs characterisation
  via vector-boson fusion and associated production: NLO and parton-shower
  effects}'',} \textit{ Eur. Phys. J.} \textbf{ C74} (2014), no.~1, 2710,
  \href{http://dx.doi.org/10.1140/epjc/s10052-013-2710-5}{\doi{10.1140/epjc/s10052-013-2710-5}},
\href{http://www.arXiv.org/abs/1311.1829}{\texttt{ arXiv:1311.1829}}.

\bibitem{Azatov:2014jga}
\hrefCMSnoop {}{A.~Azatov, C.~Grojean, A.~Paul, and E.~Salvioni, ``{Taming the
  off-shell Higgs boson}'',} \textit{ Zh. Eksp. Teor. Fiz.} \textbf{ 147}
  (2015) 410--425, \href{http://dx.doi.org/10.1134/S1063776115030140,
  10.7868/S0044451015030039}{\doi{10.1134/S1063776115030140,
  10.7868/S0044451015030039}},
  \href{http://www.arXiv.org/abs/1406.6338}{\texttt{ arXiv:1406.6338}}.
[J. Exp. Theor. Phys.120,354(2015)].

\bibitem{Cacciapaglia:2014rla}
\hrefCMSnoop {}{G.~Cacciapaglia, A.~Deandrea, G.~Drieu La~Rochelle, and J.-B.
  Flament, ``{Higgs couplings: disentangling New Physics with off-shell
  measurements}'',} \textit{ Phys. Rev. Lett.} \textbf{ 113} (2014) 201802,
  \href{http://dx.doi.org/10.1103/PhysRevLett.113.201802}{\doi{10.1103/PhysRevLett.113.201802}},
\href{http://www.arXiv.org/abs/1406.1757}{\texttt{ arXiv:1406.1757}}.

\bibitem{Denner:2014cla}
\hrefCMSnoop {}{A.~Denner, S.~Dittmaier, S.~Kallweit, and A.~M{\"u}ck, ``{HAWK
  2.0: A Monte Carlo program for Higgs production in vector-boson fusion and
  Higgs strahlung at hadron colliders}'',} \textit{ Comput. Phys. Commun.}
  \textbf{ 195} (2015) 161--171,
  \href{http://dx.doi.org/10.1016/j.cpc.2015.04.021}{\doi{10.1016/j.cpc.2015.04.021}},
\href{http://www.arXiv.org/abs/1412.5390}{\texttt{ arXiv:1412.5390}}.

\bibitem{Dolan:2014upa}
\hrefCMSnoop {}{M.~J. Dolan, P.~Harris, M.~Jankowiak, and M.~Spannowsky,
  ``{Constraining $CP$-violating Higgs sectors at the LHC using gluon
  fusion}'',} \textit{ Phys. Rev. D} \textbf{ 90} (2014) 073008,
  \href{http://dx.doi.org/10.1103/PhysRevD.90.073008}{\doi{10.1103/PhysRevD.90.073008}},
\href{http://www.arXiv.org/abs/1406.3322}{\texttt{ arXiv:1406.3322}}.

\bibitem{Englert:2014ffa}
\hrefCMSnoop {}{C.~Englert, Y.~Soreq, and M.~Spannowsky, ``{Off-Shell Higgs
  Coupling Measurements in BSM scenarios}'',} \textit{ JHEP} \textbf{ 05}
  (2015) 145,
  \href{http://dx.doi.org/10.1007/JHEP05(2015)145}{\doi{10.1007/JHEP05(2015)145}},
\href{http://www.arXiv.org/abs/1410.5440}{\texttt{ arXiv:1410.5440}}.

\bibitem{Gonzalez-Alonso:2014eva}
\hrefCMSnoop {}{M.~Gonzalez-Alonso, A.~Greljo, G.~Isidori, and D.~Marzocca,
  ``{Pseudo-observables in Higgs decays}'',} \textit{ Eur. Phys. J. C} \textbf{
  75} (2015) 128,
  \href{http://dx.doi.org/10.1140/epjc/s10052-015-3345-5}{\doi{10.1140/epjc/s10052-015-3345-5}},
\href{http://www.arXiv.org/abs/1412.6038}{\texttt{ arXiv:1412.6038}}.

\bibitem{Ballestrero:2015jca}
\hrefCMSnoop {}{A.~Ballestrero and E.~Maina, ``{Interference Effects in Higgs
  production through Vector Boson Fusion in the Standard Model and its Singlet
  Extension}'',} \textit{ JHEP} \textbf{ 01} (2016) 045,
  \href{http://dx.doi.org/10.1007/JHEP01(2016)045}{\doi{10.1007/JHEP01(2016)045}},
\href{http://www.arXiv.org/abs/1506.02257}{\texttt{ arXiv:1506.02257}}.

\bibitem{Greljo:2015sla}
\hrefCMSnoop {}{A.~Greljo, G.~Isidori, J.~M. Lindert, and D.~Marzocca,
  ``{Pseudo-observables in electroweak Higgs production}'',} \textit{ Eur.
  Phys. J. C} \textbf{ 76} (2016) 158,
  \href{http://dx.doi.org/10.1140/epjc/s10052-016-4000-5}{\doi{10.1140/epjc/s10052-016-4000-5}},
\href{http://www.arXiv.org/abs/1512.06135}{\texttt{ arXiv:1512.06135}}.

\bibitem{Hespel:2015zea}
\hrefCMSnoop {}{B.~Hespel, F.~Maltoni, and E.~Vryonidou, ``{Higgs and Z boson
  associated production via gluon fusion in the SM and the 2HDM}'',} \textit{
  JHEP} \textbf{ 06} (2015) 065,
  \href{http://dx.doi.org/10.1007/JHEP06(2015)065}{\doi{10.1007/JHEP06(2015)065}},
\href{http://www.arXiv.org/abs/1503.01656}{\texttt{ arXiv:1503.01656}}.

\bibitem{Kauer:2015dma}
\hrefCMSnoop {}{N.~Kauer, C.~O'Brien, and E.~Vryonidou, ``{Interference effects
  for $ H\to W\;W\to \ell \nu q{\overline{q}}^{\prime } $ and $ H\to ZZ\to \ell
  \overline{\ell}q\overline{q} $ searches in gluon fusion at the LHC}'',}
  \textit{ JHEP} \textbf{ 10} (2015) 074,
  \href{http://dx.doi.org/10.1007/JHEP10(2015)074}{\doi{10.1007/JHEP10(2015)074}},
\href{http://www.arXiv.org/abs/1506.01694}{\texttt{ arXiv:1506.01694}}.

\bibitem{Kauer:2015hia}
\hrefCMSnoop {}{N.~Kauer and C.~O'Brien, ``{Heavy Higgs signal–background
  interference in $gg\rightarrow VV$ in the Standard Model plus real
  singlet}'',} \textit{ Eur. Phys. J. C} \textbf{ 75} (2015) 374,
  \href{http://dx.doi.org/10.1140/epjc/s10052-015-3586-3}{\doi{10.1140/epjc/s10052-015-3586-3}},
\href{http://www.arXiv.org/abs/1502.04113}{\texttt{ arXiv:1502.04113}}.

\bibitem{Kilian:2015opv}
\hrefCMSnoop {}{W.~Kilian, T.~Ohl, J.~Reuter, and M.~Sekulla, ``{Resonances at
  the LHC beyond the Higgs boson: The scalar/tensor case}'',} \textit{ Phys.
  Rev. D} \textbf{ 93} (2016) 036004,
  \href{http://dx.doi.org/10.1103/PhysRevD.93.036004}{\doi{10.1103/PhysRevD.93.036004}},
\href{http://www.arXiv.org/abs/1511.00022}{\texttt{ arXiv:1511.00022}}.

\bibitem{Mimasu:2015nqa}
\hrefCMSnoop {}{K.~Mimasu, V.~Sanz, and C.~Williams, ``{Higher Order QCD
  predictions for Associated Higgs production with anomalous couplings to gauge
  bosons}'',} \textit{ JHEP} \textbf{ 08} (2016) 039,
  \href{http://dx.doi.org/10.1007/JHEP08(2016)039}{\doi{10.1007/JHEP08(2016)039}},
\href{http://www.arXiv.org/abs/1512.02572}{\texttt{ arXiv:1512.02572}}.

\bibitem{Degrande:2016dqg}
C.~Degrande\hrefCMSnoop {}{ {et~al.}, ``{Electroweak Higgs boson production in
  the standard model effective field theory beyond leading order in QCD}'',}
  \textit{ Eur. Phys. J.} \textbf{ C77} (2017) 262,
  \href{http://dx.doi.org/10.1140/epjc/s10052-017-4793-x}{\doi{10.1140/epjc/s10052-017-4793-x}},
\href{http://www.arXiv.org/abs/1609.04833}{\texttt{ arXiv:1609.04833}}.

\bibitem{Dwivedi:2016xwm}
\hrefCMSnoop {}{S.~Dwivedi, D.~K. Ghosh, B.~Mukhopadhyaya, and A.~Shivaji,
  ``{Distinguishing $CP$-odd couplings of the Higgs boson to weak boson
  pairs}'',} \textit{ Phys. Rev. D} \textbf{ 93} (2016) 115039,
  \href{http://dx.doi.org/10.1103/PhysRevD.93.115039}{\doi{10.1103/PhysRevD.93.115039}},
\href{http://www.arXiv.org/abs/1603.06195}{\texttt{ arXiv:1603.06195}}.

\bibitem{deFlorian:2016spz}
\hrefCMSnoop {}{{LHC Higgs Cross Section Working Group} Collaboration,
  ``{Handbook of LHC Higgs Cross Sections: 4. Deciphering the Nature of the
  Higgs Sector}'',}
  \href{http://dx.doi.org/10.23731/CYRM-2017-002}{\doi{10.23731/CYRM-2017-002}},
\href{http://www.arXiv.org/abs/1610.07922}{\texttt{ arXiv:1610.07922}}.

\bibitem{Azatov:2016xik}
\hrefCMSnoop {}{A.~Azatov, C.~Grojean, A.~Paul, and E.~Salvioni, ``{Resolving
  gluon fusion loops at current and future hadron colliders}'',} \textit{ JHEP}
  \textbf{ 09} (2016) 123,
  \href{http://dx.doi.org/10.1007/JHEP09(2016)123}{\doi{10.1007/JHEP09(2016)123}},
\href{http://www.arXiv.org/abs/1608.00977}{\texttt{ arXiv:1608.00977}}.

\bibitem{Denner:2017vms}
\hrefCMSnoop {}{A.~Denner, J.-N. Lang, and S.~Uccirati, ``{NLO electroweak
  corrections in extended Higgs Sectors with RECOLA2}'',} \textit{ JHEP}
  \textbf{ 07} (2017) 087,
  \href{http://dx.doi.org/10.1007/JHEP07(2017)087}{\doi{10.1007/JHEP07(2017)087}},
\href{http://www.arXiv.org/abs/1705.06053}{\texttt{ arXiv:1705.06053}}.

\bibitem{Deutschmann:2017qum}
\hrefCMSnoop {}{N.~Deutschmann, C.~Duhr, F.~Maltoni, and E.~Vryonidou,
  ``{Gluon-fusion Higgs production in the Standard Model Effective Field
  Theory}'',} \textit{ JHEP} \textbf{ 12} (2017) 063,
  \href{http://dx.doi.org/10.1007/JHEP12(2017)063,
  10.1007/JHEP02(2018)159}{\doi{10.1007/JHEP12(2017)063,
  10.1007/JHEP02(2018)159}},
  \href{http://www.arXiv.org/abs/1708.00460}{\texttt{ arXiv:1708.00460}}.
[Erratum: JHEP02,159(2018)].

\bibitem{Greljo:2017spw}
A.~Greljo\hrefCMSnoop {}{ {et~al.}, ``{Electroweak Higgs production with
  HiggsPO at NLO QCD}'',} \textit{ Eur. Phys. J. C} \textbf{ 77} (2017) 838,
  \href{http://dx.doi.org/10.1140/epjc/s10052-017-5422-4}{\doi{10.1140/epjc/s10052-017-5422-4}},
\href{http://www.arXiv.org/abs/1710.04143}{\texttt{ arXiv:1710.04143}}.

\bibitem{Goncalves:2017gzy}
\hrefCMSnoop {}{D.~Goncalves, T.~Plehn, and J.~M. Thompson, ``{Weak boson
  fusion at 100 TeV}'',} \textit{ Phys. Rev. D} \textbf{ 95} (2017) 095011,
  \href{http://dx.doi.org/10.1103/PhysRevD.95.095011}{\doi{10.1103/PhysRevD.95.095011}},
\href{http://www.arXiv.org/abs/1702.05098}{\texttt{ arXiv:1702.05098}}.

\bibitem{Jager:2017owh}
\hrefCMSnoop {}{B.~J{\"a}ger, L.~Salfelder, M.~Worek, and D.~Zeppenfeld,
  ``{Physics opportunities for vector-boson scattering at a future 100 TeV
  hadron collider}'',} \textit{ Phys. Rev. D} \textbf{ 96} (2017) 073008,
  \href{http://dx.doi.org/10.1103/PhysRevD.96.073008}{\doi{10.1103/PhysRevD.96.073008}},
\href{http://www.arXiv.org/abs/1704.04911}{\texttt{ arXiv:1704.04911}}.

\bibitem{Brass:2018hfw}
S.~Brass\hrefCMSnoop {}{ {et~al.}, ``{Transversal Modes and Higgs Bosons in
  Electroweak Vector-Boson Scattering at the LHC}'',} \textit{ Eur. Phys. J. C}
  \textbf{ 78} (2018) 931,
  \href{http://dx.doi.org/10.1140/epjc/s10052-018-6398-4}{\doi{10.1140/epjc/s10052-018-6398-4}},
\href{http://www.arXiv.org/abs/1807.02512}{\texttt{ arXiv:1807.02512}}.

\bibitem{Gomez-Ambrosio:2018pnl}
\hrefCMSnoop {}{R.~Gomez-Ambrosio, ``{Studies of Dimension-Six EFT effects in
  Vector Boson Scattering}'',} \textit{ Eur. Phys. J. C} \textbf{ 79} (2019)
  389,
  \href{http://dx.doi.org/10.1140/epjc/s10052-019-6893-2}{\doi{10.1140/epjc/s10052-019-6893-2}},
  \href{http://www.arXiv.org/abs/1809.04189}{\texttt{ arXiv:1809.04189}}.

\bibitem{Goncalves:2018pkt}
\hrefCMSnoop {}{D.~Gonçalves, T.~Han, and S.~Mukhopadhyay, ``{Higgs Couplings
  at High Scales}'',} \textit{ Phys. Rev. D} \textbf{ 98} (2018) 015023,
  \href{http://dx.doi.org/10.1103/PhysRevD.98.015023}{\doi{10.1103/PhysRevD.98.015023}},
\href{http://www.arXiv.org/abs/1803.09751}{\texttt{ arXiv:1803.09751}}.

\bibitem{Harlander:2018yns}
\hrefCMSnoop {}{R.~V. Harlander, J.~Klappert, C.~Pandini, and
  A.~Papaefstathiou, ``{Exploiting the WH/ZH symmetry in the search for New
  Physics}'',} \textit{ Eur. Phys. J. C} \textbf{ 78} (2018) 760,
  \href{http://dx.doi.org/10.1140/epjc/s10052-018-6234-x}{\doi{10.1140/epjc/s10052-018-6234-x}},
\href{http://www.arXiv.org/abs/1804.02299}{\texttt{ arXiv:1804.02299}}.

\bibitem{Harlander:2018yio}
\hrefCMSnoop {}{R.~V. Harlander, J.~Klappert, S.~Liebler, and L.~Simon,
  ``{vh@nnlo-v2: New physics in Higgs Strahlung}'',} \textit{ JHEP} \textbf{
  05} (2018) 089,
  \href{http://dx.doi.org/10.1007/JHEP05(2018)089}{\doi{10.1007/JHEP05(2018)089}},
\href{http://www.arXiv.org/abs/1802.04817}{\texttt{ arXiv:1802.04817}}.

\bibitem{Lee:2018fxj}
\hrefCMSnoop {}{S.~J. Lee, M.~Park, and Z.~Qian, ``{Probing unitarity violation
  in the tail of the off-shell Higgs boson in $V_LV_L$ mode}'',} \textit{ Phys.
  Rev. D} \textbf{ 100} (2019) 011702,
  \href{http://dx.doi.org/10.1103/PhysRevD.100.011702}{\doi{10.1103/PhysRevD.100.011702}},
\href{http://www.arXiv.org/abs/1812.02679}{\texttt{ arXiv:1812.02679}}.

\bibitem{Kalinowski:2018oxd}
J.~Kalinowski\hrefCMSnoop {}{ {et~al.}, ``{Same-sign WW scattering at the LHC:
  can we discover BSM effects before discovering new states?}'',} \textit{ Eur.
  Phys. J. C} \textbf{ 78} (2018) 403,
  \href{http://dx.doi.org/10.1140/epjc/s10052-018-5885-y}{\doi{10.1140/epjc/s10052-018-5885-y}},
\href{http://www.arXiv.org/abs/1802.02366}{\texttt{ arXiv:1802.02366}}.

\bibitem{Perez:2018kav}
\hrefCMSnoop {}{G.~Perez, M.~Sekulla, and D.~Zeppenfeld, ``{Anomalous quartic
  gauge couplings and unitarization for the vector boson scattering process
  $pp\rightarrow W^+W^+jjX\rightarrow \ell ^+\nu _\ell \ell ^+\nu _\ell
  jjX$}'',} \textit{ Eur. Phys. J. C} \textbf{ 78} (2018) 759,
  \href{http://dx.doi.org/10.1140/epjc/s10052-018-6230-1}{\doi{10.1140/epjc/s10052-018-6230-1}},
\href{http://www.arXiv.org/abs/1807.02707}{\texttt{ arXiv:1807.02707}}.

\bibitem{Jaquier:2019bfs}
\hrefCMSnoop {}{M.~Jaquier and R.~R{\"o}ntsch, ``{Mixed scalar-pseudoscalar
  Higgs boson production through next-to-next-to-leading order at the LHC}'',}
  \textit{ JHEP} \textbf{ 06} (2020) 005,
  \href{http://dx.doi.org/10.1007/JHEP06(2020)005}{\doi{10.1007/JHEP06(2020)005}},
\href{http://www.arXiv.org/abs/1911.10631}{\texttt{ arXiv:1911.10631}}.

\bibitem{Denner:2019fcr}
\hrefCMSnoop {}{A.~Denner, S.~Dittmaier, and A.~M{\"u}ck, ``{PROPHECY4F 3.0: A
  Monte Carlo program for Higgs-boson decays into four-fermion final states in
  and beyond the Standard Model}'',} \textit{ Comput. Phys. Commun.} \textbf{
  254} (2020) 107336,
  \href{http://dx.doi.org/10.1016/j.cpc.2020.107336}{\doi{10.1016/j.cpc.2020.107336}},
\href{http://www.arXiv.org/abs/1912.02010}{\texttt{ arXiv:1912.02010}}.

\bibitem{Banerjee:2019twi}
S.~Banerjee\hrefCMSnoop {}{ {et~al.}, ``{Towards the ultimate differential
  SMEFT analysis}'',} \textit{ JHEP} \textbf{ 09} (2020) 170,
  \href{http://dx.doi.org/10.1007/JHEP09(2020)170}{\doi{10.1007/JHEP09(2020)170}},
  \href{http://www.arXiv.org/abs/1912.07628}{\texttt{ arXiv:1912.07628}}.

\bibitem{CMS:2012vby}
\hrefCMSnoop {}{{CMS} Collaboration, ``{Study of the Mass and Spin-Parity of
  the Higgs Boson Candidate Via Its Decays to Z Boson Pairs}'',} \textit{ Phys.
  Rev. Lett.} \textbf{ 110} (2013) 081803,
  \href{http://dx.doi.org/10.1103/PhysRevLett.110.081803}{\doi{10.1103/PhysRevLett.110.081803}},
  \href{http://www.arXiv.org/abs/1212.6639}{\texttt{ arXiv:1212.6639}}.

\bibitem{CMS:2013fjq}
\hrefCMSnoop {}{{CMS} Collaboration, ``{Measurement of the Properties of a
  Higgs Boson in the Four-Lepton Final State}'',} \textit{ Phys. Rev. D}
  \textbf{ 89} (2014) 092007,
  \href{http://dx.doi.org/10.1103/PhysRevD.89.092007}{\doi{10.1103/PhysRevD.89.092007}},
  \href{http://www.arXiv.org/abs/1312.5353}{\texttt{ arXiv:1312.5353}}.

\bibitem{CMS:2014nkk}
\hrefCMSnoop {}{{CMS} Collaboration, ``{Constraints on the spin-parity and
  anomalous HVV couplings of the Higgs boson in proton collisions at 7 and 8
  TeV}'',} \textit{ Phys. Rev. D} \textbf{ 92} (2015) 012004,
  \href{http://dx.doi.org/10.1103/PhysRevD.92.012004}{\doi{10.1103/PhysRevD.92.012004}},
  \href{http://www.arXiv.org/abs/1411.3441}{\texttt{ arXiv:1411.3441}}.

\bibitem{Sirunyan:2017tqd}
\hrefCMSnoop {}{{CMS} Collaboration, ``{Constraints on anomalous Higgs boson
  couplings using production and decay information in the four-lepton final
  state}'',} \textit{ Phys. Lett. B} \textbf{ 775} (2017) 1,
  \href{http://dx.doi.org/10.1016/j.physletb.2017.10.021}{\doi{10.1016/j.physletb.2017.10.021}},
\href{http://www.arXiv.org/abs/1707.00541}{\texttt{ arXiv:1707.00541}}.

\bibitem{Sirunyan:2018qlb}
\hrefCMSnoop {}{{CMS} Collaboration, ``{Search for a new scalar resonance
  decaying to a pair of Z bosons in proton-proton collisions at $\sqrt{s}=13$
  TeV}'',} \textit{ JHEP} \textbf{ 06} (2018) 127,
  \href{http://dx.doi.org/10.1007/JHEP06(2018)127,
  10.1007/JHEP03(2019)128}{\doi{10.1007/JHEP06(2018)127,
  10.1007/JHEP03(2019)128}},
  \href{http://www.arXiv.org/abs/1804.01939}{\texttt{ arXiv:1804.01939}}.
[Erratum: JHEP03,128(2019)].

\bibitem{Sirunyan:2019pqw}
\hrefCMSnoop {}{{CMS} Collaboration, ``{Search for a heavy Higgs boson decaying
  to a pair of W bosons in proton-proton collisions at $\sqrt{s} =$ 13 TeV}'',}
\href{http://www.arXiv.org/abs/1912.01594}{\texttt{ arXiv:1912.01594}}.

\bibitem{Sirunyan:2019nbs}
\hrefCMSnoop {}{{CMS} Collaboration, ``{Constraints on anomalous $HVV$
  couplings from the production of Higgs bosons decaying to $\tau$ lepton
  pairs}'',} \textit{ Phys. Rev. D} \textbf{ 100} (2019) 112002,
  \href{http://dx.doi.org/10.1103/PhysRevD.100.112002}{\doi{10.1103/PhysRevD.100.112002}},
\href{http://www.arXiv.org/abs/1903.06973}{\texttt{ arXiv:1903.06973}}.

\bibitem{Sirunyan:2019twz}
\hrefCMSnoop {}{{CMS} Collaboration, ``{Measurements of the Higgs boson width
  and anomalous HVV couplings from on-shell and off-shell production in the
  four-lepton final state}'',} \textit{ Phys. Rev. D} \textbf{ 99} (2019)
  112003,
  \href{http://dx.doi.org/10.1103/PhysRevD.99.112003}{\doi{10.1103/PhysRevD.99.112003}},
\href{http://www.arXiv.org/abs/1901.00174}{\texttt{ arXiv:1901.00174}}.

\bibitem{Sirunyan:2020sum}
\hrefCMSnoop {}{{CMS} Collaboration, ``{Measurements of $\mathrm{t\bar{t}}$H
  production and the CP structure of the Yukawa interaction between the Higgs
  boson and top quark in the diphoton decay channel}'',} \textit{ Phys. Rev.
  Lett.} \textbf{ 125} (2020) 061801,
  \href{http://dx.doi.org/10.1103/PhysRevLett.125.061801}{\doi{10.1103/PhysRevLett.125.061801}},
\href{http://www.arXiv.org/abs/2003.10866}{\texttt{ arXiv:2003.10866}}.

\bibitem{CMS:2021ugl}
\hrefCMSnoop {}{{CMS} Collaboration, ``{Measurements of production cross
  sections of the Higgs boson in the four-lepton final state in
  proton{\textendash}proton collisions at $\sqrt{s} = 13\,\text {Te}\text {V}
  $}'',} \textit{ Eur. Phys. J. C} \textbf{ 81} (2021) 488,
  \href{http://dx.doi.org/10.1140/epjc/s10052-021-09200-x}{\doi{10.1140/epjc/s10052-021-09200-x}},
  \href{http://www.arXiv.org/abs/2103.04956}{\texttt{ arXiv:2103.04956}}.

\bibitem{CMS:2021nnc}
\hrefCMSnoop {}{{CMS} Collaboration, ``{Constraints on anomalous Higgs boson
  couplings to vector bosons and fermions in its production and decay using the
  four-lepton final state}'',} \textit{ Phys. Rev. D} \textbf{ 104} (2021)
  052004,
  \href{http://dx.doi.org/10.1103/PhysRevD.104.052004}{\doi{10.1103/PhysRevD.104.052004}},
  \href{http://www.arXiv.org/abs/2104.12152}{\texttt{ arXiv:2104.12152}}.

\bibitem{CMS:2022ley}
\hrefCMSnoop {}{{CMS} Collaboration, ``{Measurement of the Higgs boson width
  and evidence of its off-shell contributions to ZZ production}'',} \textit{
  Nature Phys.} \textbf{ 18} (2022) 1329--1334,
  \href{http://dx.doi.org/10.1038/s41567-022-01682-0}{\doi{10.1038/s41567-022-01682-0}},
  \href{http://www.arXiv.org/abs/2202.06923}{\texttt{ arXiv:2202.06923}}.

\bibitem{CMS:2022uox}
\hrefCMSnoop {}{{CMS} Collaboration, ``{Constraints on anomalous Higgs boson
  couplings to vector bosons and fermions from the production of Higgs bosons
  using the {\ensuremath{\tau}}{\ensuremath{\tau}} final state}'',} \textit{
  Phys. Rev. D} \textbf{ 108} (2023) 032013,
  \href{http://dx.doi.org/10.1103/PhysRevD.108.032013}{\doi{10.1103/PhysRevD.108.032013}},
  \href{http://www.arXiv.org/abs/2205.05120}{\texttt{ arXiv:2205.05120}}.

\bibitem{CMS:2024eka}
\hrefCMSnoop {}{{CMS} Collaboration, ``{Measurement of the Higgs boson mass and
  width using the four-lepton final state in proton-proton collisions at
  s=13{\,}{\,}TeV}'',} \textit{ Phys. Rev. D} \textbf{ 111} (2025) 092014,
  \href{http://dx.doi.org/10.1103/PhysRevD.111.092014}{\doi{10.1103/PhysRevD.111.092014}},
  \href{http://www.arXiv.org/abs/2409.13663}{\texttt{ arXiv:2409.13663}}.

\bibitem{CMS:2025xkn}
\hrefCMSnoop {}{{CMS} Collaboration, ``{Search for {\ensuremath{\gamma}}H
  production and constraints on the Yukawa couplings of light quarks to the
  Higgs boson}'',} \textit{ Phys. Rev. D} \textbf{ 112} (2025) 112001,
  \href{http://dx.doi.org/10.1103/6s8n-pvmy}{\doi{10.1103/6s8n-pvmy}},
  \href{http://www.arXiv.org/abs/2502.05665}{\texttt{ arXiv:2502.05665}}.

\bibitem{CMS:2025fpt}
\hrefCMSnoop {}{{CMS} Collaboration, ``{Determination of the spin and parity of
  all-charm tetraquarks}'',} \textit{ Nature} \textbf{ 648} (2025) 58--63,
  \href{http://dx.doi.org/10.1038/s41586-025-09711-7}{\doi{10.1038/s41586-025-09711-7}},
  \href{http://www.arXiv.org/abs/2506.07944}{\texttt{ arXiv:2506.07944}}.

\bibitem{mela}
\hrefCMSnoop {}{``{The framework supporting JHUGen, MELA, MiLoMerge.}'',}.
  \url{https://spin.pha.jhu.edu}.

\bibitem{Alwall:2006yp}
\href
  {http://www.slac.stanford.edu/spires/find/hep/www?eprint=hep-ph/0609017}{J.~Alwall
  {et~al.}, ``{A standard format for Les Houches Event Files}'',} \textit{
  Comput. Phys. Commun.} \textbf{ 176} (2007) 300--304,
  \href{http://www.arXiv.org/abs/hep-ph/0609017}{\texttt{
  arXiv:hep-ph/0609017}}.

\bibitem{Sjostrand:2014zea}
T.~Sj{\"o}strand\hrefCMSnoop {}{ {et~al.}, ``An introduction to {PYTHIA}
  8.2'',} \textit{ Comput. Phys. Commun.} \textbf{ 191} (2015) 159,
  \href{http://dx.doi.org/10.1016/j.cpc.2015.01.024}{\doi{10.1016/j.cpc.2015.01.024}},
\href{http://www.arXiv.org/abs/1410.3012}{\texttt{ arXiv:1410.3012}}.

\bibitem{GEANT4:2002zbu}
\hrefCMSnoop {}{{GEANT4} Collaboration, ``{GEANT4--a simulation toolkit}'',}
  \textit{ Nucl. Instrum. Meth. A} \textbf{ 506} (2003) 250--303,
  \href{http://dx.doi.org/10.1016/S0168-9002(03)01368-8}{\doi{10.1016/S0168-9002(03)01368-8}}.

\bibitem{Buchmuller:1985jz}
\hrefCMSnoop {}{W.~Buchmuller and D.~Wyler, ``{Effective Lagrangian Analysis of
  New Interactions and Flavor Conservation}'',} \textit{ Nucl. Phys.} \textbf{
  B268} (1986) 621--653,
\href{http://dx.doi.org/10.1016/0550-3213(86)90262-2}{\doi{10.1016/0550-3213(86)90262-2}}.

\bibitem{Grzadkowski:2010es}
\hrefCMSnoop {}{B.~Grzadkowski, M.~Iskrzynski, M.~Misiak, and J.~Rosiek,
  ``{Dimension-Six Terms in the Standard Model Lagrangian}'',} \textit{ JHEP}
  \textbf{ 10} (2010) 085,
  \href{http://dx.doi.org/10.1007/JHEP10(2010)085}{\doi{10.1007/JHEP10(2010)085}},
\href{http://www.arXiv.org/abs/1008.4884}{\texttt{ arXiv:1008.4884}}.

\bibitem{Dedes:2017zog}
A.~Dedes\hrefCMSnoop {}{ {et~al.}, ``{Feynman rules for the Standard Model
  Effective Field Theory in R$_{\xi}$ -gauges}'',} \textit{ JHEP} \textbf{ 06}
  (2017) 143,
  \href{http://dx.doi.org/10.1007/JHEP06(2017)143}{\doi{10.1007/JHEP06(2017)143}},
  \href{http://www.arXiv.org/abs/1704.03888}{\texttt{ arXiv:1704.03888}}.

\bibitem{Brivio:2017vri}
\hrefCMSnoop {}{I.~Brivio and M.~Trott, ``{The Standard Model as an Effective
  Field Theory}'',} \textit{ Phys. Rept.} \textbf{ 793} (2019) 1--98,
  \href{http://dx.doi.org/10.1016/j.physrep.2018.11.002}{\doi{10.1016/j.physrep.2018.11.002}},
\href{http://www.arXiv.org/abs/1706.08945}{\texttt{ arXiv:1706.08945}}.

\bibitem{Falkowski:2015wza}
A.~Falkowski\hrefCMSnoop {}{ {et~al.}, ``{Rosetta: an operator basis translator
  for Standard Model effective field theory}'',} \textit{ Eur. Phys. J.}
  \textbf{ C75} (2015), no.~12, 583,
  \href{http://dx.doi.org/10.1140/epjc/s10052-015-3806-x}{\doi{10.1140/epjc/s10052-015-3806-x}},
\href{http://www.arXiv.org/abs/1508.05895}{\texttt{ arXiv:1508.05895}}.

\bibitem{Barducci:2025ati}
D.~Barducci\hrefCMSnoop {}{ {et~al.}, ``{Parametrisation and dictionary for CP
  violating Higgs boson interactions}'',}
  \href{http://www.arXiv.org/abs/2511.07905}{\texttt{ arXiv:2511.07905}}.

\bibitem{Kondo:1988yd}
\hrefCMSnoop {}{K.~Kondo, ``{Dynamical Likelihood Method for Reconstruction of
  Events With Missing Momentum. 1: Method and Toy Models}'',} \textit{ J. Phys.
  Soc. Jap.} \textbf{ 57} (1988) 4126--4140,
  \href{http://dx.doi.org/10.1143/JPSJ.57.4126}{\doi{10.1143/JPSJ.57.4126}}.

\bibitem{Atwood:1991ka}
\hrefCMSnoop {}{D.~Atwood and A.~Soni, ``{Analysis for magnetic moment and
  electric dipole moment form-factors of the top quark via $e^+ e^- \to t
  \bar{t}$}'',} \textit{ Phys. Rev. D} \textbf{ 45} (1992) 2405--2413,
  \href{http://dx.doi.org/10.1103/PhysRevD.45.2405}{\doi{10.1103/PhysRevD.45.2405}}.

\bibitem{Diehl:1993br}
\hrefCMSnoop {}{M.~Diehl and O.~Nachtmann, ``{Optimal observables for the
  measurement of three gauge boson couplings in $e^+ e^- \to W^+ W^-$}'',}
  \textit{ Z. Phys. C} \textbf{ 62} (1994) 397--412,
  \href{http://dx.doi.org/10.1007/BF01555899}{\doi{10.1007/BF01555899}}.

\bibitem{Brehmer:2018kdj}
\hrefCMSnoop {}{J.~Brehmer, K.~Cranmer, G.~Louppe, and J.~Pavez,
  ``{Constraining Effective Field Theories with Machine Learning}'',} \textit{
  Phys. Rev. Lett.} \textbf{ 121} (2018), no.~11, 111801,
  \href{http://dx.doi.org/10.1103/PhysRevLett.121.111801}{\doi{10.1103/PhysRevLett.121.111801}},
  \href{http://www.arXiv.org/abs/1805.00013}{\texttt{ arXiv:1805.00013}}.

\bibitem{Brehmer:2019bvj}
J.~Brehmer\hrefCMSnoop {}{ {et~al.}, ``{Effective LHC measurements with matrix
  elements and machine learning}'',} \textit{ J. Phys. Conf. Ser.} \textbf{
  1525} (2020) 012022,
  \href{http://dx.doi.org/10.1088/1742-6596/1525/1/012022}{\doi{10.1088/1742-6596/1525/1/012022}},
\href{http://www.arXiv.org/abs/1906.01578}{\texttt{ arXiv:1906.01578}}.

\bibitem{Neyman:1933wgr}
\hrefCMSnoop {}{J.~Neyman and E.~S. Pearson, ``{On the Problem of the Most
  Efficient Tests of Statistical Hypotheses}'',} \textit{ Phil. Trans. Roy.
  Soc. Lond. A} \textbf{ 231} (1933), no.~694-706, 289--337,
  \href{http://dx.doi.org/10.1098/rsta.1933.0009}{\doi{10.1098/rsta.1933.0009}}.

\bibitem{RM-753-PR}
J.~I. Marcum, ``A Statistical Theory of Target Detection by Pulsed Radar:
  Mathematical Appendix''.
\newblock RAND Corporation, Santa Monica, CA, 1948.

\bibitem{Peterson1954SignalDetectability}
\hrefCMSnoop {}{W.~W. Peterson, T.~G. Birdsall, and W.~C. Fox, ``The theory of
  signal detectability'',} \textit{ Transactions of the IRE Professional Group
  on Information Theory} \textbf{ 4} (1954) 171--212,
  \href{http://dx.doi.org/10.1109/TIT.1954.1057460}{\doi{10.1109/TIT.1954.1057460}}.

\bibitem{Bradley_ROC}
\hrefCMSnoop {}{A.~P. "Bradley, ``The use of the area under the ROC curve in
  the evaluation of machine learning algorithms'',} \textit{ Pattern
  Recognition} \textbf{ 30} (1997).

\bibitem{CMS:2024onh}
\hrefCMSnoop {}{{CMS} Collaboration, ``{The CMS Statistical Analysis and
  Combination Tool: Combine}'',} \textit{ Comput. Softw. Big Sci.} \textbf{ 8}
  (2024) 19,
  \href{http://dx.doi.org/10.1007/s41781-024-00121-4}{\doi{10.1007/s41781-024-00121-4}},
  \href{http://www.arXiv.org/abs/2404.06614}{\texttt{ arXiv:2404.06614}}.

\bibitem{Chen_2016}
\hrefCMSnoop {}{T.~Chen and C.~Guestrin, ``XGBoost: A Scalable Tree Boosting
  System'',} in \textit{ Proceedings of the 22nd ACM SIGKDD International
  Conference on Knowledge Discovery and Data Mining}, KDD '16, pp.~785--794.
\newblock ACM, August, 2016.
\newblock
  \href{http://dx.doi.org/10.1145/2939672.2939785}{\doi{10.1145/2939672.2939785}}.

\bibitem{TMVA}
A.~Hoecker\href {https://arxiv.org/abs/physics/0703039}{ {et~al.}, ``TMVA -
  Toolkit for Multivariate Data Analysis'',} 2009.

\end{thebibliography}
